\documentclass{aa}
\usepackage{bm}
\bibpunct{(}{)}{;}{a}{}{,} 
\usepackage{graphicx}
\usepackage{txfonts}
\def\ltb{$L_{0}$ -- $t_{\rm b}$ }

\begin{document} 
        
\titlerunning{High-redshift cosmography}
\authorrunning{Hu \& Wang}

   \title{High-redshift cosmography: Application and comparison with different methods}

   \author{J. P. Hu\inst{1} 
          \and F. Y. Wang\inst{1,2} 
          }

   \institute{School of Astronomy and Space Science, Nanjing University, Nanjing 210093, China\\
             \email{fayinwang@nju.edu.cn}
         \and
             Key Laboratory of Modern Astronomy and Astrophysics (Nanjing University), Ministry of Education, Nanjing 210093, China 
             }


 
  \abstract
 { Cosmography is used in cosmological data processing in order to constrain the kinematics of the universe in a model-independent way. In this paper, we first investigate the effect of the ultraviolet (UV) and X-ray relation of a quasar on cosmological constraints. By fitting the quasar relation and cosmographic parameters simultaneously, we find that the 4$\sigma$ deviation from the cosmological constant cold dark matter ($\Lambda$CDM) model disappears. Next, utilizing the Pantheon sample and 31 long gamma-ray bursts (LGRBs), we make a comparison among the different cosmographic expansions ($z$-redshift, $y$-redshift, $E(y)$, $\log(1+z)$, $\log(1+z)+k_{ij}$, and Pad$\rm \acute{e}$ approximations) with the third-order and fourth-order expansions. The expansion order can significantly affect the results, especially for the $y$-redshift method. Through analysis from the same sample, the lower-order expansion is preferable, except the $y$-redshift and $E(y)$ methods. For the $y$-redshift and $E(y)$ methods, despite adopting the same parameterization of $y=z/(1+z)$, the performance of the latter is better than that of the former. Logarithmic polynomials, $\log(1+z)$ and $\log(1+z) + k_{ij}$, perform significantly better than $z$-redshift, $y$-redshift, and $E(y)$ methods, but worse than Pad$\rm \acute{e}$ approximations. Finally, we comprehensively analyze the results obtained from different samples. We find that the Pad$\rm \acute{e}_{(2,1)}$ method is suitable for both low and high redshift cases. The Pad$\rm \acute{e}_{(2,2)}$ method performs well in a high-redshift situation. For the $y$-redshift and $E(y)$ methods, the only constraint on the first two parameters ($q_{0}$ and $j_{0}$) is reliable. }
\maketitle
%

\section{Introduction}\label{sec:intro}
Cosmography has been widely used to restrict the state of the kinematics of our universe employing measured distances \citep{2015CQGra..32m5007V,2016IJGMM..1330002D,2019IJMPD..2830016C,2020MNRAS.494.2576C}. It is a model-independent strategy which only relies on the assumption of a homogeneous and isotropic universe which is described by the Friedman-Lemaitre-Robertson-Walker (FLRW) metric well \citep{1972gcpa.book.....W}. Its methodology is essentially based on expanding a measurable cosmological quantity into the Taylor series around the present time. In the beginning, the luminosity distance was expanded into a $z$-redshift series which estimated the cosmic evolution at $z$ $\backsim$ 0 well, but it failed at high redshifts \citep{1998tx19.confE.276C,2004JCAP...09..009C,2004ApJ...607..665R,2004CQGra..21.2603V}. The poor performance at high redshifts of the $z$-redshift expansion strongly affects the results \citep{2010JCAP...03..005V}. \cite{2007CQGra..24.5985C} pointed out that the lack of validity of the Taylor expansions for the luminosity distance could settle down at approximately $z$ $\backsim$ 1. However, there are two main problems associated with the use of the above expansion and analysis of the cosmic data \citep{2020MNRAS.494.2576C}. One is the inability to distinguish the evolution of dark energy from the cosmological constant $\Lambda$. Higher precise redshift data found from future surveys will be helpful to address this problem \citep{2019PhRvD.100d4041D,2020JCAP...03..015B,2020FrASS...7....8L}. The other issue is the convergence problem caused by data that are far from the limits of Taylor expansions, leading to a severe error propagation, which reduces the cosmography predictions \citep{2007CQGra..24.5985C,2015PhRvD..92l3512B}. In order to solve this problem, several approaches have been proposed. According to the construction method, it can be divided into two categories. One would be using the auxiliary variables to reparameterize the redshift variable through functions of $z$, for example $y$-redshift \citep{2001IJMPD..10..213C,2003PhRvL..90i1301L} and $E(y)$ \citep{2020ApJ...900...70R}. The other way would be to consider a smooth evolution of the involved observables by expanding them in terms of rational approximation, for example Pad$\rm \acute{e}$ \citep{2014PhRvD..89j3506G,2014JCAP...01..045W,2020MNRAS.494.2576C} and Chebyshev rational polynomials \citep{2012JCAP...08..002S,2018MNRAS.476.3924C}.

In order to avoid the convergent problem of the series at high redshifts, it is better to replace the parameter $z$ with $y = z/(1+z)$ \citep{2001IJMPD..10..213C,2003PhRvL..90i1301L,2010JCAP...03..005V}. By making such a transformation, $z$ $\in$ (0, $\infty$) can be mapped into $y \in$ (0, 1). In theory, employing the $y$-redshift series method, high-redshift observations could be used for cosmology, that is, type Ia supernovae (SNe Ia), gamma-ray bursts (GRBs), and quasars. However, \cite{2017EPJC...77..434Z} found that both the $z$-redshift and $y$-redshift methods have a small biased estimation by employing the Joint Light-curve Analysis (JLA) sample \citep{2014A&A...568A..22B} and combining the bias-variance tradeoff. It means that Taylor expansion can describe the SNe Ia well. Whilst they also found that a $y$-redshift produces larger variances beyond the second-order expansion. Then based on the $y$-redshift expansion, \cite{2020ApJ...900...70R} attempted to recast $E(z)$ as a function of $y= z/(1+z)$ and they adopted the new series expansion of the $E(y)$ function to compare dark energy models. From the numerical results, they found that the restriction on two of the cosmological parameters, deceleration parameter $q_0$ and jerk parameter $j_0$, are much tighter than that of other two parameters $s_0$ and $l_0$. This point is in line with the previous claim of \cite{2017EPJC...77..434Z}. Thus replacing a $z$-redshift with a $y$-redshift may only alleviate the divergence when $z$ tends to infinity and it could not completely solve the convergence problem. In recent works \citep{2019A&A...628L...4L,2019NatAs...3..272R}, a new expansion of the luminosity distance-redshift relation in terms of the logarithmic polynomials has been used for cosmography. In such a way, a $z$-redshift was replaced by $\log(1+z)$. Employing the logarithmic polynomials, \cite{2019NatAs...3..272R} found a ~4$\sigma$ tension with the $\Lambda$CDM model from a high-redshift Hubble diagram of SNe Ia and quasars. \cite{2019A&A...628L...4L} confirmed this tension with SNe Ia, GRBs and quasars using two model-independent methods ($y$-redshift and the logarithmic polynomials). After that, \cite{2021A&A...649A..65B} improved a new logarithmic polynomial expansion by introducing orthogonal terms. This new method and the corresponding results have received enough attention \citep{2021PhRvD.103h3526S}. At the same time, different opinions were voiced \citep{2020PhRvD.102l3532Y,2021PhLB..81836366B}. 

Involving the rational approximation, the Pad$\rm \acute{e}$ approximation being the most usual rational approximation method can be regarded as a generalization of the Taylor polynomial. However, it often gives a better approximation of the function than truncating its Taylor series, and it may still work where the Taylar series do not converge \citep{2014JCAP...01..045W}. Until now, due to its excellent convergence properties, the Pad$\rm \acute{e}$ approximation has been widely used in cosmology \citep{2014PhRvD..90d3531A,2017A&A...598A.113D,2019MNRAS.484.4484C,2020MNRAS.494.2576C}. Recently, \citet{2020MNRAS.494.2576C} critically compared the two main solutions of the convergence problem for the first time, that is to say the auxiliary variables ($y_1 = 1 - a$ and $y_2 = \arctan(a^{-1} -1 )$) and the Pad$\rm \acute{e}$ approximations. They found that even though $y_2$ overcomes the issues of $y_1$, the most viable approximation of the luminosity distance $d_{L}(z)$ is still given by adopting Pad$\rm \acute{e}$ approximations. In addition, they also investigated two distinct domains involving Monte Carlo analysis on the Pantheon sample, $H(z)$, and shift parameter measurements, and they conclude that the (2,1) Pad$\rm \acute{e}$ approximation is statistically the optimal approach to explain low- and high-redshift data, together with the fifth-order $y_{2}$-parameterization. At high redshifts, the (2,2) Pad$\rm \acute{e}$ approximation can be essentially ruled out. In addition to the Pad$\rm \acute{e}$ approximation, there are some other methods that can be used in cosmology instead of Taylor expansions, for example Chebyshev series \citep{2018MNRAS.476.3924C,2020JCAP...12..007Z}. Until now, there has not been a detailed numerical analysis for the abovementioned six methods ($z$-redshift, $y$-redshift, $E(y)$, logarithmic polynomials, orthogonalized logarithmic polynomials, and Pad$\rm \acute{e}$ approximations) using high-redshift data.  

In this paper, we try to confirm the effect of the quasar relation on the cosmographic constraint using the Pantheon sample and 1598 quasars in terms of the logarithmic polynomials and the Pad$\rm \acute{e}_{(2,1)}$ approximation. First, we reconsider the parameters $\gamma$, $\beta$, and $\delta$, which were obtained from the observed quasar relation as free parameters, to explore the situation of 4$\sigma$ tension. Based on previous research on the Pad$\rm \acute{e}$ approximation \citep{2020MNRAS.494.2576C}, we chose to use the Pad$\rm \acute{e}_{(2,1)}$ method which has excellent convergence to carry out cross checks. A brief introduction to Pad$\rm \acute{e}$'s excellent convergence has been provided in the previous section. We also compare the Pad$\rm \acute{e}_{(2,1)}$ method with other cosmographic expansions including the logarithmic polynomials combining the Pantheon sample and a new high-redsihft GRB sample including 31 LGRBs \citep{2022ApJ...924...97W}. We also provide a detailed numerical analysis for the different expansions in terms of the values for the Akaike information criterion (AIC) and the Bayesian information criterion (BIC). A fiducial value $H_{0}$ = 70 km s$^{-1}$ Mpc$^{-1}$ is adopted in this paper.

The outline of this paper is as follows. In Sect. 2, we describe the observational data sets that are used in our analyses. Sect. 3 briefly introduces the different cosmographic approaches and the model selection method. We discuss the effect of the quasar relation on the cosmographic constraints in Sect. 4. The comparison results of different cosmographic approaches are depicted in Sect. 5. Finally, conclusions and discussion are presented in the last section.

\section{Observational data sets}

\subsection{Pantheon}. This sample consists 1048 sources covering the redshift range 0.01 < $z$ < 2.26 \citep{2018ApJ...859..101S}. This sample, as a set of latest SNe Ia data points, was provided by a number of SN surveys including CfA1-4 \citep{1999AJ....117..707R,2006AJ....131..527J,2009ApJ...700..331H,2009ApJ...700.1097H,2012ApJS..200...12H}, the Carnegie supernova (SN) project \citep[CSP;][] {2010AJ....139..519C,2010AJ....139..120F,2011AJ....142..156S}, the Sloan Digital Sky Survey \citep[SDSS;][]{2008AJ....135..338F,2009ApJS..185...32K,2018PASP..130f4002S}, the Pan-STARRS1 \citep[PS1;][]{2014ApJ...795...44R,2014ApJ...795...45S}, the supernova legacy survey \citep[SNLS;][]{2011ApJS..192....1C,2011ApJ...737..102S}, and the Hubble Space Telescope \citep[HST;][]{2004ApJ...607..665R,2007ApJ...659...98R,2012ApJ...746...85S,2014ApJ...783...28G,2014AJ....148...13R,2018ApJ...853..126R} SN surveys. According to the redshift range, these surveys have overlapping redshift ranges of 0.01 $\lesssim$ $z$ $\lesssim$ 0.10 for CfA1-4 and CSP; 0.1 $\lesssim$ $z$ $\lesssim$ 0.4 for SDSS; 0.103 $\lesssim$ $z$ $\lesssim$ 0.68 for PS1; 0.3 $\lesssim$ $z$ $\lesssim$ 1.1 for SNLS; and 1.0 $<$ $z$ for HST. We note that SNe Ia have a nearly uniform intrinsic luminosity with an absolute magnitude around $M$ $\sim$ -19.5 \citep{2001LRR.....4....1C} which promote it to a well-established class of {standard candles}. The difficulty in the cosmological applications lies in the identification of absolute magnitude $M$, due to different sources of systematic and statistical errors. In this work, we only consider the systematic error. The observations used, redshift $z$, distance modulus $\mu$, and corresponding 1$\sigma$ error $\sigma_{\mu}$ are provided by \citet{2018ApJ...859..101S} and can be publicly obtained from the website\footnote{https://github.com/dscolnic/Pantheon/}. The observational distance modulus $\mu_{obs}$(SN) was calculated with the following formula \citep{1998A&A...331..815T},
\begin{equation}
        \label{eq:SNobs}
        \mu_{obs}(SN) = m_{B} - M + \alpha x_{1} - \beta c + \Delta_{M} + \Delta_{B},
\end{equation}  
where $\alpha$ is the coefficient of the relation between the luminosity and stretch, and $\beta$ is the coefficient of the relation between the luminosity and color. These two parameters were retrieved by using the beams with bias correction (BBC) method. We note that $x_{1}$ and $c$ are the light-curve shape parameter and the color of the SN, respectively; $\Delta_{M}$ is a distance correction based on the host galaxy mass of SN; and $\Delta_{B}$ is a distance correlation based on predicted biases from simulations. Furthermore, $M$ is the absolute B-band magnitude of a fiducial SN Ia with $x_{1}$ = 0 and $c$ = 0. Generally, a standard marginalization over $M$ was performed \citep{2001LRR.....4....1C,2014A&A...568A..22B,2018ApJ...859..101S}.

\subsection{GRBs}. Gamma-ray bursts are promising high-redshift probes \citep[i.e.,][]{2015NewAR..67....1W,Wang2016,Wei2017,2021MNRAS.tmp.3230C,Demianski2021,2021MNRAS.507..730H}. This sample is a compilation of 31 LGRBs ranging from 1.45 $<$ $z$ $<$ 5.91. Similar to SNe Ia, this sample shows a plateau phase caused by the same physical mechanism \citep{2022ApJ...924...97W}. \citet{2022ApJ...924...97W} consider a special case in which the energy injection from electromagnetic dipole emission of millisecond magnetars is larger than the external shock emission. The corresponding light curves of X-ray afterglow show a plateau with a constant luminosity followed by a decay index of about -2 \citep{1998A&A...333L..87D,2001ApJ...552L..35Z}. The relation between the luminosity $L_{0}$ and end time $t_{b}$ of the plateau can be well described by the correlation  
\begin{equation}
        \label{eq:logL}
        \log \left (\frac{L_0}{10^{47}~\rm erg/s} \right) = k_{1} \times \log \frac{t_{b}}{10^3(1+z)~ \rm s}  + b_{1},
\end{equation}
where $L_{0}$ is the plateau luminosity, $t_{b}$ is the end time of the plateau, and $k_{1}$ and $b_{1}$ are two free parameters. Due to the lack of GRBs at low redshifts, the Hubble parameter data constructed by \citet{2018ApJ...856....3Y} were used to calibrate the \ltb correlation via the Gaussian process (GP) method. The calibrated results of the \ltb correlation are $k_{1}$=-1.02$\pm0.12$ and $b_{1}$ = 1.69$\pm$0.13. Considering the systematic error $\sigma_{int}$ = 0.22 for the GRB Gold sample, \citet{2022ApJ...924...97W} used the calibrated \ltb correlation to derive the observational distance modulus $\mu_{obs}$ and the corresponding errors. In this work, we directly used the results in Table 2 of \citet{2022ApJ...924...97W}. We would like to specify that $\mu_{obs}$(GRB) and its uncertainty can be derived from \citep{2022ApJ...924...97W}
\begin{eqnarray}
        \label{grb:muobs}
        \mu_{obs}(GRB) &=& \frac{5}{2} (\log{L_0} - \log \frac{4\pi F_0}{(1+z)^{1-\beta}} - 24.49) + 25,
\end{eqnarray}
and
\begin{eqnarray}
        \label{grb:muerr}
        \sigma_{obs} &=& \frac{5}{2} ((\log^{2} (\frac{t_{b}}{1+z}) -3) \sigma_{k_{1}}^{2} + k_{1}^{2}  (\frac{\sigma_{t_{b}}}{t_{b} \ln{10}})^{2} \nonumber \\
        &+& \sigma_{b_{1}}^{2}+ (\frac{\sigma_{F_0}}{F_0 \ln{10}})^{2} + \sigma_{int}^{2})^{1/2}.
\end{eqnarray}

\subsection{Quasars}. The quasar sample consists of 1598 objects in the redshift range (0.04, 5.1) with high-quality UV and X-ray flux measurements. The quasar relation between the UV and X-ray emission of quasars can be employed to turn quasars into standard candles, which can be easily described by a formula $\log{L_{\rm X}}$ = $\gamma$$log{L_{\rm UV}}$ + $\beta$, where $\gamma$ and $\beta$ are two free parameters. This quasar relation has been found from optically and X-ray active galactic nucleus samples with the slope parameter $\gamma$ around 0.5 to 0.7 \citep{2003AJ....125..433V,2005AJ....130..387S,2006AJ....131.2826S,2007ApJ...665.1004J,2009ApJ...690..644G,2009ApJS..183...17Y,2010ApJ...708.1388Y,2012MNRAS.422.3268J,2015ApJ...815...33R,2016ApJ...819..154L,2021EPJC...81..948Z}. More descriptions about adopting the UV and X-ray correlation of quasars to constrain cosmological parameters can be found in Sect. \ref{sec:444}. Due to the lack of quasars at low redshifts, the calibration of parameters $\gamma$ and $\beta$ uses other low-redshift observations, that is SNe Ia. Here, we regard them as free parameters and fit them simultaneously with cosmological parameters. The values for $z$, $F_{UV}$, $F_{X}$, and $\sigma_{F_{UV}}$ provided by \citet{2019NatAs...3..272R} are used. 

\begin{figure}[htbp]
        \centering
        \includegraphics[width=0.4\textwidth,angle=0]{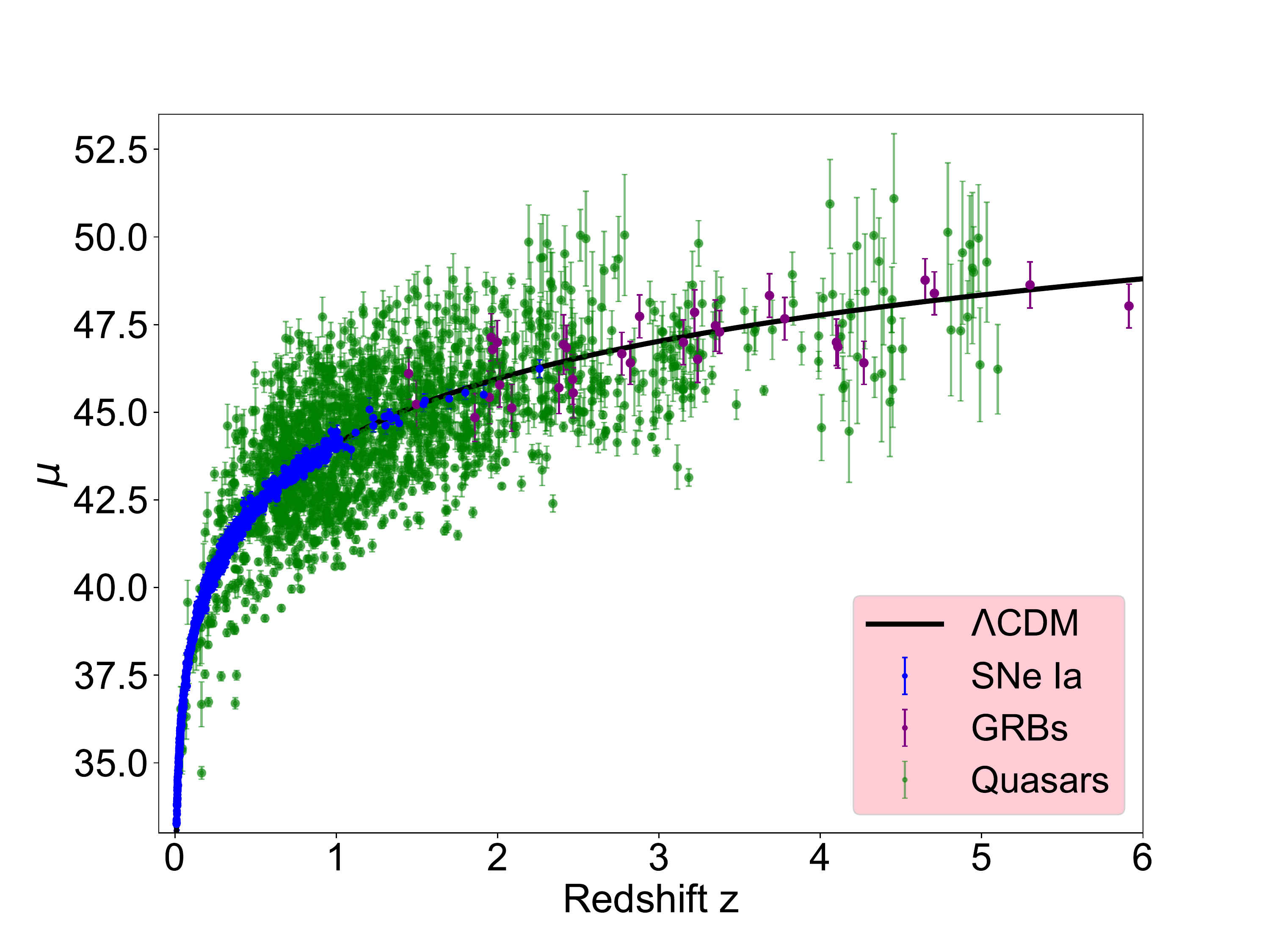}
        \caption{Hubble diagram of SNe Ia, GRBs, and quasars. The picture includes 1048 SNe Ia of the Pantheon sample, 31 LGRBs, and 1598 quasars. The black solid line is a flat $\Lambda$CDM model with $\Omega_{m}$=0.3 and Hubble constant $H_{0}$ = 70 km s$^{-1}$ Mpc$^{-1}$. Distance moduli and corresponding errors are provided by \citet{2019NatAs...3..272R} which were calculated by the calibrated UV and X-ray correlation using the low-redshift SNe Ia.}
        \label{F0}       
\end{figure}

We plot the Hubble diagram of SNe Ia, GRBs, and quasars, as shown in Fig. \ref{F0}. In discussing the impact of the quasar relation on cosmological constraints, we used the same sample of \citet{2019NatAs...3..272R}, including SNe Ia and quasars, referred to as the SN-Q sample. When comparing different cosmological expansion methods, we use the combined sample. The maximum redshift of 31 LGRBs is higher than that of the Pantheon sample. Therefore, compared to the Pantheon sample, the combined sample can be regarded as a high-redshift sample. 

\section{Method}\label{sec:method}
In this section, we briefly introduce the different  cosmographic techniques and the model selection method. Cosmography is an artful combination of kinematic parameters by the Taylor expansion with an assumption of large-scale homogeneity and isotropy. In this framework, the evolution of the universe can be described by some cosmographic parameters, such as Hubble parameter $H$, deceleration $q$, jerk $j$, snap $s$, and lerk $l$ parameters. The definition of them can be expressed as follows:
\begin{eqnarray}
        \label{eq:q0j0}
        H = \frac{\dot{a}}{a}, q = -\frac{1}{H^{2}}\frac{\ddot{a}}{a}, j=\frac{1}{H^3}\frac{\dot{\ddot{a}}}{a},
        s=\frac{1}{H^4}\frac{\ddot{\ddot{a}}}{a}, l=\frac{1}{H^5}\frac{\dot{\ddot{\ddot{a}}}}{a}\label{j}.
\end{eqnarray} 

Initially, the luminosity distance is expanded as $z$-redshift series \citep{1972gcpa.book.....W,1998tx19.confE.276C,2004CQGra..21.2603V}. The $z$-redshift method is completely applicable at $z \ll 1$. As the observed redshift increases, the $z$-redshift exhibits poor behavior which prompts researchers to find new ways to this problem. Many cosmographic approaches have been proposed in order to obtain as much information as possible from higher-redshift observations directly, among which the representative ones are $y$-redshift, $E(y)$, $\log(1+z)$, $\log(1+z)$+$k_{ij}$, and Pad$\rm \acute{e}$ approximations. These methods have been widely used in cosmological applications \citep{2009A&A...507...53W,2014MNRAS.443.1680W,2014JCAP...01..045W,2017A&A...598A.113D,2019MNRAS.484.4484C,2019EPJC...79..698Y,2020MNRAS.494.2576C,2020MNRAS.491.4960L,2020ApJ...900...70R,2022PhRvD.105b1301C}. Next, we briefly introduction these methods.

We first introduce the standard model as a guide. Considering the flat $\Lambda$CDM model, the luminosity distance $d_{L}$ can be calculated from
\begin{equation}
        d_{L} = \frac{c(1+z)}{H_{0}} \int_{0}^{z} 
        \frac{dz'}{\sqrt{\Omega_{m} (1+z')^{3} + (1-\Omega_{m})}},
        \label{q2}
\end{equation}
where $c$ is the speed of light, $H_{0}$ is the Hubble constant, and $\Omega_{m}$ is the matter density. 

\textbf{$z$-redshift}. The $z$-redshift method is the earliest Taylor series used in cosmology. The luminosity distance can be conveniently expressed as 
\begin{eqnarray}
        \label{eq:dlz}
        d_{L}(z) &=& \frac{c}{H_{0}}[z + \frac{1}{2} (1-q_{0})z^{2} -\frac{1}{6}(1-q_{0}-3q_{0}^{2}+j_{0})z^{3} \nonumber \\
        &+& \frac{1}{24}(2-2q_{0}-15q_{0}^{2}-15q_{0}^{3}+5j_{0}+10q_{0}j_{0}+s_{0})z^{4}\nonumber \\
        &+& \frac{1}{120}(-6+6q_{0}+81q_{0}^{2}+165q_{0}^{3}+105q_{0}^{4}+10j_{0}^{2}\nonumber \\
        &-&27j_{0}-110q_{0}j_{0}-105q_{0}^{2}j_{0}-15q_{0}s_{0}-11s_{0}-l_{0})z^{5}\nonumber \\
        &+&\textit{O}(z^6)],
\end{eqnarray}
\citep{2007CQGra..24.5985C,2011PhRvD..84l4061C}, where $H_{0}$, $q_{0}$, $j_{0}$, $s_{0}$, and $l_{0}$ are the current values. The first two terms above are Weinberg's version of the Hubble law which can be found from equation (14.6.8) in the book by \citet{1972gcpa.book.....W}. The third term and the fourth term are equivalent to that obtained by \citet{1998tx19.confE.276C} and \citet{2004CQGra..21.2603V}, respectively.     

\textbf{$y$-redshift}. Mathematically, $z$-redshift expansion should be performed near a small quantity, that is the low redshift $z \backsim$ 0. At that time, with the increase in SN data, the highest redshift was already greater than 1.00. \citet{2007CQGra..24.5985C} pointed out that the use of the $z$-redshift for $z > 1$ is likely to lead to a convergent problem, that is to say any Taylor series in $z$ is guaranteed to diverge for $z > 1$. In order to relieve this mathematical problem, they introduced an improved parameterization $y = z/(1+z)$. The luminosity distance was written as follows:                      
\begin{eqnarray}
        \label{eq:dly}
        d_{L}(y) &=& \frac{c}{H_{0}}[y - \frac{1}{2}(q_{0} - 3)y^{2} + \frac{1}{6}(11-5q_{0} + 3q_{0}^{2} - j_{0})y^{3}\nonumber \\
        &+& \frac{1}{24}(50 - 7j_{0}- 26q_{0} + 10q_{0}j_{0} + 21 q_{0}^{2} - 15q_{0}^{3} + s_{0})y^{4} \nonumber \\
        &+& \frac{1}{120}(274 - 154q_{0}+141q_{0}^{2}- 135q_{0}^{3} + 105q_{0}^{4} - 47j_{0} \nonumber \\
        &+& 10j_{0}^{2}+90q_{0}j_{0}-105q_{0}^{2}j_{0}-15q_{0}s_{0}+9s_{0}-l_{0})y^5 \nonumber \\
        &+&\textit{O}(y^6)].
\end{eqnarray}

\bm{$E(y)$}. Also using the improved parameterization $y = z/(1+z)$, \citet{2020ApJ...900...70R} reconstructed the $E(z)$ as a function of $y$-redshift. The form of the function $E(z)$ can be described as follows:
\begin{eqnarray}
        \label{eq:Ey}
        E(y) &=& 1 + (1+q_{0})y+\frac{1}{2}(2-q_{0}^{2}+2q_{0}+j_{0})y^{2}\nonumber \\
        &+& \frac{1}{6}(6+3q_{0}^{3}-3q_{0}^{2}+6q_{0}-4q_{0}j_{0}+3j_{0}-s_{0})y^{3} \nonumber \\
        &+& \frac{1}{24}(-15q_{0}^{4}+12q_{0}^{3}+25q_{0}^{2}j_{0} +7q_{0}s_{0}-4j_{0}^{2}-16q_{0}j_{0} \nonumber \\
        &-& 12q_{0}^{2}+l_{0}-4s_{0}+12j_{0}+24q_{0}+24)y^{4} \nonumber \\
        &+&\textit{O}(y^5).
\end{eqnarray}
The corresponding luminosity distance can be obtained by using follow formula: 
\begin{eqnarray}
        \label{eq:dlEy}
        d_{L}(z) = \frac{c(1+z)}{H_{0}} \int_{0}^{z} \frac{dz'}{E(y)}.
\end{eqnarray}

\begin{figure*}
        \centering
        \includegraphics[width=0.8\textwidth,angle=0]{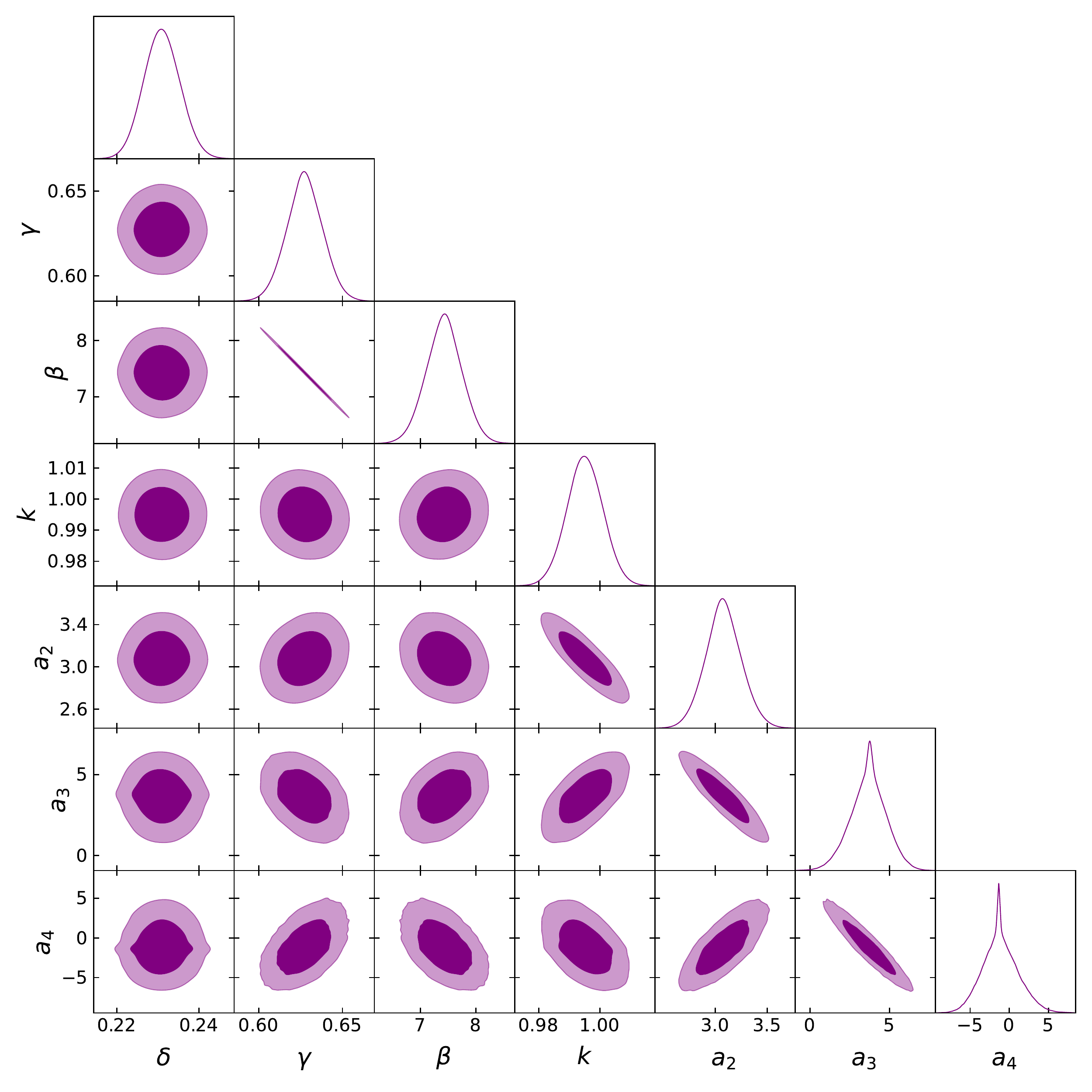}
        \caption{Confidence contours ($1\sigma$ and $2\sigma$) for the parameters space ($\delta$, $\gamma$, $\beta$, $k$, $a_2$, $a_3$, and $a_4$) from the SN-Q sample utilizing the $\log(1+z)$ method.}
        \label{F2}       
\end{figure*}

\textbf{Pad$\rm \acute{e}$ polynomials}. The Pad$\rm \acute{e}$ polynomials \citep{1992A} was built up from the standard Taylor definition and can be used to lower divergences at $z \geq$ 1. It often gives a better approximation for the function than truncating its Taylor series, and it may still work where the Taylar series does not converge \citep{2014JCAP...01..045W}. Due to its excellent convergence properties, the Pad$\rm \acute{e}$ polynomials have been considered at high redshifts in cosmography \citep{2014PhRvD..89j3506G,2014JCAP...01..045W,2017A&A...598A.113D,2018JCAP...05..008C,2020MNRAS.494.2576C}. The Taylor expansion of a generic function $f(z)$ can be described by a given function $f(z) = \int_{i=0}^{\infty} c_{i}z^{i}$, where $c_{i} = f^{i}(0)/i!$, which is approximated by means of a $(n, m)$ Pad$\rm \acute{e}$ approximation by the radio polynomial \citep{2020MNRAS.494.2576C}  
\begin{eqnarray}
        \label{eq:Pnm}
        P_{n, m}(z) = \frac{\sum_{i = 0}^{n} a_{i}z^{i}}{1 + \sum_{j = 1}^{m} b_{j}z^{j}};\end{eqnarray} 
there are a total $(n + m + 1)$ number of independent coefficients. In the numerator, we have $n + 1$ independent coefficients, whereas in the denominator there is $m$. Since, by construction,  $b_{0} = 1$ is required, we have
\begin{eqnarray}
        \label{eq:ff}
        f(z) -P_{n , m} (z) = \textit{O}(z^{n+m+1}). 
\end{eqnarray}
The coefficients $b_{i}$ in Eq. (\ref{eq:Pnm}) were thus determined by solving the follow homogeneous system of linear equations \citep{1993Litvinov}:
\begin{eqnarray}
        \label{eq:bb0}
        \sum_{j=1}^{m} b_{j}c_{n+k+j} = -b_{0}c_{n+k},
\end{eqnarray} 
which is valid for $k = 1,...,m$. All coefficients $a_{i}$ in Eq. (\ref{eq:Pnm}) can be computed using the formula
\begin{eqnarray}
        \label{eq:ai}
        a_{i} = \sum_{k=0}^{i} b_{i-k}c_{n+k}. 
\end{eqnarray} 
In terms of the investigations of \citet{2020MNRAS.494.2576C} on the Pad$\rm \acute{e}$ polynomials, we finally chose to use the Pad$\rm \acute{e}$$_{(2,1)}$ approximation and Pad$\rm \acute{e}$$_{(2,2)}$ approximation to represent the third-order polynomials and the fourth-order polynomials, respectively. More detailed information about the selections of the specific polynomials can be found in \citet{2020MNRAS.494.2576C}. We subsequently provide the corresponding luminosity distances \citep{2020MNRAS.494.2576C}: \\
(1) Pad$\rm \acute{e}$$_{(2,1)}$: 
\begin{eqnarray}
        \label{eq:dlp21}
        d_{L}(z) = \frac{c}{H_{0}}[ \frac{z(6(-1+q_{0})+(-5-2j_{0}+q_{0}(8+3q_{0}))z)}{-2(3+z+j_{0}z)+2q_{0}(3+z+3q_{0}z)}].
\end{eqnarray}
(2) Pad$\rm \acute{e}$$_{(2,2)}$: 
\begin{eqnarray}
        \label{eq:dlp22}
        d_{L}(z) &=& \frac{c}{H_0}[6z(10 + 9 z - 6 q_0^3 z + s_0 z - 2 q_0^2 (3 + 7 z) \nonumber \\
        &-& q_0 (16 + 19 z) +
        j_0 (4 + (9 + 6 q_0) z))\Big/(60 + 24 z \nonumber \\
        &+& 6 s_0 z - 2 z^2     + 4 j_0^2 z^2 - 9 q_0^4 z^2 - 3 s_0 z^2 \nonumber \\
        &+& 6 q_0^3 z (-9 + 4 z) + q_0^2 (-36 - 114 z + 19 z^2)  \nonumber \\
        &+&j_0 (24 + 6 (7 + 8 q_0) z + (-7 - 23 q_0 + 6 q_0^2) z^2) \nonumber \\
        &+&  q_0 (-96 - 36 z + (4 + 3 s_0) z^2))].
\end{eqnarray}

\bm{$\log(1+z)$}. 
The logarithmic polynomials, $\log(1+z)$, were recently proposed by \citet{2019NatAs...3..272R} and then adopted to test the flat $\Lambda $CDM tension \citep{2019A&A...628L...4L,2020PhRvD.102l3532Y}. The new form of the relation between the luminosity distance and the redshift should be written as
\begin{eqnarray}
        \label{eq:dllog}
        d_{L}(z) &=& k\ln(10)\frac{c}{H_{0}}\times[\log(1+z)+a_{2}\log^{2}(1+z) \nonumber \\
        &+&a_{3}\log^{3}(1+z)+a_{4}\log^{4}(1+z)] + \textit{O}[\log^{5}(1+z)]; \nonumber \\
\end{eqnarray}
here, $k$, $a_2$, $a_3$, and $a_4$ are free parameters. 

\bm{$\log(1+z) + k_{ij}$}. According to the previous logarithmic polynomials, \citet{2021A&A...649A..65B} modified the expression of the luminosity distance to make cosmograpgic coefficients with no covariance. They named this method orthogonalized logarithmic polynomials. The modified expression of the luminosity distance is
\begin{eqnarray}
        \label{eq:dllogk}
        d_{L}(z) &=& k\ln(10)\frac{c}{H_{0}}\times\{\log(1+z)+a_{2}\log^{2}(1+z) \nonumber \\
        &+&a_{3}[k_{32}\log^{2}(1+z)+\log^{3}(1+z)]\nonumber \\
        &+&a_{4}[k_{42}\log^{2}(1+z)+k_{43}\log^{3}(1+z)+\log^{4}(1+z)]\} \nonumber \\
        &+& \textit{O}[\log^{5}(1+z)], \nonumber \\
\end{eqnarray}
where the coefficients $k_{ij}$ are not free parameters and they are associated with our used data set. They were determined through the procedure described in Appendix A of \citet{2021A&A...649A..65B}. 

\subsection{Statistical analysis and selection criteria}
Using the luminosity distance derived from the different cosmographic techniques, we can obtain the corresponding theoretical distance modulus 
\begin{eqnarray}
        \mu_{th} = 5 \log_{10} \frac{d_{\rm L}}{\textnormal{Mpc}} + 25.
        \label{eq:mu}
\end{eqnarray}
The best fitting values of the free parameters are achieved by minimizing the value of
$\chi^{2}$,
\begin{eqnarray}
        \chi^{2} = \sum_{i=1}^{N}\frac{(\mu_{\rm obs}(z_{i}) - \mu_{\rm th}(P_{i},z_{i}))^{2}}{\sigma_{i}^{2}},
        \label{eq:chi}
\end{eqnarray}
where $\sigma_{i}(z_{i})$ represents the observational uncertainties of the distance modulus and $P_{i}$ represents the free parameters to be fitted. For the Pantheon sample, the parameter $\sigma_{i}$ only includes the statistical error. In the GRBs sample, the parameter $\sigma_{i}$ not only contains statistical errors but also takes the intrinsic dispersion $\sigma_{int}$ of the \ltb correlation into account. For the combination of different datasets, the corresponding $\chi_{Total}^{2}$ is equal to the sum of $\chi^{2}$ of each dataset. Using the combination of SNe Ia and GRBs as an example, the $\chi_{Total}^{2}$ is  
\begin{eqnarray}
        \chi_{\rm Total}^{2} = \chi_{\rm SNe Ia}^{2}+\chi_{\rm GRBs}^{2}.
        \label{eq:chi_2}
\end{eqnarray}

The likelihood analysis was performed employing a Bayesian Monte Carlo Markov Chain (MCMC) \citep{2013PASP..125..306F} method with the $emcee$\footnote{https://emcee.readthedocs.io/en/stable/} package. We used the $getdist$ package \citep{2019arXiv191013970L} to plot the MCMC samples.


\begin{figure*}
        \centering
        \includegraphics[width=0.34\textwidth]{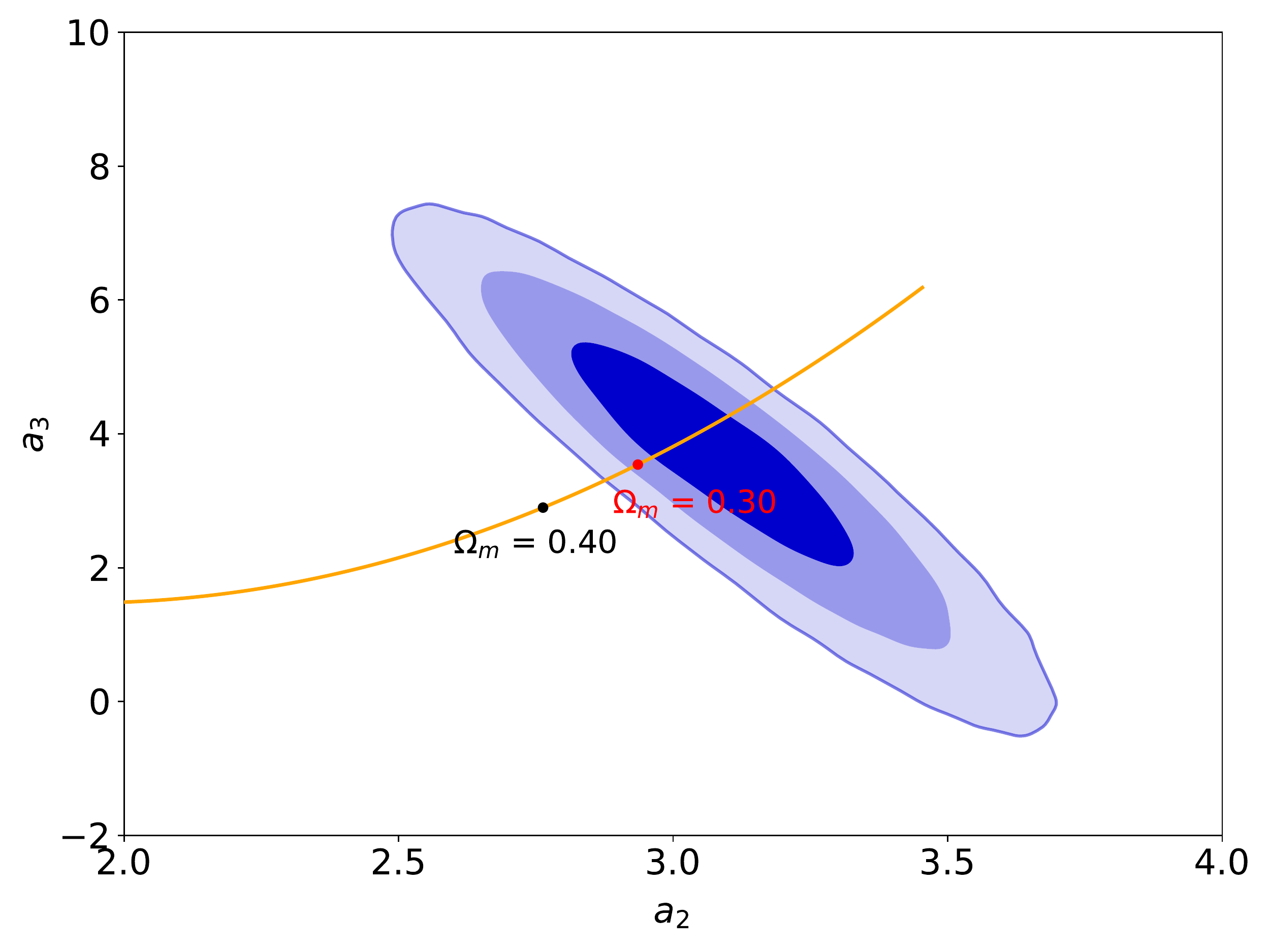}
        \includegraphics[width=0.34\textwidth]{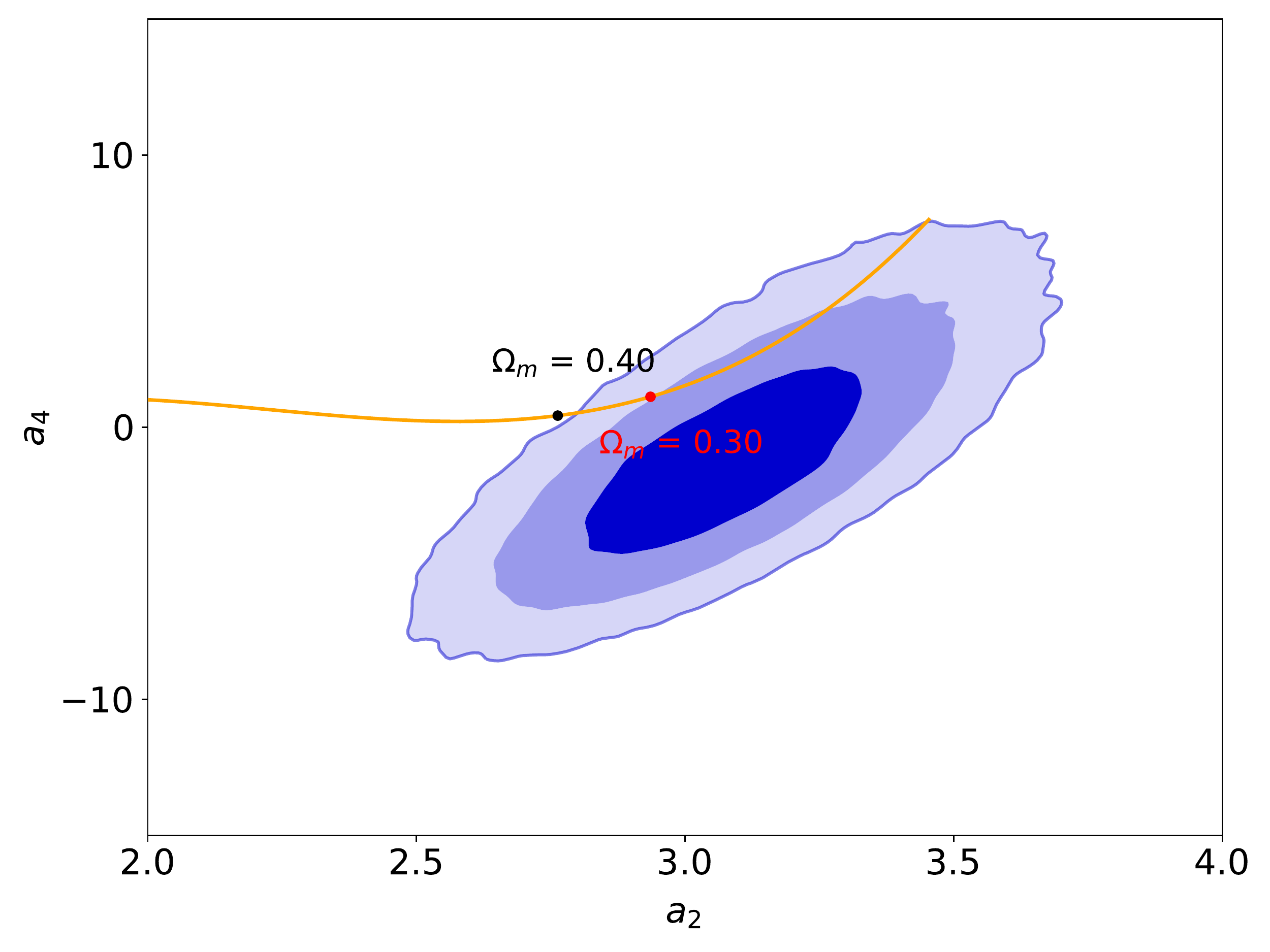}

        \includegraphics[width=0.34\textwidth]{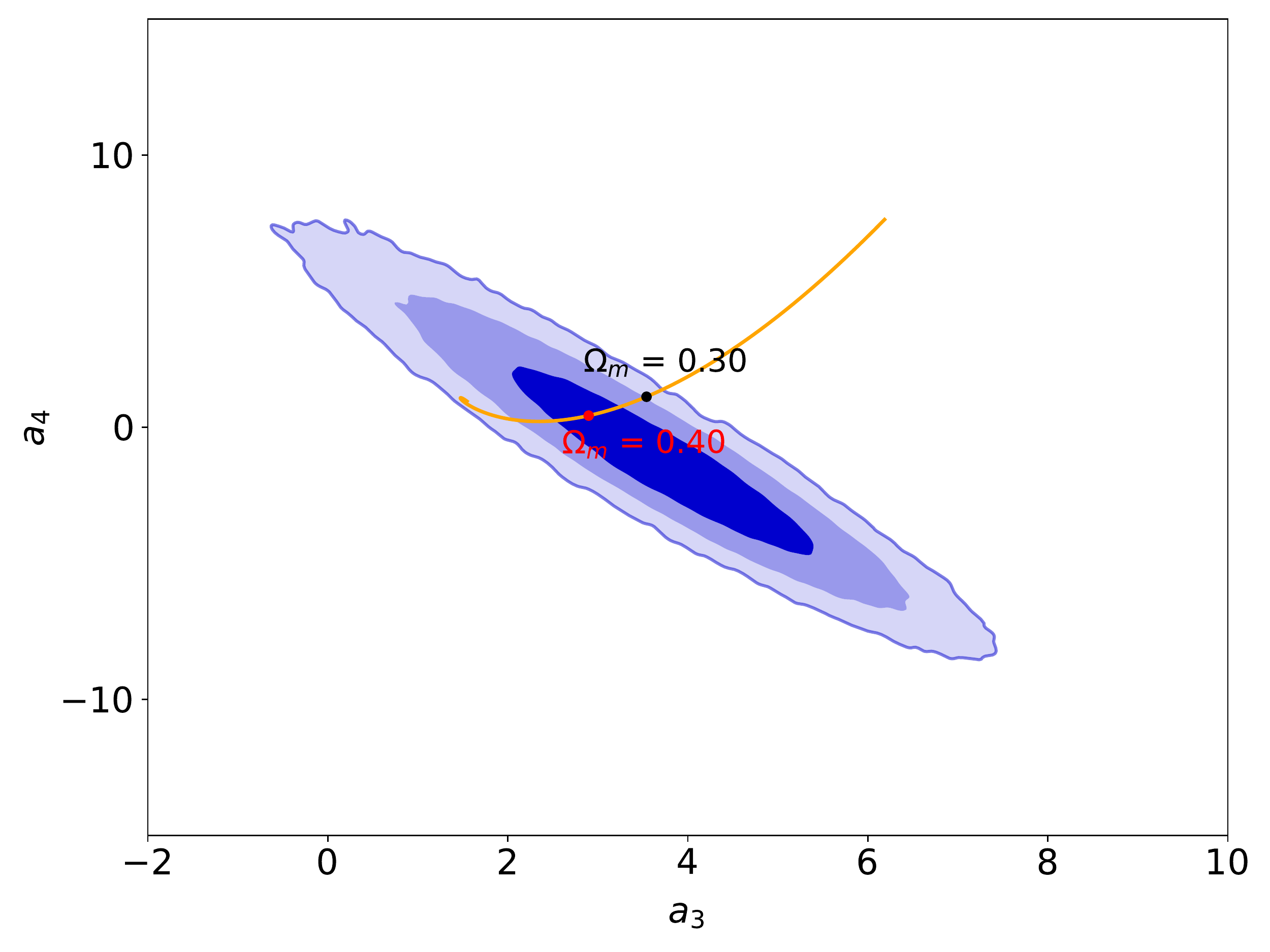}
        \includegraphics[width=0.34\textwidth]{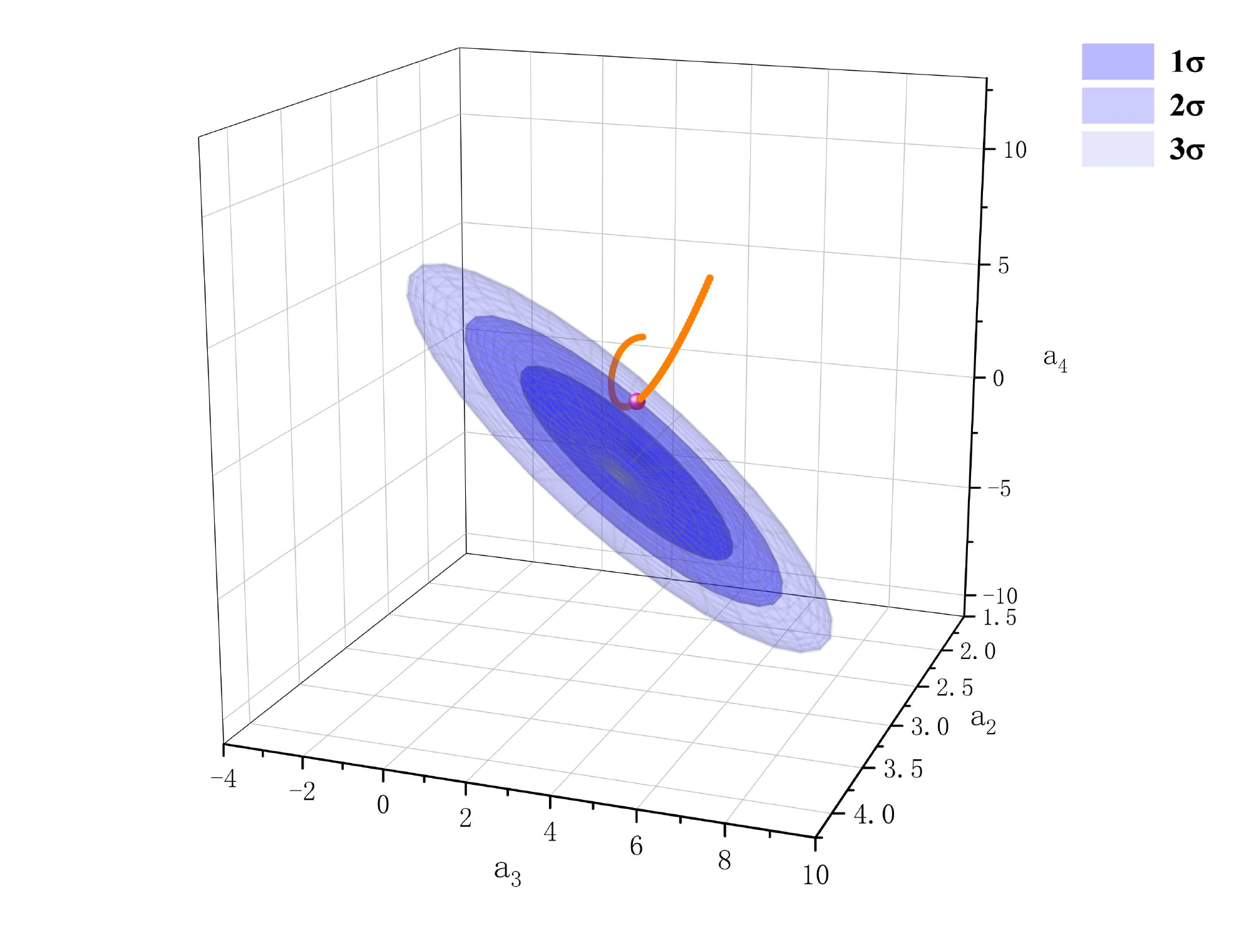}
        \caption{3D confidence ellipsoid ($1\sigma$, $2\sigma$, and $3\sigma$) for the parameters space ($a_2$, $a_3$, and $a_4$) and the corresponding 2D projections on panels $a_2-a_3$, $a_2-a_4$, and $a_3-a_4$ from the SN-Q sample using the $\log(1+z)$ method. The solid orange line shows the relation between two and three parameters in the $\log(1+z)$ model. In 2D projections, red and black points correspond to $\Omega_{m} = 0.30$ and $\Omega_{m} = 0.40$, respectively. In the 3D confidence ellipsoid, the purple point corresponds to $\Omega_{m} = 0.30$. }
        \label{F3}       
\end{figure*}

Out of necessity, we briefly introduce the model comparison methods used. We note that AIC and BIC are the last set of techniques that can be employed for model comparison based on information theory, and they are widely used in cosmological model comparisons; AIC and BIC are defined by \citet{10.1214/aos/1176344136} and \citet{1100705}, respectively. The corresponding definitions read as follows:  
\begin{eqnarray}
        AIC &=& 2n + \chi_{min}^{2},\\
        BIC &=& n\log{N} + \chi_{min}^{2},
        \label{eq:AIC}
\end{eqnarray}
where $n$ is the number of free parameters, $N$ is the total number of data points, and $\chi_{min}^{2}$ is the value of $\chi^{2}$ calculated with the best fitting parameters. The model that has lower values of AIC and BIC will be the suitable model for the employed data set. Moreover, we also calculated the differences between $\Delta$AIC and $\Delta$BIC with respect to the corresponding flat $\Lambda$CDM values to measure the amount of information lost by adding extra parameters in the statistical fitting. Negative values of $\Delta$AIC and $\Delta$BIC suggest that the model under investigation performs better than the reference model. For positive values of $\Delta$AIC and $\Delta$BIC, we adopt the judgment criteria of the literature \citep{2020MNRAS.494.2576C}:
\begin{itemize}\label{AICintervals}
	\item $\Delta\text{AIC(BIC)}\in [0,2]$ indicates weak evidence in favor of the reference model, leaving the question on which model is the most suitable open;
	\item $\Delta\text{AIC(BIC)}\in (2,6]$ indicates mild evidence against the given model with respect to the reference paradigm;
	and        \item $\Delta\text{AIC(BIC)}> 6$ indicates strong evidence against the given model, which should be rejected.
\end{itemize}

\section{Potential deviation from $\Lambda$CDM from quasar data}\label{sec:444}
\citet{2019NatAs...3..272R} found a 4$\sigma$ deviation from $ \Lambda$CDM via the Hubble diagram of SNe Ia and 1598 quasars. Then \citet{2019A&A...628L...4L} confirmed the tension between the $\Lambda$CDM model and the best cosmographic parameters at 4$\sigma$ with SNe Ia from the Pantheon sample and quasars, and at $>$ 4 $\sigma$ with SNe Ia, quasars, and GRBs. These two works both used a calibrated quasar relation between the UV and X-ray luminosity which can be parameterized as \citep{1986ApJ...305...83A}
\begin{eqnarray}
        \log_{10}(L_{X}) = \gamma\log_{10}(L_{UV}) + \beta,
        \label{eq:LUV}
\end{eqnarray}
where $L_{X}$ is the rest-frame monochromatic luminosity at 2 keV
and $L_{UV}$ is the luminosity at 2,500 {\AA}. We note that $\gamma$ and $\beta$ are fixed \citep{2019NatAs...3..272R}, and then they were used for the cosmological constraints. The details on the calibration of quasar relation between the UV and X-ray luminosity are provided in \citet{2019NatAs...3..272R}. In this section, we fit the quasar relation and cosmographic parameters simultaneously.

Utilizing Eq. (\ref{eq:LUV}), we derived the theorietical X-ray flux  \citep{2020A&A...643A..93H,2020MNRAS.497..263K}:
\begin{eqnarray}
        \phi([F_{UV}]_{i},d_{L}[z_{i}]) &=&\log_{10}(F_{X}) \nonumber \\ 
        &=& \gamma(\log{F_{UV}}) +(\gamma-1)\log{4\pi} \nonumber\\ 
        &+& 2(\gamma-1)\log_{10}{d_{L}} + \beta.
        \label{eq:luv}
\end{eqnarray}
Here, $F_x$ and $F_{uv}$ represent the X-ray and UV flux, respectively. For the Pad$\rm \acute{e}_{(2,1)}$ approximation and logarithmic polynomials, the corresponding luminosity distances $d_{L}$ were calculated by utilizing Eqs. (\ref{eq:dlp21}) and (\ref{eq:dllog}), respectively. For the quasar sample, the best fitting values of our used parameters can be obtained by minimizing the corresponding $\chi^{2}_{Q}$, 
\begin{eqnarray}
        \chi^{2}_{Q} &=&
        \sum_{i=1}^{1598}(\frac{(\log_{10}(F_{X})_{i}-\phi([F_{UV}]_{i},d_{L}[z_{i}]))^{2}}{s_{i}^{2}} + \ln(2\pi s_{i}^{2})), \label{q4}
\end{eqnarray}
where the variance $s_{i}^{2}$ consists of the global intrinsic scatter $\delta$ and the measurement error $\sigma_{i}$ in $(F_{X})_{i}$,
that is $s_{i}^{2} \equiv \delta^{2} + \sigma_{i}^{2}$. Compared to $\delta$ and $\sigma_{i}^{2}$, the error of $(F_{UV})_{i}$ is negligible. The function $\phi$ corresponds to the theoretical X-ray flux obtained by Eq. (\ref{eq:luv}). Combining Eqs. (\ref{eq:dllog}), (\ref{eq:luv}), and (\ref{q4}), we can obtain the fitting results, that is $\delta$ = 0.23$\pm$0.01, $\gamma$=0.63$\pm$0.01, $\beta$=7.42$\pm$0.32, $k$ = 0.99$\pm$0.01, $a_2$=3.08$\pm$0.17, $a_3$= 3.70$\pm$1.10, and $a_4$=-1.20$^{+1.90}_{-2.20}$. The corresponding confidence contours are shown in Fig. \ref{F2}. In Fig. \ref{F3}, we plotted the 3D confidence ellipsoid for parameters space ($a_2$, $a_3$, and $a_4$) and the corresponding 2D projections on panels $a_2-a_3$, $a_2-a_4$, and $a_3-a_4$. The solid orange lines show the relation between parameters $a_2$, $a_3$, and $a_4$. It is easy to find that the constrained results of the 2D projections are consistent with the flat $\Lambda$CDM model within the 2$\sigma$ level. The results of the 3D confidence ellipsoid are consistent with the flat $\Lambda$CDM model within the 3$\sigma$ level. Through analyzing Figs. \ref{F2} and \ref{F3}, we confirm that considering correlation parameters $\delta$, $\gamma$, and $\beta$ as free parameters, the 4$\sigma$ tension between the $\Lambda$CDM model and the best cosmographic parameters from the SN-Q sample reduce to 3$\sigma$.

\begin{figure}[htbp]
        \centering
        \includegraphics[width=0.4\textwidth,angle=0]{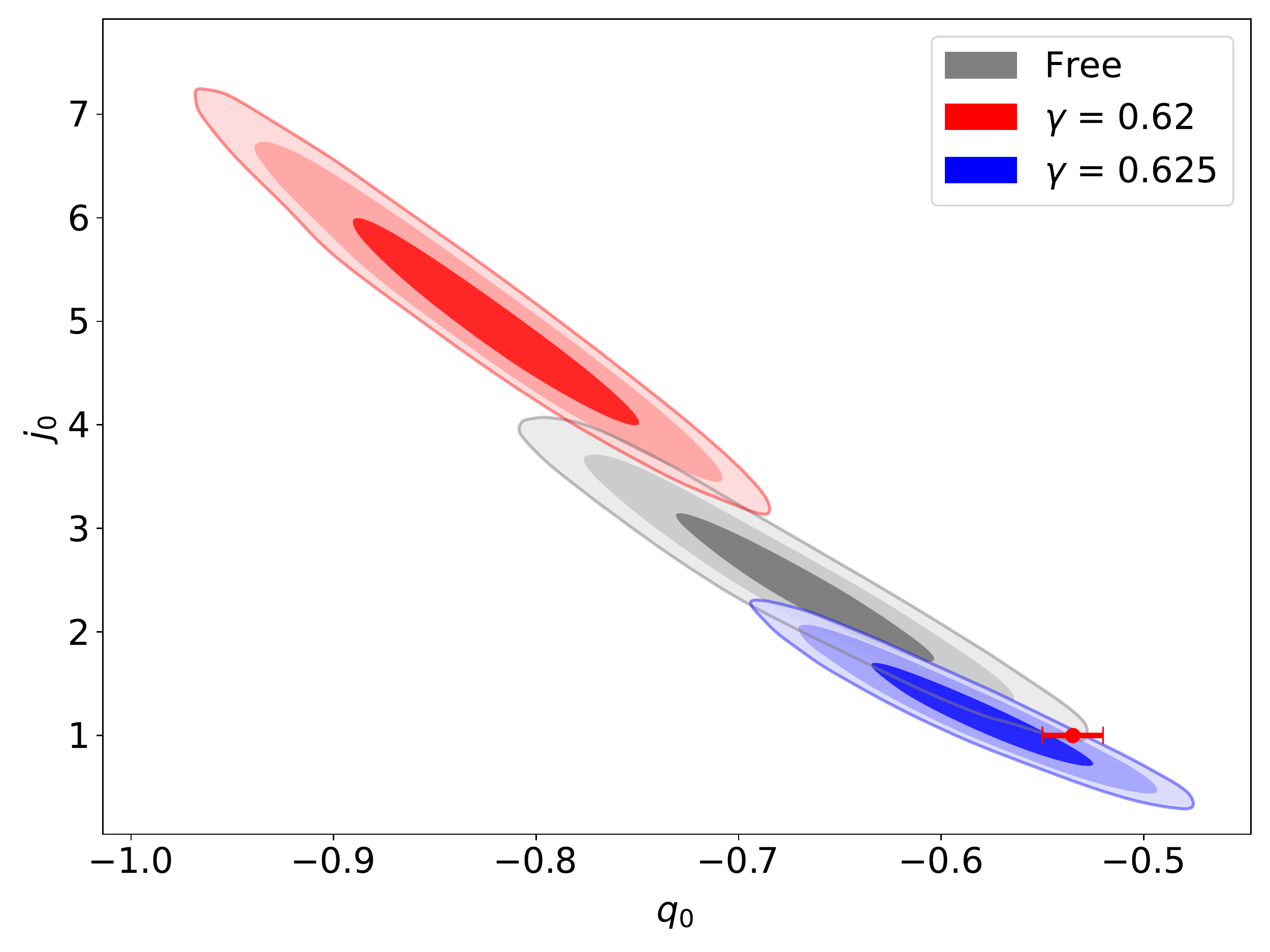}
        \caption{Confidence contours ($1\sigma$, $2\sigma$, and 3$\sigma$) for the parameters space ($q_0$ and $j_0$) from the SN-Q sample utilizing the Pad$\rm \acute{e}_{(2,1)}$ method. The red point represents (-0.54, 1) the value given by the Planck 2018 results \citep{2020A&A...641A...7P}.}
        \label{F5}       
\end{figure}

While using the Pad$\rm \acute{e}_{(2,1)}$ approximation, we checked the influence of the quasar relation on the cosmological constraints. We first provide the constraints considering a free quasar relation. Then combining that with the following two special cases, the effect of the slope parameter $\gamma$ is studied. One case is the fitting parameters $q_0$ and $j_0$ in the case of fixed parameters $\delta$, $\gamma$, and $\beta$. The other is to refit $q_0$ and $j_0$ by changing the setting of the fixed parameter $\gamma$ within a 1$\sigma$ error (0.005). The fixed parameters $\delta$, $\gamma$, and $\beta$ are given in terms of the best fitting results from 1598 quasars in a flat $\Lambda$CDM model. By substituting Eq. (\ref{eq:luv}) into Eq. (\ref{q4}), and replacing $d_{L}$ with Eq. (\ref{q2}), we obtained the best fits, that is  $\Omega_{m}$ = 0.65$^{+0.16}_{-0.19}$, $\delta$ = 0.23$\pm$0.01, $\gamma$=0.62$\pm$0.01, and $\beta$ = 7.60$\pm$0.28. This result is consistent with previous research within a 1$\sigma$ error \citep{2016ApJ...819..154L,2019MNRAS.489..517M,2019A&A...631A.120S,2020MNRAS.492.4456K,2020ApJ...899...71L,2020ApJ...888...99W,2022MNRAS.510.2753K}. Here, it is worth noting that the 1$\sigma$ errors of the $\gamma$ best fit from previous research are both larger than 0.01.
\begin{figure}[htbp]
        \centering
        \includegraphics[width=0.5\textwidth,angle=0]{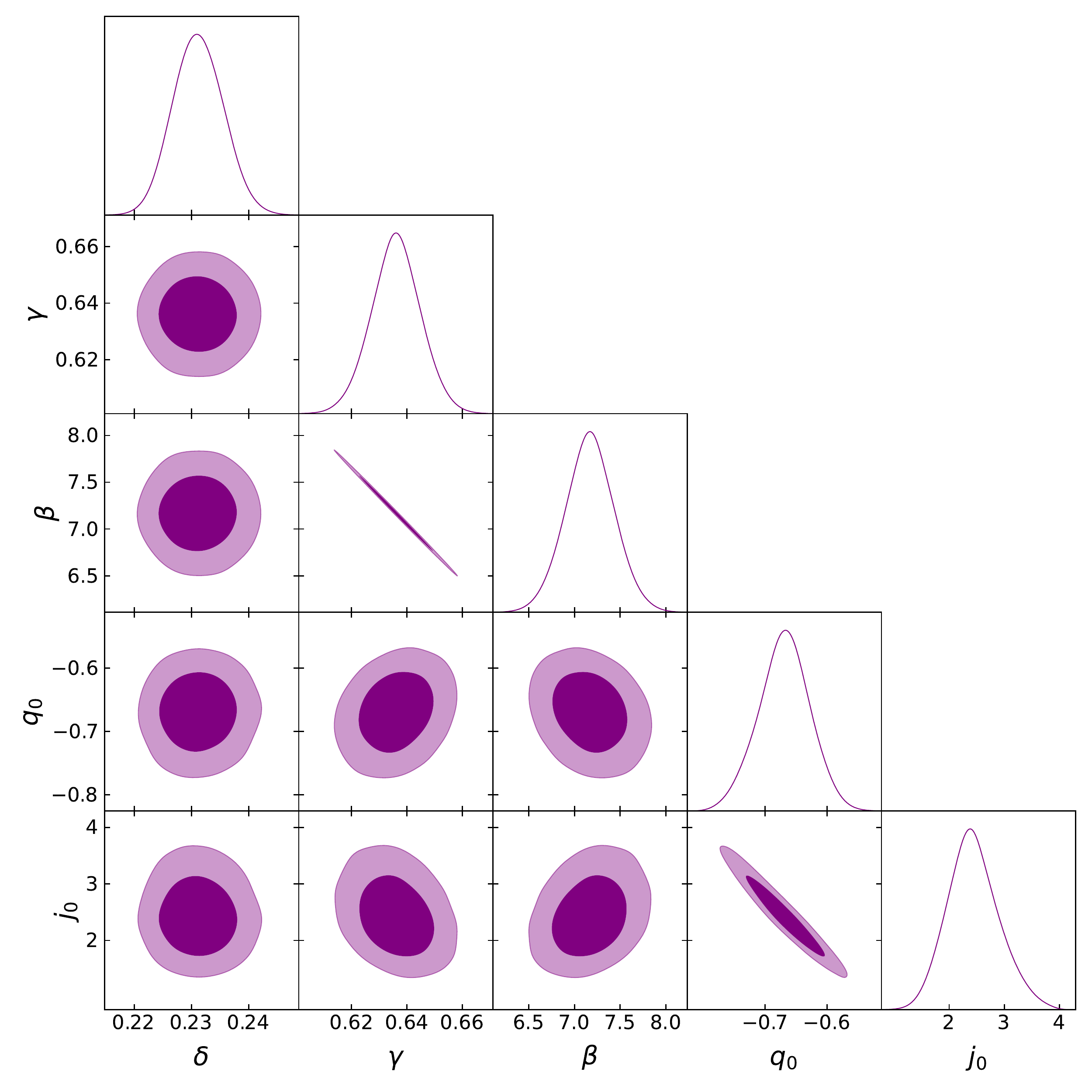}
        \caption{Confidence contours ($1\sigma$ and $2\sigma$) for the parameters space ($\delta$, $\gamma$, $\beta$, $q_0$, and $j_0$) from the SN-Q sample using the Pad$\rm \acute{e}_{(2,1)}$ method.}
        \label{F4}       
\end{figure}
Fig. \ref{F4} shows the fitting results with a free quasar relation. The best fits are $\delta$=0.23$\pm$0.01, $\gamma$=0.64$\pm$0.01, $\beta$=7.17$^{+0.28}_{-0.27}$, $q_0$=-0.67$\pm$0.04, and $j_{0}$=2.43$^{+0.51}_{-0.46}$. This result is in line with the previous results obtained by the logarithmic polynomials. The corresponding $q_0-j_0$ projection is also plotted in Fig. \ref{F5} as a gray contour. In addition, we also show the confidence contours of the other two special cases. (1) When we chose $\delta$ = 0.23 , $\gamma$ =0.62, and $\beta$ = 7.60 (best fits from 1598 quasars in a flat $\Lambda$CDM model), the results are $q_0$ = -0.82$\pm$0.05 and $j_{0}$ = 4.99$^{+0.69}_{-0.63}$, which is represented by the red contours in Fig. \ref{F5}. There exists more than a 4$\sigma$ tension with the flat $\Lambda$CDM. (2) In only changing $\gamma$ = 0.625, the new result is $q_0$=-0.58$\pm$0.03 and $j_{0}$=1.19$^{+0.34}_{-0.31}$ which is described by the blue contours of Fig. \ref{F5}. We find that the $\gamma$ value changes from 0.62 to 0.625 (within a 1$\sigma$ error of 0.01), and the  4$\sigma$ tension disappears. This result is consistent with the $\Lambda$CDM model within a $1\sigma$ level. We find that the quasar relation can obviously affect the cosmographic constraint, especially the slope parameter $\gamma$. 

\section{Results}
\begin{figure*}[htbp]
        \centering
        \includegraphics[width=0.4\textwidth]{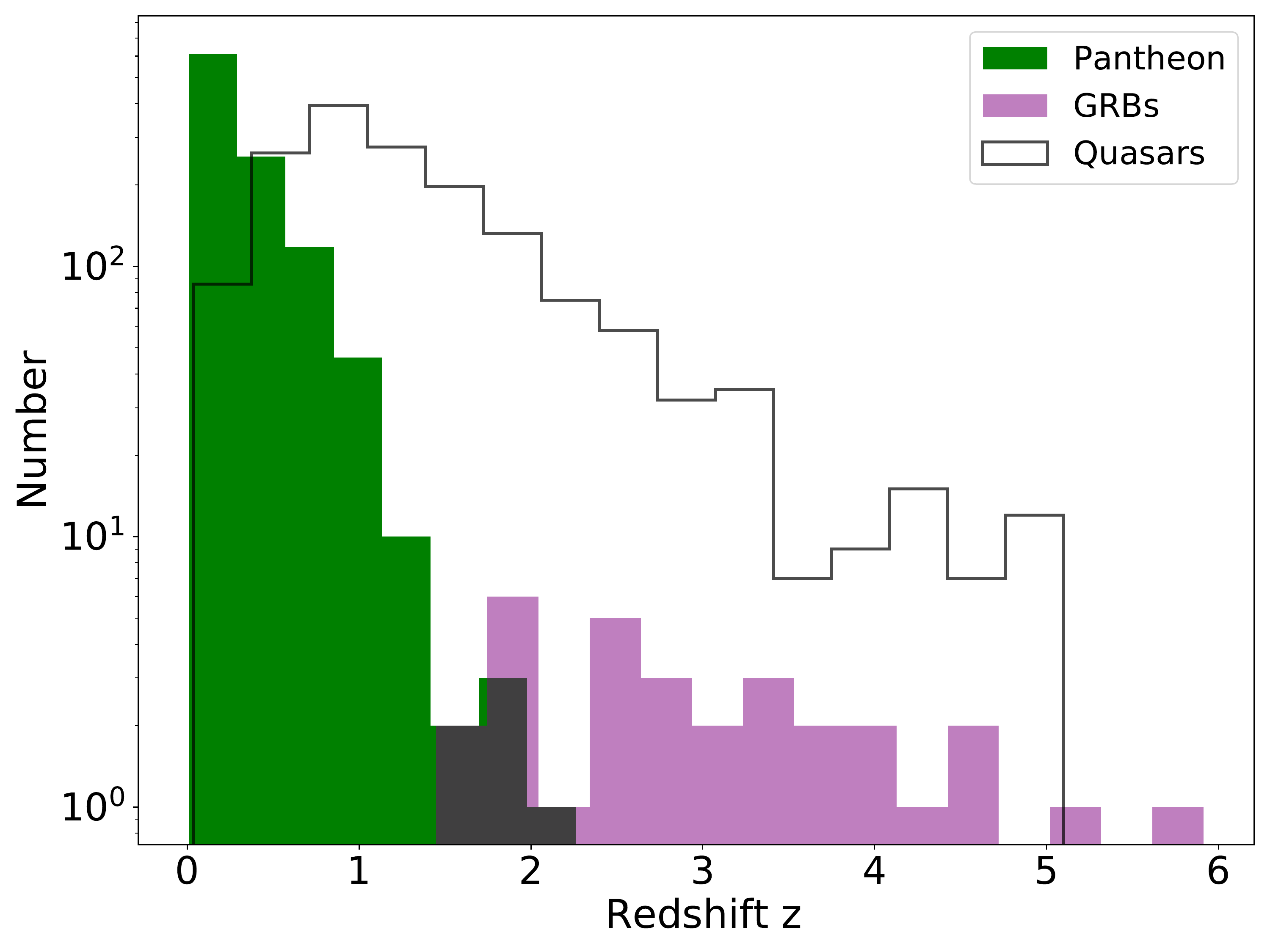}
        \includegraphics[width=0.4\textwidth]{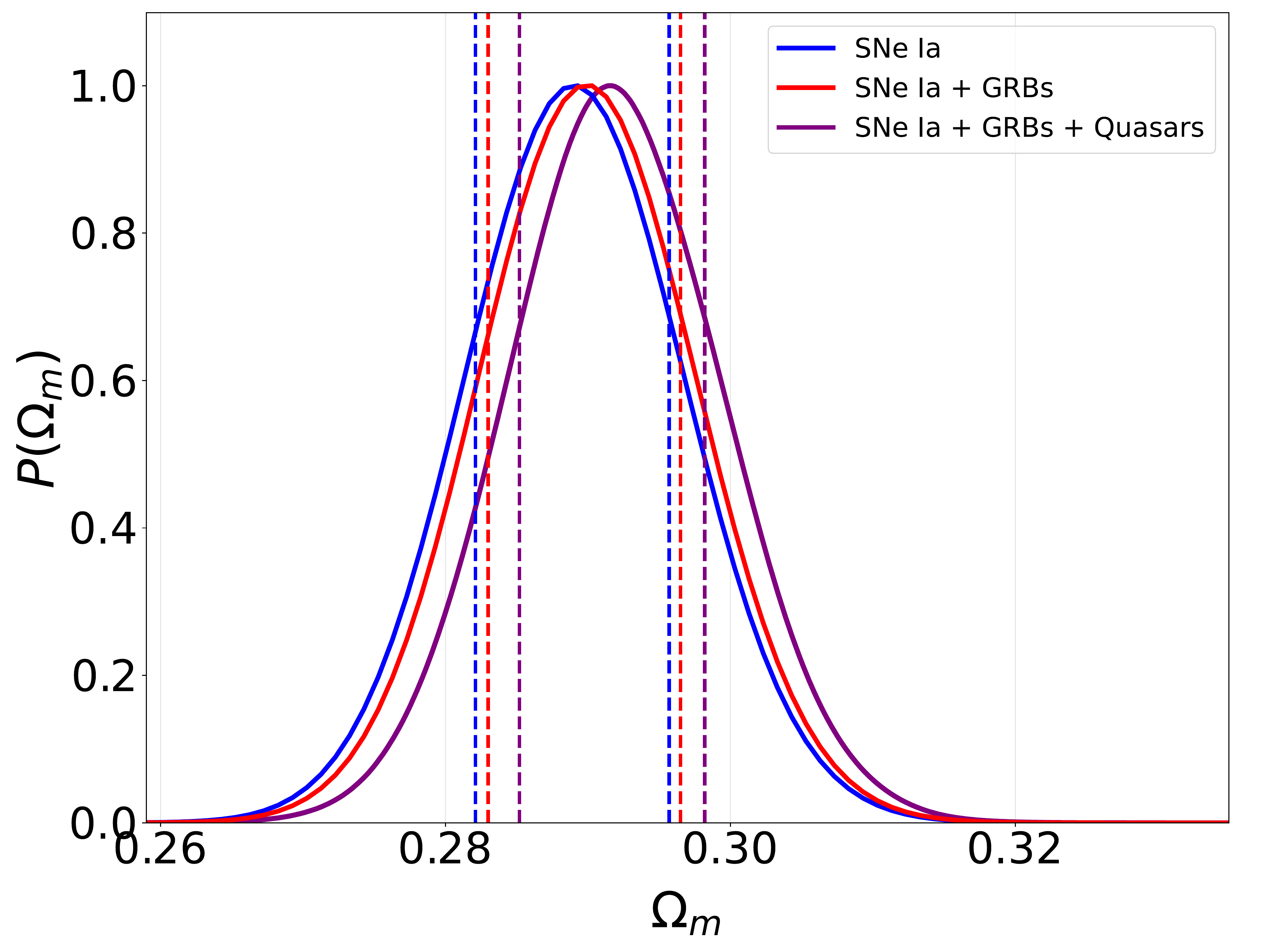}
        \caption{Fundamental information about the used data sets. Left panel: Redshift distribution of SNe Ia, GRBs, and quasars. Right panel: Constrained results of $\Omega_{m}$ from different combinations of SNe Ia, GRBs, and quasars in the spatially flat $\rm \Lambda$CDM model.}
        \label{F1}       
\end{figure*}

Firstly, we performed an analysis of the used samples, including the redshift distribution and the constraints on $\Omega_{m}$ in the spatially flat $\Lambda$CDM model, as shown in Fig. \ref{F1}. From the redshift distribution, we found that SNe Ia are mainly distributed in low redshift ($z < 1$), and GRBs and quasars are mainly distributed in high redshift ($z > 1$). The best fits of $\Omega_{m}$ are all near 0.29 with a 1$\sigma$ error of 0.01. This result will be used in subsequent comparisons. Below, we show the main results.

\begin{figure*}[htbp]
        \centering
        \includegraphics[width=0.4\textwidth]{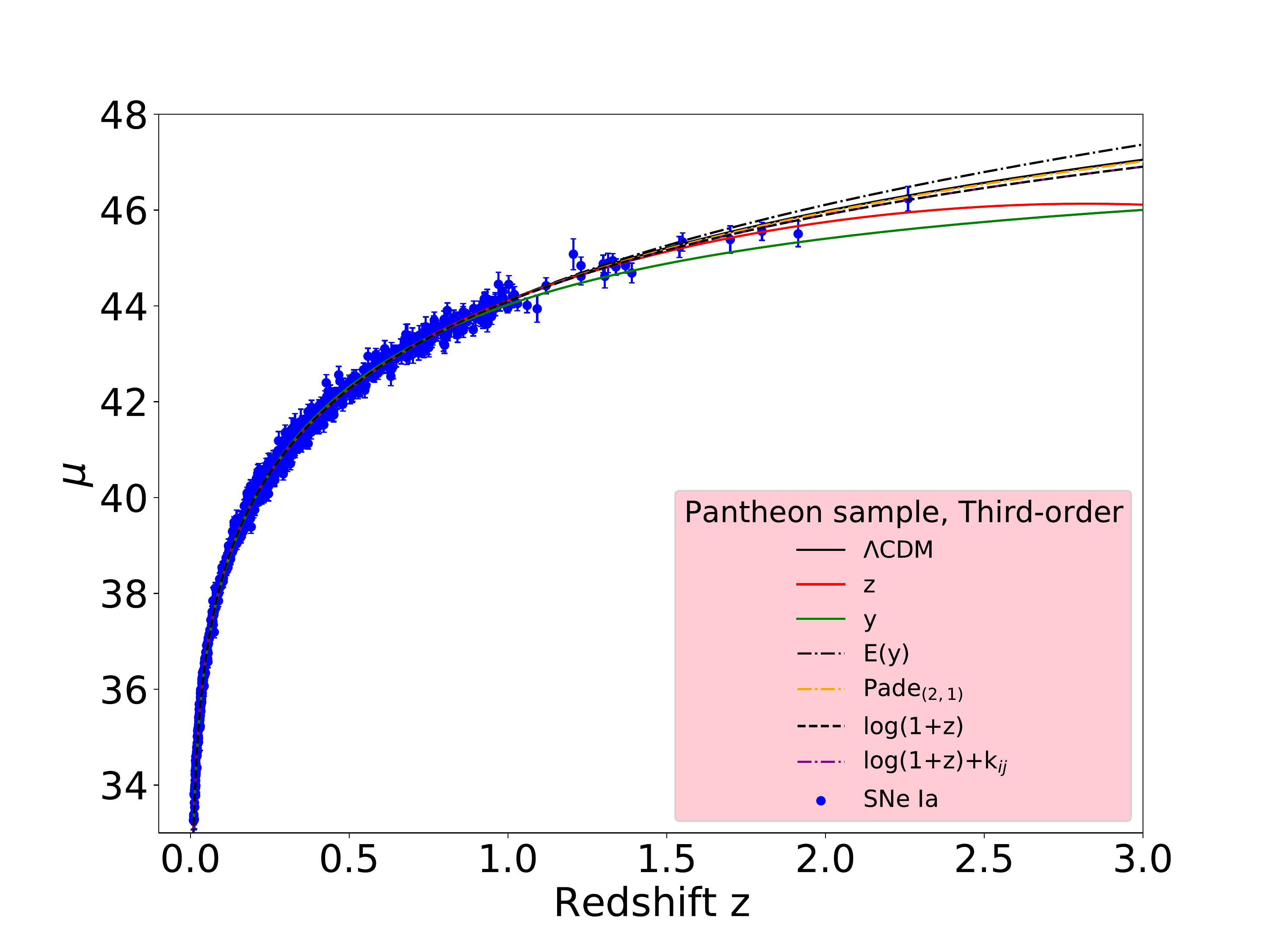}
        \includegraphics[width=0.4\textwidth]{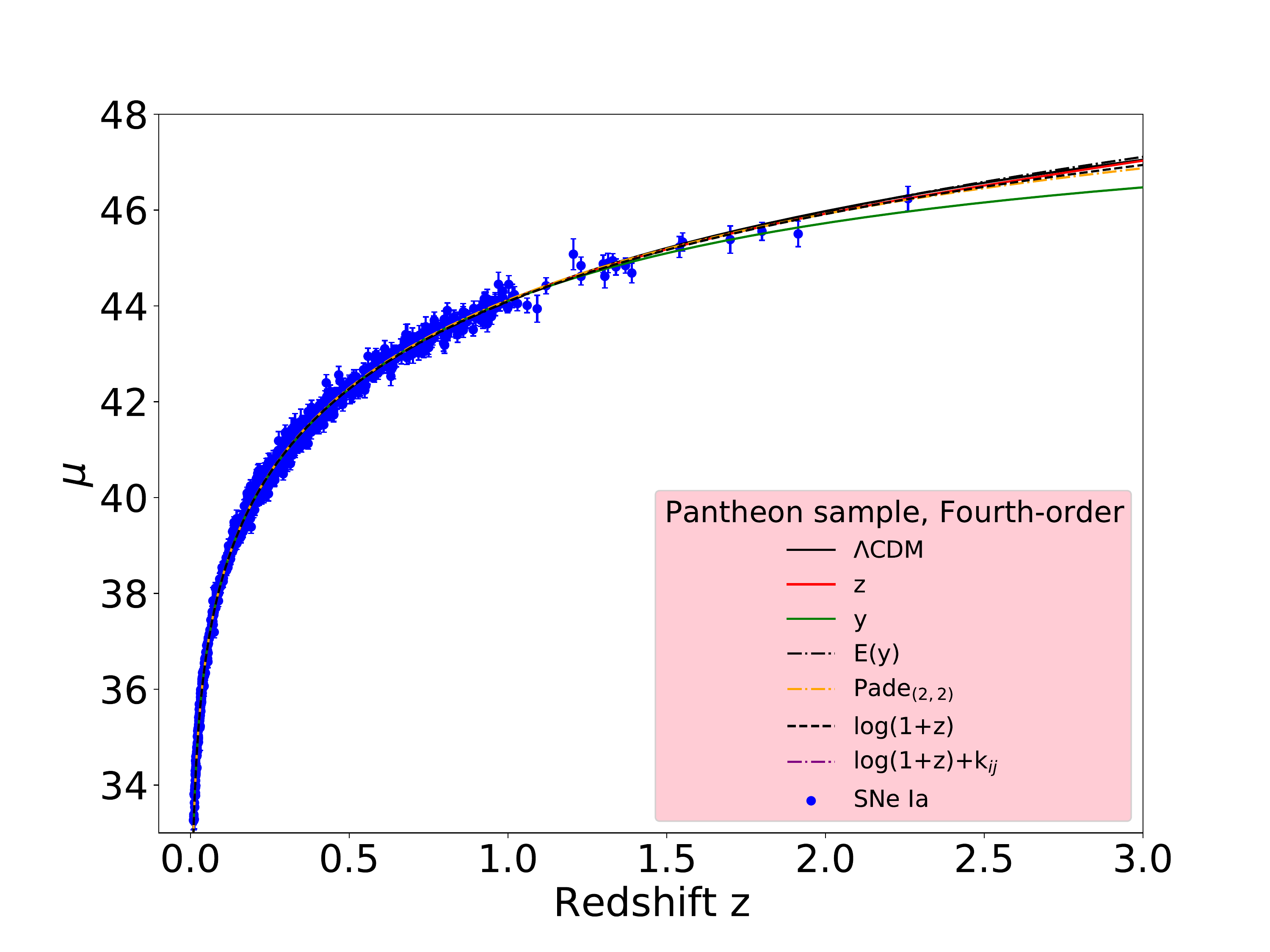} 

        \includegraphics[width=0.4\textwidth]{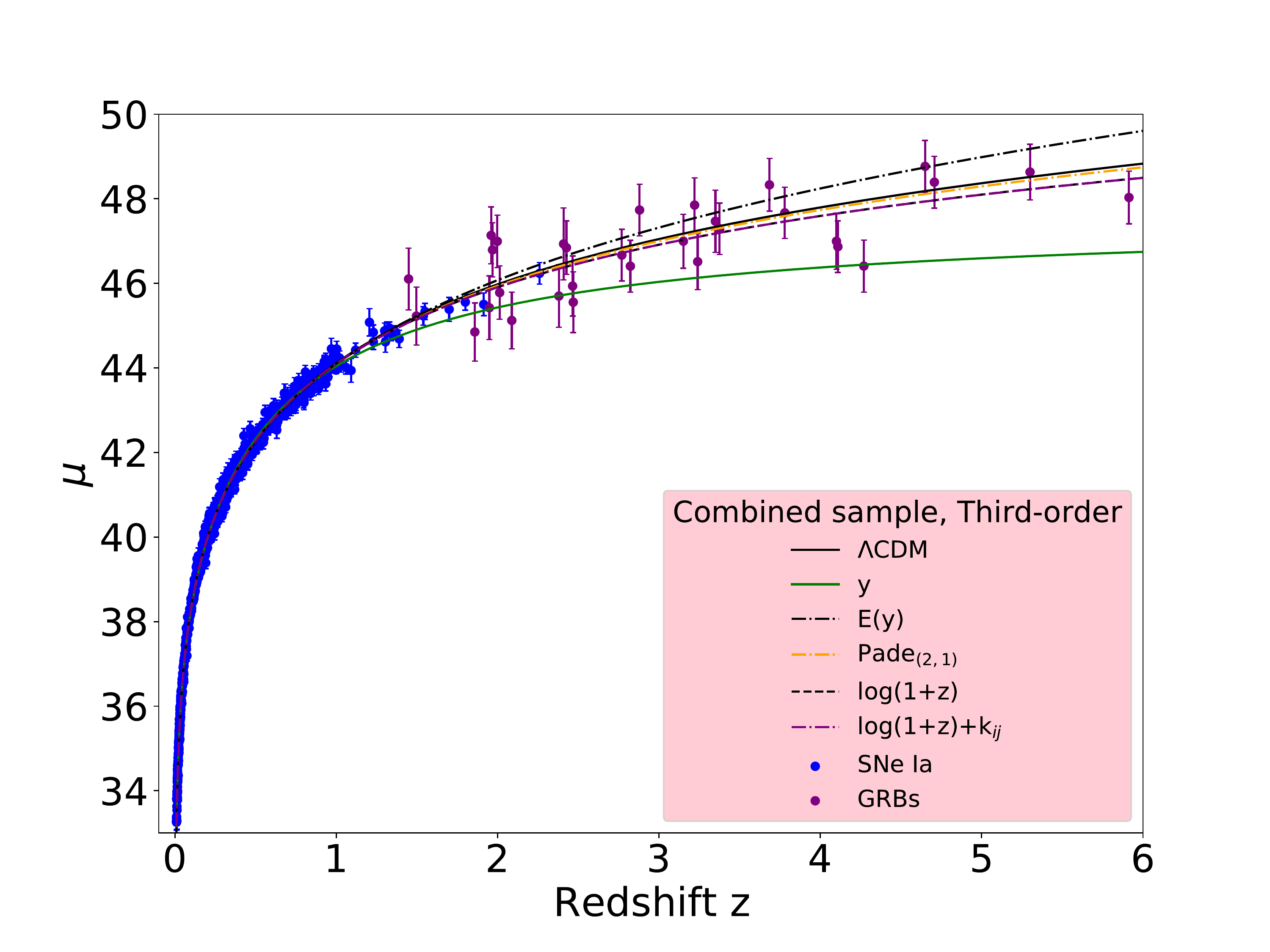}
        \includegraphics[width=0.4\textwidth]{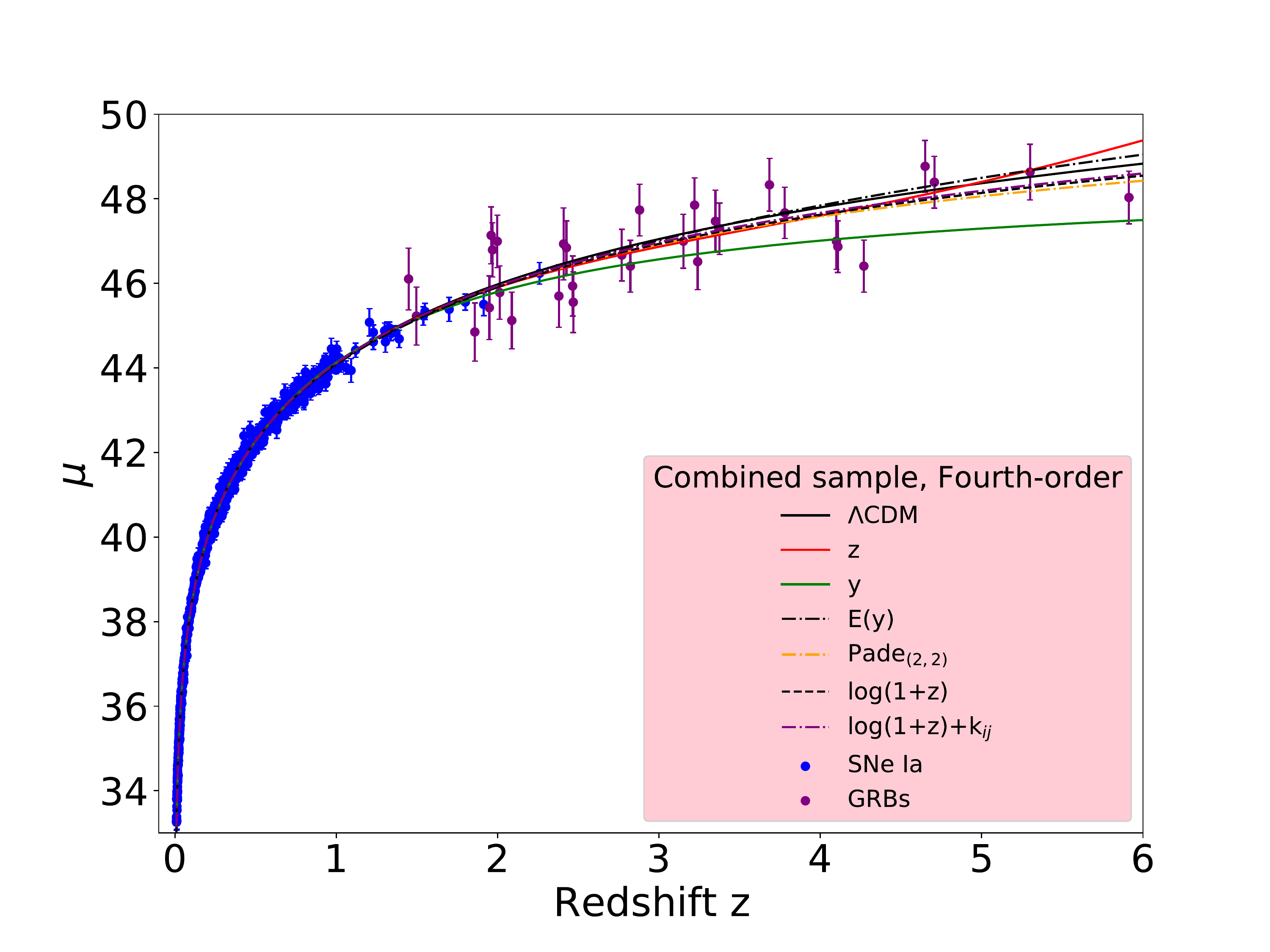}
        \caption{Hubble diagram of SNe Ia and GRBs. The upper two panels and the lower two panels use the Pantheon sample and the combined sample, respectively. The expansion order of the right two panels and the left two panels is different. The former is the third order and the latter is the fourth order. Blue points are SNe Ia from the Pantheon sample. Purple points are 31 LGRBs constructed by \citet{2022ApJ...924...97W}.  }
        \label{F6}       
\end{figure*}

A comparison with different cosmographic methods as another main work also yields many meaningful results. We plotted the Hubble diagram of SNe Ia and GRBs, as well as the corresponding theoretical lines with different methods or different expansion orders, as shown in Fig. \ref{F6}. The upper two panels and the lower two panels use the Pantheon sample and a combination of the Pantheon sample and 31 LGRBs, respectively. Expansion orders of the right two panels and the left two panels are the third order and the fourth order, respectively. By analyzing Fig. \ref{F6}, we found that as the expansion order increases, the difference between the prediction lines of different methods becomes blurred. However, the results of the $y$-refshift method are obviously different from other methods. This is one of the reasons why we studied the $y$-redshift with different expansion orders separately. The corresponding best fitting results and statistical information ($\chi^{2}_{\rm min}$, AIC, BIC, $\Delta$AIC, and $\Delta$BIC) of the Pantheon sample and the combined sample are listed in Tables \ref{T1} and \ref{T2}, respectively.

\begin{table*}[htbp]
        \begin{spacing}{1.2}
                \small
                \caption{ Best fitting results obtained from the Pantheon sample using different cosmographic techniques. }               \label{T1}
                \centering
                \begin{tabular}{p{1.9cm}p{1.7cm}p{1.7cm}p{1.7cm}p{1.6cm}p{1cm}p{1cm}p{1cm}p{1cm}p{1cm}} 
                        \hline\hline
                        Method  & Parameters &  & &  & $\chi^{2}_{min}$  &  AIC   & BIC &  $\Delta$AIC   & $\Delta$BIC \\ \hline
                        $\Lambda$CDM &$\Omega_{m}$ = 0.289 &&& & 1067.49 &  1069.49 & 1070.51 & 0 & 0 \\ \hline
                        Third-order & $q_{0}$ & $j_{0}$ &  & & & & & & \\ \hline
                        $z$ &-0.55$\pm$0.03& 0.71$\pm$0.17&&&1066.44 & 1070.44&1072.48 &0.95&1.97 \\ \hline
                        $y$ &0.27$\pm$0.08&-21.27$^{+1.05}_{-1.03}$&&&1137.97 & 1141.97& 1144.01& 72.48&73.50 \\ \hline
                        $E(y)$  &-0.76$\pm$0.05&4.50$^{+0.59}_{-0.56}$&&&1073.78 & 1077.78& 1079.82 & 8.29&9.31\\ \hline
                        Pad$\rm \acute{e}_{(2,1)}$ &-0.65$\pm$0.04&2.18$^{+0.50}_{-0.47}$&&& 1066.32  & 1070.32 &1072.36 & 0.83& 1.85\\ \hline
                        &k&$a_{2}$&$a_{3}$&& && & & \\ \hline
                        $\log(1+z)$  &1.00$\pm$0.01&3.09$\pm$0.12&3.33$\pm$0.41&& 1064.84 &1070.84&1073.90 & 1.35&3.39 \\ \hline
                        $\log(1+z)+k_{ij}$ &1.00$\pm$0.01&3.99$\pm0.05$&3.34$^{+0.40}_{-0.39}$&& 1064.82 &1070.82&1073.88 & 1.33& 3.37 \\ \hline \hline
                        Fourth-order &$q_{0}$ & $j_{0}$ & $s_{0}$ & & & &&&\\ \hline
                        $z$ &-0.59$\pm$ 0.04&1.18$^{+0.43}_{-0.42}$&0.88$^{+1.69}_{-1.36}$&&1065.12 & 1071.12&1074.18 &1.63&3.67 \\ \hline
                        $y$ &-0.99$^{+.17}_{-0.16}$&17.73$^{+5.37}_{-5.36}$&389.80$^{+130.00}_{-117.20}$&&1067.05 & 1073.05&1076.11 & 3.56& 5.60\\ \hline
                        $E(y)$ &-0.51$\pm0.11$&-1.47$^{+2.25}_{-2.40}$&-37.08$^{+22.44}_{-21.75}$&&1067.78 & 1073.78&1076.84 & 4.29& 6.33\\ \hline
                        Pad$\rm \acute{e}_{(2,2)}$ &-0.62$\pm$0.06&1.58$^{+1.00}_{-0.83}$&3.04$^{+7.34}_{-4.29}$&&1071.47 & 1077.47&1080.53 & 7.89& 10.02\\ \hline
                        &k&$a_{2}$&$a_{3}$&$a_{4}$& && & & \\ \hline
                        $\log(1+z)$ &0.99$\pm$0.01&3.14$^{+0.22}_{-0.21}$&2.95$^{+1.67}_{-1.64}$&0.84$\pm$3.73&1064.79 & 1072.79&1076.87 & 3.30&6.36 \\ \hline
                        $\log(1+z)+k_{ij}$ &0.99$\pm$0.01&3.99$\pm$0.06&3.31$\pm$0.42&0.89$\pm$3.64& 1064.78 &1072.78& 1076.86 & 3.29&6.35\\
                        \hline \hline
                \end{tabular}   
        \end{spacing}
\end{table*}

 In Table \ref{T1}, among the results of the third-order expansion, the smallest values for $\Delta$AIC and $\Delta$BIC were all obtained by the Pad$\rm \acute{e}_{(2,1)}$ method, that is $\Delta$AIC = 0.83 and $\Delta$BIC = 1.85.  Also, $\Delta$AIC and $\Delta$BIC of the $z$-redshift method are close to the values of the former. The ones of the $y$-redshift and the $E(y)$ methods are $\Delta$AIC = 72.48 and $\Delta$BIC = 73.50, and $\Delta$AIC = 8.29 and $\Delta$BIC = 9.31, respectively, which are both larger than 6. The latter values are significantly better than the former. From the results of the fourth-order expansion, the smallest values for $\Delta$AIC and $\Delta$BIC are 1.63 and 3.67, which are given by the $z$-redshift method. Utilizing the Pad$\rm \acute{e}_{(2,2)}$ method, we found that $\Delta$AIC = 7.89 and $\Delta$BIC = 10.02, which are larger than what was found for the other methods. The results for the $y$-redshift and $E(y)$ methods are similar. Combining the results of third-order expansion with the results of firth-order expansion, the minimum values of $\Delta$AIC and $\Delta$BIC from Table \ref{T1} are $\Delta$AIC = 0.83 and $\Delta$BIC = 1.85, which were obtained by the Pad$\rm \acute{e}_{(2,1)}$ method. Through analyzing statistical information in Table \ref{T1}, it is easy to find that, except for the $y$-redshift and $E(y)$ methods, the values of $\Delta$AIC and $\Delta$BIC increase with the expansion order. In comparison to other methods, the influence of the expansion order on the results for the $y$-redshift method is more obvious. For the Pantheon sample (low redshift case), the third-order expansion of $y$-redshift and $E(y)$ methods and Pad$\rm \acute{e}_{(2,2)}$ can be rule out based on the values for $\Delta$AIC and $\Delta$BIC.

\linespread{1.0}
\begin{table*}[htbp]
        \begin{spacing}{1.2}
                \caption{Best fitting results obtained from the combined sample using different cosmographic techniques.}                \label{T2}
                \centering
                \begin{tabular}{p{1.9cm}p{1.7cm}p{1.7cm}p{1.7cm}p{1.75cm}p{1cm}p{1cm}p{1cm}p{1cm}p{1cm}} 
                        \hline\hline
                        Method&Parameters&&&    & $\chi^{2}_{min}$  &  AIC & BIC  & $\Delta$AIC& $\Delta$BIC \\ \hline
                        $\Lambda$CDM &$\Omega_{m}$ = 0.290&&&& 1105.12 &  1107.12& 1108.15  & 0& 0 \\ \hline
                        Third-order & $q_{0}$ & $j_{0}$ &  & & & & & & \\ \hline
                        $z$ &--&--&&& -- & --& -- & -- &-- \\ \hline
                        $y$ &0.36$^{+0.08}_{-0.07}$&-22.69$^{+1.00}_{-0.98}$&&& 1246.71 & 1250.71& 1252.78  & 143.59 &144.63  \\ \hline
                        $E(y)$  &-0.81$\pm$0.05&5.17$^{+0.55}_{-0.54}$&&& 1129.65 & 1133.65& 1135.72  & 26.53 & 27.57 \\ \hline
                        Pad$\rm \acute{e}_{(2,1)}$  &-0.66$\pm$0.04&2.30$^{+0.48}_{0.44}$&&& 1103.35 & 1107.35& 1109.42  & 0.23 & 1.27 \\ \hline
                        &k&$a_{2}$&$a_{3}$&& && & & \\ \hline
                        $\log(1+z)$  &1.00$\pm$0.01&3.08$\pm$0.11&3.40$^{+0.37}_{-0.36}$&& 1100.97 & 1106.97& 1110.07 & -0.15 &1.92  \\ \hline
                        $\log(1+z)+k_{ij}$  &1.00$\pm$0.01&4.02$^{+0.06}_{-0.05}$&3.39$\pm$0.37&& 1100.96 & 1106.96 & 1110.06 & -0.15  &1.91 \\ \hline \hline
                        Fourth-order &$q_{0}$ & $j_{0}$ & $s_{0}$ & & & &&&\\ \hline
                        $z$ &-0.56$\pm$0.03&0.94$^{+0.22}_{-0.21}$&0.23$^{+0.61}_{-0.48}$&& 1103.84 & 1109.84& 1112.94 & 2.72 &5.96 \\ \hline
                        $y$ &-1.19$\pm$0.15&25.42$^{+5.15}_{-5.03}$&584.90$^{+140.20}_{-126.20}$&& 1121.21 & 1127.21& 1130.30 &  20.09 &23.33 \\ \hline
                        $E(y)$  &-0.42$\pm$0.01&-3.76$^{+2.09}_{-2.12}$&-59.03$^{+18.20}_{-17.11}$&& 1108.61 & 1114.61& 1117.71 & 7.49 &10.73 \\ \hline
                        Pad$\rm \acute{e}_{(2,2)}$  &-0.63$\pm$0.05&1.77$^{+0.65}_{-0.57}$&4.45$^{+4.23}_{-3.03}$&& 1101.61 & 1107.61& 1110.71  & 0.49 &2.56 \\ \hline
                        &k&$a_{2}$&$a_{3}$&$a_{4}$& && & & \\ \hline
                        $\log(1+z)$  &0.99$\pm$0.01&3.12$^{+1.69}_{-1.73}$&3.11$^{+1.03}_{-1.07}$&0.53$^{+2.05}_{-1.90}$& 1100.94 & 1108.94& 1113.07 & 1.82& 6.48 \\ \hline
                        $\log(1+z)+k_{ij}$  &0.99$\pm$0.01&3.99$\pm$0.14&3.17$^{+0.95}_{0.97}$&4.78$^{+1.98}_{-1.92}$& 1100.93 & 1108.93& 1113.06 & 1.81 & 6.47 \\ \hline
                        Five-order &$q_{0}$ & $j_{0}$ & $s_{0}$ & $l_{0}$ & & &&&\\ \hline
                        $y$ &-0.36$^{+0.15}_{-0.09}$&-12.90$^{+3.60}_{-5.80}$&379$\pm$94&-4170$\pm$1100& 1104.97 & 1112.97& 1117.11 &  5.85 & 8.96 \\ 
                        \hline \hline
                \end{tabular}   
        \end{spacing}
\end{table*}

 In Table \ref{T2}, one can notice that there is no convergent result from the combined sample using the third-order expansion of the $z$-redshift method. From the third-order results, the values for $\Delta$AIC and $\Delta$BIC obtained by the Pad$\rm \acute{e}_{(2,1)}$, $\log(1+z)$, and $\log(1+z)+k_{ij}$ methods are very close, and they are all less than 2. For the Pad$\rm \acute{e}_{(2,1)}$ method, the values for $\Delta$AIC and $\Delta$BIC are 0.23 and 1.27, respectively. For the $\log(1+z)$ and $\log(1+z)+k_{ij}$ methods, the corresponding results are $\Delta$AIC = -0.15 and $\Delta$BIC = 1.92, and $\Delta$AIC = -0.15 and $\Delta$BIC = 1.91, respectively. The results obtained from the $y$-redshift and $E(y)$ methods are both larger than 6. For the fourth-order case, the minimum values, $\Delta$AIC = 0.49 and $\Delta$BIC = 2.56, were obtained by the Pad$\rm \acute{e}_{(2,2)}$ method. The results of the $y$-redshift and $E(y)$ methods are still greater than 6 and those for the former are worse. Minimum values of $\Delta$AIC and $\Delta$BIC in Table \ref{T2} are also given by the Pad$\rm \acute{e}_{(2,1)}$ method. Looking through the results of Table \ref{T2}, we found that the main conclusions obtained from the Pantheon sample are both confirmed by employing the combined sample including the Pantheon sample and 31 LGRBs (high-redshift case). In the case of high redshift, the fourth-order expansions of $y$-redshift and $E(y)$ are also excluded. 
 
\begin{figure*}[htbp]
        \centering
        \includegraphics[width=0.47\textwidth]{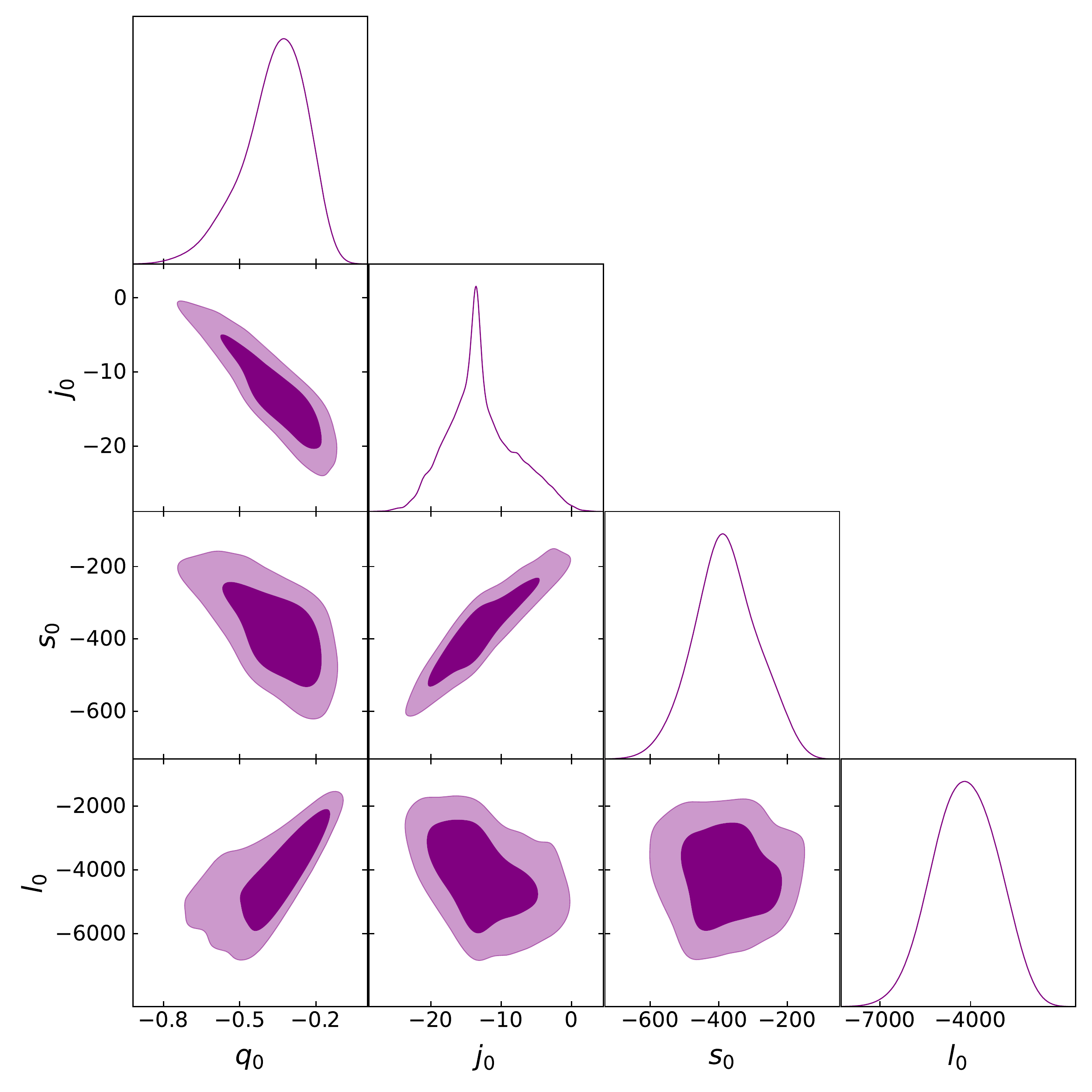}
        \includegraphics[width=0.47\textwidth]{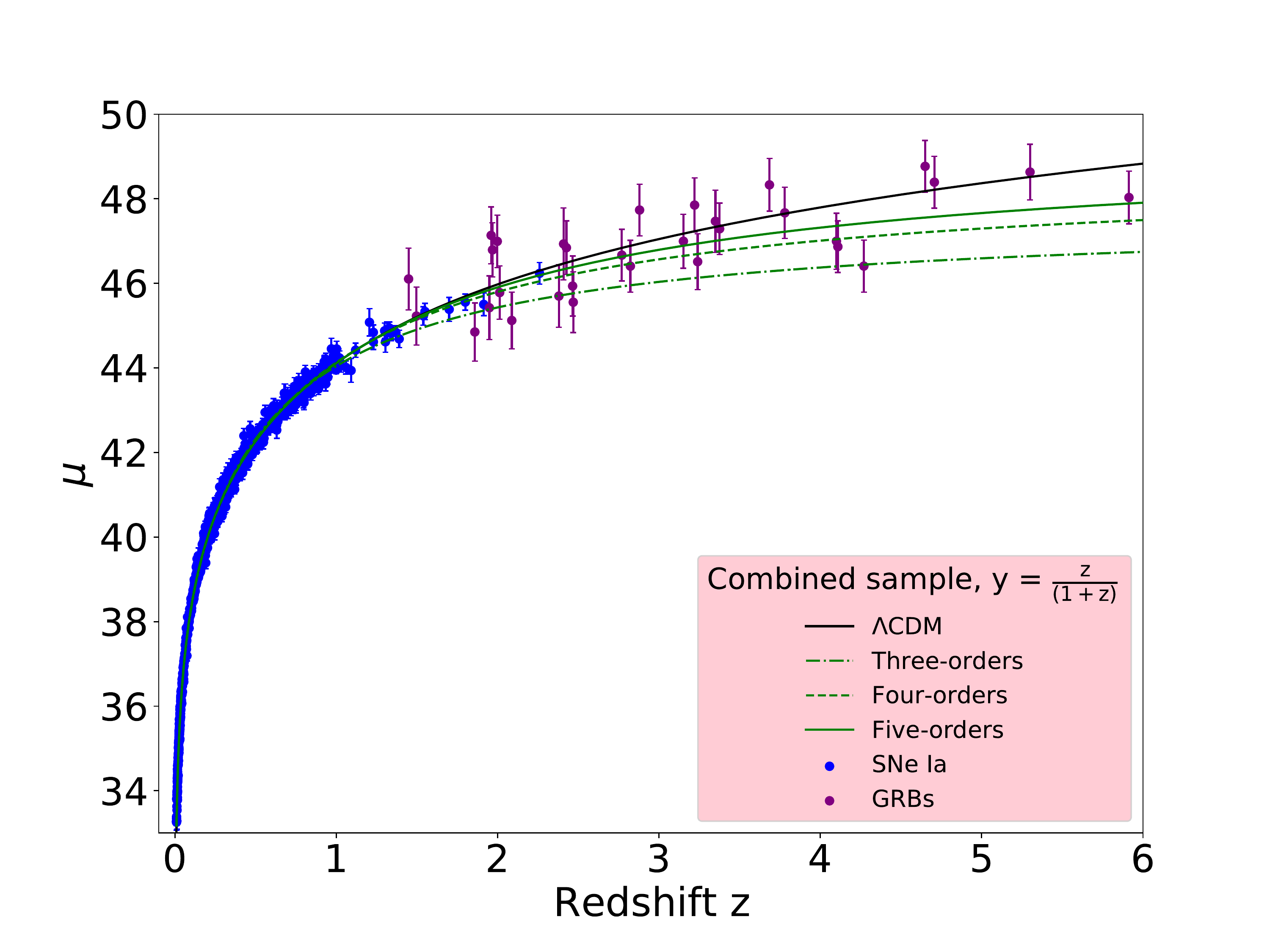}
        \caption{Comparison between different expansion orders using the $y$-redshift method. The right panel shows the confidence contours ($1\sigma$ and $2\sigma$) for the parameters space ($q_0$, $j_0$, $s_0$, and $l_0$) from the combined sample. Blue points are SNe Ia from the Pantheon sample. Purple points are 31 LGRBs constructed by \citet{2022ApJ...924...97W}.}
        \label{F7}       
 \end{figure*}

 And we also found some interesting results. There is no convergent result using the third-order expansion of the $z$-redshift method, but the fourth-order expansion can be used to obtain a good result. The Pad$\rm \acute{e}_{(2,2)}$ method, which is ruled out in the case of low redshift (Pantheon sample), also has a good result. It demonstrates that it is important to choose the suitable expansion method and order based on the sample used. In Table \ref{T2}, we present the constraints of the fifth-order expansion and the corresponding Hubble diagram from the combined sample using the $y$-redshift method, as shown in the left panel and the right panel of Fig. \ref{F7}, respectively. The best fits are $q_{0}$=-0.36$^{+0.15}_{-0.09}$, $j_{0}$=-12.90$^{+3.60}_{-5.80}$, $s_{0}$ = 379.04$\pm$94.00, and $l_{0}$=-4170$\pm$1100. The values for $\Delta$AIC and $\Delta$BIC are equal to 5.85 and 8.96, which are smaller than what was obtained by the third-order expansion and fourth-order expansion. To understand the influence of the expansion order on the results more intuitively, the third-order and the fourth-order expansions are also plotted in the right panel of Fig. \ref{F7}. We find that the theoretical line of higher order is closer to that of the flat $\Lambda$CDM model. At the same time, from the results of the $y$-redshift method in Table \ref{T2}, we confirm that the values for $\Delta$AIC and $\Delta$BIC decrease with the expansion order increase. 
 
All confidence contours corresponding to Tables \ref{T1} and \ref{T2} are shown in Figs. \ref{F8} - \ref{F11}. Figs. \ref{F8} and \ref{F9} were obtained by utilizing the Pantheon sample for the third order and the fourth order, respectively. Figs. \ref{F10} and \ref{F11} are similar to Figs. \ref{F8} and \ref{F9}, except that the combined sample is used for the former. In order to better display the results of comparison, we drew the consequences of the same parameters in a graph. The results of $y$-redshift are not tight and deviate significantly from other results. So we plotted the $y$-redshift results in a single graph.

\section{Conclusions and discussion}
In this paper, we first investigate the effect of the quasar relation between the UV and X-ray luminosities on the cosmographic constraints using the logarithmic polynomials and Pad$\rm \acute{e}_{(2,1)}$ methods. 
Then based on the Pantheon sample and 31 LGRBs, we critically compare six kinds of cosmographic methods, that is $z-$redshift, $y$-redshift, $E(y)$, $\log(1+z$), $\log(1+z)+k_{ij}$, and Pad$\rm \acute{e}$ polynomials. By investigating the effect of the quasar relation on the cosmographic constraint, we were able to find that the quasar relation can obviously affect the cosmographic constraint, especially the slope parameter $\gamma$. The change in the $\gamma$ value by 0.005 makes the 4$\sigma$ tension disappear. And until now, the best fits of the calibrated $\gamma$ have been in the range (0.612, 0.648), which was obtained from the latest quasar sample using different calibrated methods \citep{2019NatAs...3..272R,2020ApJ...888...99W,2021MNRAS.507..919L,2021SCPMA..6459511Z}. In terms of our study, the $\gamma$ value changes from 0.612 to 0.648, and the tension changes significantly. This indicates that the 4$\sigma$ or 3$\sigma$ tension between the $\Lambda$CDM model and the best cosmographic parameters from the SN-Q sample may be uncertain. In order to better constrain the cosmographic parameters, we need more precise data of quasars to yield a tighter quasar relation. A next-generation X-ray, such as eROSITA \citep{2020FrASS...7....8L}, will provide us more copious and precise data for quasars which could help us solve this puzzle. 

Our other main interest is to compare different cosmographic techniques by combining new high-redshift data. We adopted the AIC and BIC selection criteria as tools for inferring the statistical significance of a given scenario with respect to the $\Lambda$CDM model. The corresponding results are shown in Figs. \ref{F6}, \ref{F7} and Tables \ref{T1}, \ref{T2}. By analyzing the theoretical lines produced by different methods and different orders, we found that increasing the order makes the theoretical predictions obtained by different methods closer to the predictions of the $\Lambda$CDM model with $\Omega_{m} = 0.29$. Among them, the increase in the order has the most obvious influence on the $y$-redshift method. This can also be found from the changes in $\Delta$AIC and $\Delta$BIC. For the Pantheon sample, Pad$\rm \acute{e}_{(2,1)}$ turns out to be the best-performing approximation in the third- and fourth-order expansions. The third-order expansions of the $y$-redshift and $E(y)$ methods can be ruled out. For the combined sample, the Pad$\rm \acute{e}_{(2,1)}$ approximation still performs well. The third- and fourth-order expansions of the $y$-redshift and $E(y)$ methods are both excluded. The $E(y)$ method is better than the $y$-redshift method. On the contrary, the $\log(1+z)$ and $\log(1+z) + k_{ij}$ methods have good performance in these two expansion orders. Combining all of the results, we found that the Pad$\rm \acute{e}_{(2,1)}$ method is suitable for low and high redshift cases. The Pad$\rm \acute{e}_{(2,2)}$ method performs well in high redshift situations. The performance of the $E(y)$ method is better than that of the $y$-redshift method. The $\log(1+z)$ and $\log(1+z) + k_{ij}$ methods are better than the $z$-redshift, $y$-redshift, and $E(y)$ methods, but worse than Pad$\rm \acute{e}$ approximations. Except for the $y$-redshift and $E(y)$ methods, the values of $\Delta$AIC and $\Delta$BIC increase with the expansion order. But the constraints of the $y$-redshift and $E(y)$ methods may be problematic, that is, the high-order coefficients are abnormal. To minimize risk, one might only constrain the first two parameters ($q_{0}$ and $j_{0}$), and the other parameters could be marginalized in a large range \citep{2017EPJC...77..434Z,2022ApJ...924...97W}.

\section*{Acknowledgements}
We thank the anonymous referee for helpful comments. This work was supported by the National Natural Science
Foundation of China (grant No. U1831207) and the China Manned Spaced Project (CMS-CSST-2021-A12).

\bibliographystyle{aa} 
\bibliography{aaref} 

\begin{thebibliography}{103}
\expandafter\ifx\csname natexlab\endcsname\relax\def\natexlab#1{#1}\fi

\bibitem[{Akaike(1974)}]{1100705}
Akaike, H. 1974, IEEE Transactions on Automatic Control, 19, 716

\bibitem[{{Aviles} {et~al.}(2014){Aviles}, {Bravetti}, {Capozziello}, \&
  {Luongo}}]{2014PhRvD..90d3531A}
{Aviles}, A., {Bravetti}, A., {Capozziello}, S., \& {Luongo}, O. 2014, \prd,
  90, 043531

\bibitem[{{Avni} \& {Tananbaum}(1986)}]{1986ApJ...305...83A}
{Avni}, Y. \& {Tananbaum}, H. 1986, \apj, 305, 83

\bibitem[{{Banerjee} {et~al.}(2021){Banerjee}, {{\'O} Colg{\'a}in}, {Sasaki},
  {Sheikh-Jabbari}, \& {Yang}}]{2021PhLB..81836366B}
{Banerjee}, A., {{\'O} Colg{\'a}in}, E., {Sasaki}, M., {Sheikh-Jabbari}, M.~M.,
  \& {Yang}, T. 2021, Physics Letters B, 818, 136366

\bibitem[{{Bargiacchi} {et~al.}(2021){Bargiacchi}, {Risaliti}, {Benetti},
  {Capozziello}, {Lusso}, {Saccardi}, \& {Signorini}}]{2021A&A...649A..65B}
{Bargiacchi}, G., {Risaliti}, G., {Benetti}, M., {et~al.} 2021, \aap, 649, A65

\bibitem[{{Betoule} {et~al.}(2014){Betoule}, {Kessler}, {Guy}, {Mosher},
  {Hardin}, {Biswas}, {Astier}, {El-Hage}, {Konig}, {Kuhlmann}, {Marriner},
  {Pain}, {Regnault}, {Balland}, {Bassett}, {Brown}, {Campbell}, {Carlberg},
  {Cellier-Holzem}, {Cinabro}, {Conley}, {D'Andrea}, {DePoy}, {Doi}, {Ellis},
  {Fabbro}, {Filippenko}, {Foley}, {Frieman}, {Fouchez}, {Galbany}, {Goobar},
  {Gupta}, {Hill}, {Hlozek}, {Hogan}, {Hook}, {Howell}, {Jha}, {Le Guillou},
  {Leloudas}, {Lidman}, {Marshall}, {M{\"o}ller}, {Mour{\~a}o}, {Neveu},
  {Nichol}, {Olmstead}, {Palanque-Delabrouille}, {Perlmutter}, {Prieto},
  {Pritchet}, {Richmond}, {Riess}, {Ruhlmann-Kleider}, {Sako}, {Schahmaneche},
  {Schneider}, {Smith}, {Sollerman}, {Sullivan}, {Walton}, \&
  {Wheeler}}]{2014A&A...568A..22B}
{Betoule}, M., {Kessler}, R., {Guy}, J., {et~al.} 2014, \aap, 568, A22

\bibitem[{{Bonilla} {et~al.}(2020){Bonilla}, {D'Agostino}, {Nunes}, \& {de
  Araujo}}]{2020JCAP...03..015B}
{Bonilla}, A., {D'Agostino}, R., {Nunes}, R.~C., \& {de Araujo}, J. C.~N. 2020,
  \jcap, 2020, 015

\bibitem[{{Busti} {et~al.}(2015){Busti}, {de la Cruz-Dombriz}, {Dunsby}, \&
  {S{\'a}ez-G{\'o}mez}}]{2015PhRvD..92l3512B}
{Busti}, V.~C., {de la Cruz-Dombriz}, {\'A}., {Dunsby}, P. K.~S., \&
  {S{\'a}ez-G{\'o}mez}, D. 2015, \prd, 92, 123512

\bibitem[{{Cai} {et~al.}(2022){Cai}, {Guo}, {Wang}, {Yu}, \&
  {Zhou}}]{2022PhRvD.105b1301C}
{Cai}, R.-G., {Guo}, Z.-K., {Wang}, S.-J., {Yu}, W.-W., \& {Zhou}, Y. 2022,
  \prd, 105, L021301

\bibitem[{{Caldwell} \& {Kamionkowski}(2004)}]{2004JCAP...09..009C}
{Caldwell}, R.~R. \& {Kamionkowski}, M. 2004, \jcap, 2004, 009

\bibitem[{{Cao} {et~al.}(2021){Cao}, {Khadka}, \&
  {Ratra}}]{2021MNRAS.tmp.3230C}
{Cao}, S., {Khadka}, N., \& {Ratra}, B. 2021, \mnras
  [\eprint[arXiv]{2110.14840}]

\bibitem[{{Capozziello} {et~al.}(2018{\natexlab{a}}){Capozziello},
  {D'Agostino}, \& {Luongo}}]{2018MNRAS.476.3924C}
{Capozziello}, S., {D'Agostino}, R., \& {Luongo}, O. 2018{\natexlab{a}},
  \mnras, 476, 3924

\bibitem[{{Capozziello} {et~al.}(2018{\natexlab{b}}){Capozziello},
  {D'Agostino}, \& {Luongo}}]{2018JCAP...05..008C}
{Capozziello}, S., {D'Agostino}, R., \& {Luongo}, O. 2018{\natexlab{b}}, \jcap,
  2018, 008

\bibitem[{{Capozziello} {et~al.}(2019){Capozziello}, {D'Agostino}, \&
  {Luongo}}]{2019IJMPD..2830016C}
{Capozziello}, S., {D'Agostino}, R., \& {Luongo}, O. 2019, International
  Journal of Modern Physics D, 28, 1930016

\bibitem[{{Capozziello} {et~al.}(2020){Capozziello}, {D'Agostino}, \&
  {Luongo}}]{2020MNRAS.494.2576C}
{Capozziello}, S., {D'Agostino}, R., \& {Luongo}, O. 2020, \mnras, 494, 2576

\bibitem[{{Capozziello} {et~al.}(2011){Capozziello}, {Lazkoz}, \&
  {Salzano}}]{2011PhRvD..84l4061C}
{Capozziello}, S., {Lazkoz}, R., \& {Salzano}, V. 2011, \prd, 84, 124061

\bibitem[{{Capozziello} \& {Ruchika}(2019)}]{2019MNRAS.484.4484C}
{Capozziello}, S. \& {Ruchika}, Sen, A.~A. 2019, \mnras, 484, 4484

\bibitem[{{Carroll}(2001)}]{2001LRR.....4....1C}
{Carroll}, S.~M. 2001, Living Reviews in Relativity, 4, 1

\bibitem[{{Catto{\"e}n} \& {Visser}(2007)}]{2007CQGra..24.5985C}
{Catto{\"e}n}, C. \& {Visser}, M. 2007, Classical and Quantum Gravity, 24, 5985

\bibitem[{{Chevallier} \& {Polarski}(2001)}]{2001IJMPD..10..213C}
{Chevallier}, M. \& {Polarski}, D. 2001, International Journal of Modern
  Physics D, 10, 213

\bibitem[{{Chiba} \& {Nakamura}(1998)}]{1998tx19.confE.276C}
{Chiba}, T. \& {Nakamura}, T. 1998, in 19th Texas Symposium on Relativistic
  Astrophysics and Cosmology, ed. J.~{Paul}, T.~{Montmerle}, \& E.~{Aubourg},
  276

\bibitem[{{Conley} {et~al.}(2011){Conley}, {Guy}, {Sullivan}, {Regnault},
  {Astier}, {Balland}, {Basa}, {Carlberg}, {Fouchez}, {Hardin}, {Hook},
  {Howell}, {Pain}, {Palanque-Delabrouille}, {Perrett}, {Pritchet}, {Rich},
  {Ruhlmann-Kleider}, {Balam}, {Baumont}, {Ellis}, {Fabbro}, {Fakhouri},
  {Fourmanoit}, {Gonz{\'a}lez-Gait{\'a}n}, {Graham}, {Hudson}, {Hsiao},
  {Kronborg}, {Lidman}, {Mourao}, {Neill}, {Perlmutter}, {Ripoche}, {Suzuki},
  \& {Walker}}]{2011ApJS..192....1C}
{Conley}, A., {Guy}, J., {Sullivan}, M., {et~al.} 2011, \apjs, 192, 1

\bibitem[{{Contreras} {et~al.}(2010){Contreras}, {Hamuy}, {Phillips},
  {Folatelli}, {Suntzeff}, {Persson}, {Stritzinger}, {Boldt}, {Gonz{\'a}lez},
  {Krzeminski}, {Morrell}, {Roth}, {Salgado}, {Maureira}, {Burns}, {Freedman},
  {Madore}, {Murphy}, {Wyatt}, {Li}, \& {Filippenko}}]{2010AJ....139..519C}
{Contreras}, C., {Hamuy}, M., {Phillips}, M.~M., {et~al.} 2010, \aj, 139, 519

\bibitem[{{D'Agostino} \& {Nunes}(2019)}]{2019PhRvD.100d4041D}
{D'Agostino}, R. \& {Nunes}, R.~C. 2019, \prd, 100, 044041

\bibitem[{{Dai} \& {Lu}(1998)}]{1998A&A...333L..87D}
{Dai}, Z.~G. \& {Lu}, T. 1998, \aap, 333, L87

\bibitem[{{Demianski} {et~al.}(2017){Demianski}, {Piedipalumbo}, {Sawant}, \&
  {Amati}}]{2017A&A...598A.113D}
{Demianski}, M., {Piedipalumbo}, E., {Sawant}, D., \& {Amati}, L. 2017, \aap,
  598, A113

\bibitem[{{Demianski} {et~al.}(2021){Demianski}, {Piedipalumbo}, {Sawant}, \&
  {Amati}}]{Demianski2021}
{Demianski}, M., {Piedipalumbo}, E., {Sawant}, D., \& {Amati}, L. 2021, \mnras,
  506, 903

\bibitem[{{Dunsby} \& {Luongo}(2016)}]{2016IJGMM..1330002D}
{Dunsby}, P. K.~S. \& {Luongo}, O. 2016, International Journal of Geometric
  Methods in Modern Physics, 13, 1630002

\bibitem[{{Folatelli} {et~al.}(2010){Folatelli}, {Phillips}, {Burns},
  {Contreras}, {Hamuy}, {Freedman}, {Persson}, {Stritzinger}, {Suntzeff},
  {Krisciunas}, {Boldt}, {Gonz{\'a}lez}, {Krzeminski}, {Morrell}, {Roth},
  {Salgado}, {Madore}, {Murphy}, {Wyatt}, {Li}, {Filippenko}, \&
  {Miller}}]{2010AJ....139..120F}
{Folatelli}, G., {Phillips}, M.~M., {Burns}, C.~R., {et~al.} 2010, \aj, 139,
  120

\bibitem[{{Foreman-Mackey} {et~al.}(2013){Foreman-Mackey}, {Hogg}, {Lang}, \&
  {Goodman}}]{2013PASP..125..306F}
{Foreman-Mackey}, D., {Hogg}, D.~W., {Lang}, D., \& {Goodman}, J. 2013, \pasp,
  125, 306

\bibitem[{{Frieman} {et~al.}(2008){Frieman}, {Bassett}, {Becker}, {Choi},
  {Cinabro}, {DeJongh}, {Depoy}, {Dilday}, {Doi}, {Garnavich}, {Hogan},
  {Holtzman}, {Im}, {Jha}, {Kessler}, {Konishi}, {Lampeitl}, {Marriner},
  {Marshall}, {McGinnis}, {Miknaitis}, {Nichol}, {Prieto}, {Riess}, {Richmond},
  {Romani}, {Sako}, {Schneider}, {Smith}, {Takanashi}, {Tokita}, {van der
  Heyden}, {Yasuda}, {Zheng}, {Adelman-McCarthy}, {Annis}, {Assef},
  {Barentine}, {Bender}, {Blandford}, {Boroski}, {Bremer}, {Brewington},
  {Collins}, {Crotts}, {Dembicky}, {Eastman}, {Edge}, {Edmondson}, {Elson},
  {Eyler}, {Filippenko}, {Foley}, {Frank}, {Goobar}, {Gueth}, {Gunn},
  {Harvanek}, {Hopp}, {Ihara}, {Ivezi{\'c}}, {Kahn}, {Kaplan}, {Kent},
  {Ketzeback}, {Kleinman}, {Kollatschny}, {Kron}, {Krzesi{\'n}ski}, {Lamenti},
  {Leloudas}, {Lin}, {Long}, {Lucey}, {Lupton}, {Malanushenko}, {Malanushenko},
  {McMillan}, {Mendez}, {Morgan}, {Morokuma}, {Nitta}, {Ostman}, {Pan},
  {Rockosi}, {Romer}, {Ruiz-Lapuente}, {Saurage}, {Schlesinger}, {Snedden},
  {Sollerman}, {Stoughton}, {Stritzinger}, {Subba Rao}, {Tucker}, {Vaisanen},
  {Watson}, {Watters}, {Wheeler}, {Yanny}, \& {York}}]{2008AJ....135..338F}
{Frieman}, J.~A., {Bassett}, B., {Becker}, A., {et~al.} 2008, \aj, 135, 338

\bibitem[{{Graur} {et~al.}(2014){Graur}, {Rodney}, {Maoz}, {Riess}, {Jha},
  {Postman}, {Dahlen}, {Holoien}, {McCully}, {Patel}, {Strolger},
  {Ben{\'\i}tez}, {Coe}, {Jouvel}, {Medezinski}, {Molino}, {Nonino}, {Bradley},
  {Koekemoer}, {Balestra}, {Cenko}, {Clubb}, {Dickinson}, {Filippenko},
  {Frederiksen}, {Garnavich}, {Hjorth}, {Jones}, {Leibundgut}, {Matheson},
  {Mobasher}, {Rosati}, {Silverman}, {U}, {Jedruszczuk}, {Li}, {Lin},
  {Mirmelstein}, {Neustadt}, {Ovadia}, \& {Rogers}}]{2014ApJ...783...28G}
{Graur}, O., {Rodney}, S.~A., {Maoz}, D., {et~al.} 2014, \apj, 783, 28

\bibitem[{{Green} {et~al.}(2009){Green}, {Aldcroft}, {Richards}, {Barkhouse},
  {Constantin}, {Haggard}, {Karovska}, {Kim}, {Kim}, {Vikhlinin}, {Anderson},
  {Mossman}, {Kashyap}, {Myers}, {Silverman}, {Wilkes}, \&
  {Tananbaum}}]{2009ApJ...690..644G}
{Green}, P.~J., {Aldcroft}, T.~L., {Richards}, G.~T., {et~al.} 2009, \apj, 690,
  644

\bibitem[{{Gruber} \& {Luongo}(2014)}]{2014PhRvD..89j3506G}
{Gruber}, C. \& {Luongo}, O. 2014, \prd, 89, 103506

\bibitem[{{Hicken} {et~al.}(2009{\natexlab{a}}){Hicken}, {Challis}, {Jha},
  {Kirshner}, {Matheson}, {Modjaz}, {Rest}, {Wood-Vasey}, {Bakos}, {Barton},
  {Berlind}, {Bragg}, {Brice{\~n}o}, {Brown}, {Caldwell}, {Calkins}, {Cho},
  {Ciupik}, {Contreras}, {Dendy}, {Dosaj}, {Durham}, {Eriksen}, {Esquerdo},
  {Everett}, {Falco}, {Fernandez}, {Gaba}, {Garnavich}, {Graves}, {Green},
  {Groner}, {Hergenrother}, {Holman}, {Hradecky}, {Huchra}, {Hutchison},
  {Jerius}, {Jordan}, {Kilgard}, {Krauss}, {Luhman}, {Macri}, {Marrone},
  {McDowell}, {McIntosh}, {McNamara}, {Megeath}, {Mochejska}, {Munoz},
  {Muzerolle}, {Naranjo}, {Narayan}, {Pahre}, {Peters}, {Peterson}, {Rines},
  {Ripman}, {Roussanova}, {Schild}, {Sicilia-Aguilar}, {Sokoloski}, {Smalley},
  {Smith}, {Spahr}, {Stanek}, {Barmby}, {Blondin}, {Stubbs}, {Szentgyorgyi},
  {Torres}, {Vaz}, {Vikhlinin}, {Wang}, {Westover}, {Woods}, \&
  {Zhao}}]{2009ApJ...700..331H}
{Hicken}, M., {Challis}, P., {Jha}, S., {et~al.} 2009{\natexlab{a}}, \apj, 700,
  331

\bibitem[{{Hicken} {et~al.}(2012){Hicken}, {Challis}, {Kirshner}, {Rest},
  {Cramer}, {Wood-Vasey}, {Bakos}, {Berlind}, {Brown}, {Caldwell}, {Calkins},
  {Currie}, {de Kleer}, {Esquerdo}, {Everett}, {Falco}, {Fernandez},
  {Friedman}, {Groner}, {Hartman}, {Holman}, {Hutchins}, {Keys}, {Kipping},
  {Latham}, {Marion}, {Narayan}, {Pahre}, {Pal}, {Peters}, {Perumpilly},
  {Ripman}, {Sipocz}, {Szentgyorgyi}, {Tang}, {Torres}, {Vaz}, {Wolk}, \&
  {Zezas}}]{2012ApJS..200...12H}
{Hicken}, M., {Challis}, P., {Kirshner}, R.~P., {et~al.} 2012, \apjs, 200, 12

\bibitem[{{Hicken} {et~al.}(2009{\natexlab{b}}){Hicken}, {Wood-Vasey},
  {Blondin}, {Challis}, {Jha}, {Kelly}, {Rest}, \&
  {Kirshner}}]{2009ApJ...700.1097H}
{Hicken}, M., {Wood-Vasey}, W.~M., {Blondin}, S., {et~al.} 2009{\natexlab{b}},
  \apj, 700, 1097

\bibitem[{{Hu} {et~al.}(2021){Hu}, {Wang}, \& {Dai}}]{2021MNRAS.507..730H}
{Hu}, J.~P., {Wang}, F.~Y., \& {Dai}, Z.~G. 2021, \mnras, 507, 730

\bibitem[{{Hu} {et~al.}(2020){Hu}, {Wang}, \& {Wang}}]{2020A&A...643A..93H}
{Hu}, J.~P., {Wang}, Y.~Y., \& {Wang}, F.~Y. 2020, \aap, 643, A93

\bibitem[{{Jha} {et~al.}(2006){Jha}, {Kirshner}, {Challis}, {Garnavich},
  {Matheson}, {Soderberg}, {Graves}, {Hicken}, {Alves}, {Arce}, {Balog},
  {Barmby}, {Barton}, {Berlind}, {Bragg}, {Brice{\~n}o}, {Brown}, {Buckley},
  {Caldwell}, {Calkins}, {Carter}, {Concannon}, {Donnelly}, {Eriksen},
  {Fabricant}, {Falco}, {Fiore}, {Garcia}, {G{\'o}mez}, {Grogin}, {Groner},
  {Groot}, {Haisch}, {Hartmann}, {Hergenrother}, {Holman}, {Huchra},
  {Jayawardhana}, {Jerius}, {Kannappan}, {Kim}, {Kleyna}, {Kochanek},
  {Koranyi}, {Krockenberger}, {Lada}, {Luhman}, {Luu}, {Macri}, {Mader},
  {Mahdavi}, {Marengo}, {Marsden}, {McLeod}, {McNamara}, {Megeath}, {Moraru},
  {Mossman}, {Muench}, {Mu{\~n}oz}, {Muzerolle}, {Naranjo}, {Nelson-Patel},
  {Pahre}, {Patten}, {Peters}, {Peters}, {Raymond}, {Rines}, {Schild},
  {Sobczak}, {Spahr}, {Stauffer}, {Stefanik}, {Szentgyorgyi}, {Tollestrup},
  {V{\"a}is{\"a}nen}, {Vikhlinin}, {Wang}, {Willner}, {Wolk}, {Zajac}, {Zhao},
  \& {Stanek}}]{2006AJ....131..527J}
{Jha}, S., {Kirshner}, R.~P., {Challis}, P., {et~al.} 2006, \aj, 131, 527

\bibitem[{{Jin} {et~al.}(2012){Jin}, {Ward}, \& {Done}}]{2012MNRAS.422.3268J}
{Jin}, C., {Ward}, M., \& {Done}, C. 2012, \mnras, 422, 3268

\bibitem[{{Just} {et~al.}(2007){Just}, {Brandt}, {Shemmer}, {Steffen},
  {Schneider}, {Chartas}, \& {Garmire}}]{2007ApJ...665.1004J}
{Just}, D.~W., {Brandt}, W.~N., {Shemmer}, O., {et~al.} 2007, \apj, 665, 1004

\bibitem[{{Kessler} {et~al.}(2009){Kessler}, {Becker}, {Cinabro}, {Vanderplas},
  {Frieman}, {Marriner}, {Davis}, {Dilday}, {Holtzman}, {Jha}, {Lampeitl},
  {Sako}, {Smith}, {Zheng}, {Nichol}, {Bassett}, {Bender}, {Depoy}, {Doi},
  {Elson}, {Filippenko}, {Foley}, {Garnavich}, {Hopp}, {Ihara}, {Ketzeback},
  {Kollatschny}, {Konishi}, {Marshall}, {McMillan}, {Miknaitis}, {Morokuma},
  {M{\"o}rtsell}, {Pan}, {Prieto}, {Richmond}, {Riess}, {Romani}, {Schneider},
  {Sollerman}, {Takanashi}, {Tokita}, {van der Heyden}, {Wheeler}, {Yasuda}, \&
  {York}}]{2009ApJS..185...32K}
{Kessler}, R., {Becker}, A.~C., {Cinabro}, D., {et~al.} 2009, \apjs, 185, 32

\bibitem[{{Khadka} \& {Ratra}(2020{\natexlab{a}})}]{2020MNRAS.492.4456K}
{Khadka}, N. \& {Ratra}, B. 2020{\natexlab{a}}, \mnras, 492, 4456

\bibitem[{{Khadka} \& {Ratra}(2020{\natexlab{b}})}]{2020MNRAS.497..263K}
{Khadka}, N. \& {Ratra}, B. 2020{\natexlab{b}}, \mnras, 497, 263

\bibitem[{{Khadka} \& {Ratra}(2022)}]{2022MNRAS.510.2753K}
{Khadka}, N. \& {Ratra}, B. 2022, \mnras, 510, 2753

\bibitem[{{Lewis}(2019)}]{2019arXiv191013970L}
{Lewis}, A. 2019, arXiv e-prints, arXiv:1910.13970

\bibitem[{{Li} {et~al.}(2020){Li}, {Du}, \& {Xu}}]{2020MNRAS.491.4960L}
{Li}, E.-K., {Du}, M., \& {Xu}, L. 2020, \mnras, 491, 4960

\bibitem[{{Li} {et~al.}(2021){Li}, {Keeley}, {Shafieloo}, {Zheng}, {Cao},
  {Biesiada}, \& {Zhu}}]{2021MNRAS.507..919L}
{Li}, X., {Keeley}, R.~E., {Shafieloo}, A., {et~al.} 2021, \mnras, 507, 919

\bibitem[{{Linder}(2003)}]{2003PhRvL..90i1301L}
{Linder}, E.~V. 2003, \prl, 90, 091301

\bibitem[{Litvinov(1993)}]{1993Litvinov}
Litvinov, G. 1993, Russian J. Math. Phys., 36, 313

\bibitem[{{Liu} {et~al.}(2020){Liu}, {Cao}, {Biesiada}, {Liu}, {Geng}, \&
  {Lian}}]{2020ApJ...899...71L}
{Liu}, T., {Cao}, S., {Biesiada}, M., {et~al.} 2020, \apj, 899, 71

\bibitem[{{Lusso}(2020)}]{2020FrASS...7....8L}
{Lusso}, E. 2020, Frontiers in Astronomy and Space Sciences, 7, 8

\bibitem[{{Lusso} {et~al.}(2019){Lusso}, {Piedipalumbo}, {Risaliti},
  {Paolillo}, {Bisogni}, {Nardini}, \& {Amati}}]{2019A&A...628L...4L}
{Lusso}, E., {Piedipalumbo}, E., {Risaliti}, G., {et~al.} 2019, \aap, 628, L4

\bibitem[{{Lusso} \& {Risaliti}(2016)}]{2016ApJ...819..154L}
{Lusso}, E. \& {Risaliti}, G. 2016, \apj, 819, 154

\bibitem[{{Melia}(2019)}]{2019MNRAS.489..517M}
{Melia}, F. 2019, \mnras, 489, 517

\bibitem[{{Planck Collaboration} {et~al.}(2020){Planck Collaboration},
  {Akrami}, {Ashdown}, {Aumont}, {Baccigalupi}, {Ballardini}, {Banday},
  {Barreiro}, {Bartolo}, {Basak}, {Benabed}, {Bersanelli}, {Bielewicz}, {Bock},
  {Bond}, {Borrill}, {Bouchet}, {Boulanger}, {Bucher}, {Burigana}, {Butler},
  {Calabrese}, {Cardoso}, {Casaponsa}, {Chiang}, {Colombo}, {Combet},
  {Contreras}, {Crill}, {de Bernardis}, {de Zotti}, {Delabrouille}, {Delouis},
  {Di Valentino}, {Diego}, {Dor{\'e}}, {Douspis}, {Ducout}, {Dupac},
  {Efstathiou}, {Elsner}, {En{\ss}lin}, {Eriksen}, {Fantaye},
  {Fernandez-Cobos}, {Finelli}, {Frailis}, {Fraisse}, {Franceschi}, {Frolov},
  {Galeotta}, {Galli}, {Ganga}, {G{\'e}nova-Santos}, {Gerbino}, {Ghosh},
  {Gonz{\'a}lez-Nuevo}, {G{\'o}rski}, {Gruppuso}, {Gudmundsson}, {Hamann},
  {Handley}, {Hansen}, {Herranz}, {Hivon}, {Huang}, {Jaffe}, {Jones},
  {Keih{\"a}nen}, {Keskitalo}, {Kiiveri}, {Kim}, {Krachmalnicoff}, {Kunz},
  {Kurki-Suonio}, {Lagache}, {Lamarre}, {Lasenby}, {Lattanzi}, {Lawrence}, {Le
  Jeune}, {Levrier}, {Liguori}, {Lilje}, {Lindholm}, {L{\'o}pez-Caniego}, {Ma},
  {Mac{\'\i}as-P{\'e}rez}, {Maggio}, {Maino}, {Mandolesi}, {Mangilli},
  {Marcos-Caballero}, {Maris}, {Martin}, {Mart{\'\i}nez-Gonz{\'a}lez},
  {Matarrese}, {Mauri}, {McEwen}, {Meinhold}, {Mennella}, {Migliaccio},
  {Miville-Desch{\^e}nes}, {Molinari}, {Moneti}, {Montier}, {Morgante}, {Moss},
  {Natoli}, {Pagano}, {Paoletti}, {Partridge}, {Perrotta}, {Pettorino},
  {Piacentini}, {Polenta}, {Puget}, {Rachen}, {Reinecke}, {Remazeilles},
  {Renzi}, {Rocha}, {Rosset}, {Roudier}, {Rubi{\~n}o-Mart{\'\i}n},
  {Ruiz-Granados}, {Salvati}, {Savelainen}, {Scott}, {Shellard}, {Sirignano},
  {Sunyaev}, {Suur-Uski}, {Tauber}, {Tavagnacco}, {Tenti}, {Toffolatti},
  {Tomasi}, {Trombetti}, {Valenziano}, {Valiviita}, {Van Tent}, {Vielva},
  {Villa}, {Vittorio}, {Wandelt}, {Wehus}, {Zacchei}, {Zibin}, \&
  {Zonca}}]{2020A&A...641A...7P}
{Planck Collaboration}, {Akrami}, Y., {Ashdown}, M., {et~al.} 2020, \aap, 641,
  A7

\bibitem[{{Rest} {et~al.}(2014){Rest}, {Scolnic}, {Foley}, {Huber}, {Chornock},
  {Narayan}, {Tonry}, {Berger}, {Soderberg}, {Stubbs}, {Riess}, {Kirshner},
  {Smartt}, {Schlafly}, {Rodney}, {Botticella}, {Brout}, {Challis}, {Czekala},
  {Drout}, {Hudson}, {Kotak}, {Leibler}, {Lunnan}, {Marion}, {McCrum},
  {Milisavljevic}, {Pastorello}, {Sanders}, {Smith}, {Stafford}, {Thilker},
  {Valenti}, {Wood-Vasey}, {Zheng}, {Burgett}, {Chambers}, {Denneau}, {Draper},
  {Flewelling}, {Hodapp}, {Kaiser}, {Kudritzki}, {Magnier}, {Metcalfe},
  {Price}, {Sweeney}, {Wainscoat}, \& {Waters}}]{2014ApJ...795...44R}
{Rest}, A., {Scolnic}, D., {Foley}, R.~J., {et~al.} 2014, \apj, 795, 44

\bibitem[{{Rezaei} {et~al.}(2020){Rezaei}, {Pour-Ojaghi}, \&
  {Malekjani}}]{2020ApJ...900...70R}
{Rezaei}, M., {Pour-Ojaghi}, S., \& {Malekjani}, M. 2020, \apj, 900, 70

\bibitem[{{Riess} {et~al.}(1999){Riess}, {Kirshner}, {Schmidt}, {Jha},
  {Challis}, {Garnavich}, {Esin}, {Carpenter}, {Grashius}, {Schild}, {Berlind},
  {Huchra}, {Prosser}, {Falco}, {Benson}, {Brice{\~n}o}, {Brown}, {Caldwell},
  {dell'Antonio}, {Filippenko}, {Goodman}, {Grogin}, {Groner}, {Hughes},
  {Green}, {Jansen}, {Kleyna}, {Luu}, {Macri}, {McLeod}, {McLeod}, {McNamara},
  {McLean}, {Milone}, {Mohr}, {Moraru}, {Peng}, {Peters}, {Prestwich},
  {Stanek}, {Szentgyorgyi}, \& {Zhao}}]{1999AJ....117..707R}
{Riess}, A.~G., {Kirshner}, R.~P., {Schmidt}, B.~P., {et~al.} 1999, \aj, 117,
  707

\bibitem[{{Riess} {et~al.}(2018){Riess}, {Rodney}, {Scolnic}, {Shafer},
  {Strolger}, {Ferguson}, {Postman}, {Graur}, {Maoz}, {Jha}, {Mobasher},
  {Casertano}, {Hayden}, {Molino}, {Hjorth}, {Garnavich}, {Jones}, {Kirshner},
  {Koekemoer}, {Grogin}, {Brammer}, {Hemmati}, {Dickinson}, {Challis}, {Wolff},
  {Clubb}, {Filippenko}, {Nayyeri}, {U}, {Koo}, {Faber}, {Kocevski}, {Bradley},
  \& {Coe}}]{2018ApJ...853..126R}
{Riess}, A.~G., {Rodney}, S.~A., {Scolnic}, D.~M., {et~al.} 2018, \apj, 853,
  126

\bibitem[{{Riess} {et~al.}(2007){Riess}, {Strolger}, {Casertano}, {Ferguson},
  {Mobasher}, {Gold}, {Challis}, {Filippenko}, {Jha}, {Li}, {Tonry}, {Foley},
  {Kirshner}, {Dickinson}, {MacDonald}, {Eisenstein}, {Livio}, {Younger}, {Xu},
  {Dahl{\'e}n}, \& {Stern}}]{2007ApJ...659...98R}
{Riess}, A.~G., {Strolger}, L.-G., {Casertano}, S., {et~al.} 2007, \apj, 659,
  98

\bibitem[{{Riess} {et~al.}(2004){Riess}, {Strolger}, {Tonry}, {Casertano},
  {Ferguson}, {Mobasher}, {Challis}, {Filippenko}, {Jha}, {Li}, {Chornock},
  {Kirshner}, {Leibundgut}, {Dickinson}, {Livio}, {Giavalisco}, {Steidel},
  {Ben{\'\i}tez}, \& {Tsvetanov}}]{2004ApJ...607..665R}
{Riess}, A.~G., {Strolger}, L.-G., {Tonry}, J., {et~al.} 2004, \apj, 607, 665

\bibitem[{{Risaliti} \& {Lusso}(2015)}]{2015ApJ...815...33R}
{Risaliti}, G. \& {Lusso}, E. 2015, \apj, 815, 33

\bibitem[{{Risaliti} \& {Lusso}(2019)}]{2019NatAs...3..272R}
{Risaliti}, G. \& {Lusso}, E. 2019, Nature Astronomy, 3, 272

\bibitem[{{Rodney} {et~al.}(2014){Rodney}, {Riess}, {Strolger}, {Dahlen},
  {Graur}, {Casertano}, {Dickinson}, {Ferguson}, {Garnavich}, {Hayden}, {Jha},
  {Jones}, {Kirshner}, {Koekemoer}, {McCully}, {Mobasher}, {Patel}, {Weiner},
  {Cenko}, {Clubb}, {Cooper}, {Filippenko}, {Frederiksen}, {Hjorth},
  {Leibundgut}, {Matheson}, {Nayyeri}, {Penner}, {Trump}, {Silverman}, {U},
  {Azalee Bostroem}, {Challis}, {Rajan}, {Wolff}, {Faber}, {Grogin}, \&
  {Kocevski}}]{2014AJ....148...13R}
{Rodney}, S.~A., {Riess}, A.~G., {Strolger}, L.-G., {et~al.} 2014, \aj, 148, 13

\bibitem[{{Sako} {et~al.}(2018){Sako}, {Bassett}, {Becker}, {Brown},
  {Campbell}, {Wolf}, {Cinabro}, {D'Andrea}, {Dawson}, {DeJongh}, {Depoy},
  {Dilday}, {Doi}, {Filippenko}, {Fischer}, {Foley}, {Frieman}, {Galbany},
  {Garnavich}, {Goobar}, {Gupta}, {Hill}, {Hayden}, {Hlozek}, {Holtzman},
  {Hopp}, {Jha}, {Kessler}, {Kollatschny}, {Leloudas}, {Marriner}, {Marshall},
  {Miquel}, {Morokuma}, {Mosher}, {Nichol}, {Nordin}, {Olmstead}, {{\"O}stman},
  {Prieto}, {Richmond}, {Romani}, {Sollerman}, {Stritzinger}, {Schneider},
  {Smith}, {Wheeler}, {Yasuda}, \& {Zheng}}]{2018PASP..130f4002S}
{Sako}, M., {Bassett}, B., {Becker}, A.~C., {et~al.} 2018, \pasp, 130, 064002

\bibitem[{{Salvestrini} {et~al.}(2019){Salvestrini}, {Risaliti}, {Bisogni},
  {Lusso}, \& {Vignali}}]{2019A&A...631A.120S}
{Salvestrini}, F., {Risaliti}, G., {Bisogni}, S., {Lusso}, E., \& {Vignali}, C.
  2019, \aap, 631, A120

\bibitem[{Schwarz(1978)}]{10.1214/aos/1176344136}
Schwarz, G. 1978, The Annals of Statistics, 6, 461

\bibitem[{{Scolnic} {et~al.}(2014){Scolnic}, {Rest}, {Riess}, {Huber}, {Foley},
  {Brout}, {Chornock}, {Narayan}, {Tonry}, {Berger}, {Soderberg}, {Stubbs},
  {Kirshner}, {Rodney}, {Smartt}, {Schlafly}, {Botticella}, {Challis},
  {Czekala}, {Drout}, {Hudson}, {Kotak}, {Leibler}, {Lunnan}, {Marion},
  {McCrum}, {Milisavljevic}, {Pastorello}, {Sanders}, {Smith}, {Stafford},
  {Thilker}, {Valenti}, {Wood-Vasey}, {Zheng}, {Burgett}, {Chambers},
  {Denneau}, {Draper}, {Flewelling}, {Hodapp}, {Kaiser}, {Kudritzki},
  {Magnier}, {Metcalfe}, {Price}, {Sweeney}, {Wainscoat}, \&
  {Waters}}]{2014ApJ...795...45S}
{Scolnic}, D., {Rest}, A., {Riess}, A., {et~al.} 2014, \apj, 795, 45

\bibitem[{{Scolnic} {et~al.}(2018){Scolnic}, {Jones}, {Rest}, {Pan},
  {Chornock}, {Foley}, {Huber}, {Kessler}, {Narayan}, {Riess}, {Rodney},
  {Berger}, {Brout}, {Challis}, {Drout}, {Finkbeiner}, {Lunnan}, {Kirshner},
  {Sanders}, {Schlafly}, {Smartt}, {Stubbs}, {Tonry}, {Wood-Vasey}, {Foley},
  {Hand}, {Johnson}, {Burgett}, {Chambers}, {Draper}, {Hodapp}, {Kaiser},
  {Kudritzki}, {Magnier}, {Metcalfe}, {Bresolin}, {Gall}, {Kotak}, {McCrum}, \&
  {Smith}}]{2018ApJ...859..101S}
{Scolnic}, D.~M., {Jones}, D.~O., {Rest}, A., {et~al.} 2018, \apj, 859, 101

\bibitem[{{Shafieloo}(2012)}]{2012JCAP...08..002S}
{Shafieloo}, A. 2012, \jcap, 2012, 002

\bibitem[{{Speri} {et~al.}(2021){Speri}, {Tamanini}, {Caldwell}, {Gair}, \&
  {Wang}}]{2021PhRvD.103h3526S}
{Speri}, L., {Tamanini}, N., {Caldwell}, R.~R., {Gair}, J.~R., \& {Wang}, B.
  2021, \prd, 103, 083526

\bibitem[{{Steffen} {et~al.}(2006){Steffen}, {Strateva}, {Brandt}, {Alexander},
  {Koekemoer}, {Lehmer}, {Schneider}, \& {Vignali}}]{2006AJ....131.2826S}
{Steffen}, A.~T., {Strateva}, I., {Brandt}, W.~N., {et~al.} 2006, \aj, 131,
  2826

\bibitem[{{Steven} \& {Harold}(1992)}]{1992A}
{Steven}, G.~K. \& {Harold}, R.~P. 1992, Birkhuser Advanced Texts, 39, 373

\bibitem[{{Strateva} {et~al.}(2005){Strateva}, {Brandt}, {Schneider}, {Vanden
  Berk}, \& {Vignali}}]{2005AJ....130..387S}
{Strateva}, I.~V., {Brandt}, W.~N., {Schneider}, D.~P., {Vanden Berk}, D.~G.,
  \& {Vignali}, C. 2005, \aj, 130, 387

\bibitem[{{Stritzinger} {et~al.}(2011){Stritzinger}, {Phillips}, {Boldt},
  {Burns}, {Campillay}, {Contreras}, {Gonzalez}, {Folatelli}, {Morrell},
  {Krzeminski}, {Roth}, {Salgado}, {DePoy}, {Hamuy}, {Freedman}, {Madore},
  {Marshall}, {Persson}, {Rheault}, {Suntzeff}, {Villanueva}, {Li}, \&
  {Filippenko}}]{2011AJ....142..156S}
{Stritzinger}, M.~D., {Phillips}, M.~M., {Boldt}, L.~N., {et~al.} 2011, \aj,
  142, 156

\bibitem[{{Sullivan} {et~al.}(2011){Sullivan}, {Guy}, {Conley}, {Regnault},
  {Astier}, {Balland}, {Basa}, {Carlberg}, {Fouchez}, {Hardin}, {Hook},
  {Howell}, {Pain}, {Palanque-Delabrouille}, {Perrett}, {Pritchet}, {Rich},
  {Ruhlmann-Kleider}, {Balam}, {Baumont}, {Ellis}, {Fabbro}, {Fakhouri},
  {Fourmanoit}, {Gonz{\'a}lez-Gait{\'a}n}, {Graham}, {Hudson}, {Hsiao},
  {Kronborg}, {Lidman}, {Mourao}, {Neill}, {Perlmutter}, {Ripoche}, {Suzuki},
  \& {Walker}}]{2011ApJ...737..102S}
{Sullivan}, M., {Guy}, J., {Conley}, A., {et~al.} 2011, \apj, 737, 102

\bibitem[{{Suzuki} {et~al.}(2012){Suzuki}, {Rubin}, {Lidman}, {Aldering},
  {Amanullah}, {Barbary}, {Barrientos}, {Botyanszki}, {Brodwin}, {Connolly},
  {Dawson}, {Dey}, {Doi}, {Donahue}, {Deustua}, {Eisenhardt}, {Ellingson},
  {Faccioli}, {Fadeyev}, {Fakhouri}, {Fruchter}, {Gilbank}, {Gladders},
  {Goldhaber}, {Gonzalez}, {Goobar}, {Gude}, {Hattori}, {Hoekstra}, {Hsiao},
  {Huang}, {Ihara}, {Jee}, {Johnston}, {Kashikawa}, {Koester}, {Konishi},
  {Kowalski}, {Linder}, {Lubin}, {Melbourne}, {Meyers}, {Morokuma}, {Munshi},
  {Mullis}, {Oda}, {Panagia}, {Perlmutter}, {Postman}, {Pritchard}, {Rhodes},
  {Ripoche}, {Rosati}, {Schlegel}, {Spadafora}, {Stanford}, {Stanishev},
  {Stern}, {Strovink}, {Takanashi}, {Tokita}, {Wagner}, {Wang}, {Yasuda},
  {Yee}, \& {Supernova Cosmology Project}}]{2012ApJ...746...85S}
{Suzuki}, N., {Rubin}, D., {Lidman}, C., {et~al.} 2012, \apj, 746, 85

\bibitem[{{Tripp}(1998)}]{1998A&A...331..815T}
{Tripp}, R. 1998, \aap, 331, 815

\bibitem[{{Vignali} {et~al.}(2003){Vignali}, {Brandt}, \&
  {Schneider}}]{2003AJ....125..433V}
{Vignali}, C., {Brandt}, W.~N., \& {Schneider}, D.~P. 2003, \aj, 125, 433

\bibitem[{{Visser}(2004)}]{2004CQGra..21.2603V}
{Visser}, M. 2004, Classical and Quantum Gravity, 21, 2603

\bibitem[{{Visser}(2015)}]{2015CQGra..32m5007V}
{Visser}, M. 2015, Classical and Quantum Gravity, 32, 135007

\bibitem[{{Vitagliano} {et~al.}(2010){Vitagliano}, {Xia}, {Liberati}, \&
  {Viel}}]{2010JCAP...03..005V}
{Vitagliano}, V., {Xia}, J.-Q., {Liberati}, S., \& {Viel}, M. 2010, \jcap,
  2010, 005

\bibitem[{{Wang} {et~al.}(2015){Wang}, {Dai}, \& {Liang}}]{2015NewAR..67....1W}
{Wang}, F.~Y., {Dai}, Z.~G., \& {Liang}, E.~W. 2015, \nar, 67, 1

\bibitem[{{Wang} {et~al.}(2009){Wang}, {Dai}, \& {Qi}}]{2009A&A...507...53W}
{Wang}, F.~Y., {Dai}, Z.~G., \& {Qi}, S. 2009, \aap, 507, 53

\bibitem[{{Wang} {et~al.}(2022){Wang}, {Hu}, {Zhang}, \&
  {Dai}}]{2022ApJ...924...97W}
{Wang}, F.~Y., {Hu}, J.~P., {Zhang}, G.~Q., \& {Dai}, Z.~G. 2022, \apj, 924, 97

\bibitem[{{Wang} \& {Wang}(2014)}]{2014MNRAS.443.1680W}
{Wang}, J.~S. \& {Wang}, F.~Y. 2014, \mnras, 443, 1680

\bibitem[{{Wang} {et~al.}(2016){Wang}, {Wang}, {Cheng}, \& {Dai}}]{Wang2016}
{Wang}, J.~S., {Wang}, F.~Y., {Cheng}, K.~S., \& {Dai}, Z.~G. 2016, \aap, 585,
  A68

\bibitem[{{Wei} {et~al.}(2014){Wei}, {Yan}, \& {Zhou}}]{2014JCAP...01..045W}
{Wei}, H., {Yan}, X.-P., \& {Zhou}, Y.-N. 2014, \jcap, 2014, 045

\bibitem[{{Wei} \& {Melia}(2020)}]{2020ApJ...888...99W}
{Wei}, J.-J. \& {Melia}, F. 2020, \apj, 888, 99

\bibitem[{{Wei} \& {Wu}(2017)}]{Wei2017}
{Wei}, J.-J. \& {Wu}, X.-F. 2017, International Journal of Modern Physics D,
  26, 1730002

\bibitem[{{Weinberg}(1972)}]{1972gcpa.book.....W}
{Weinberg}, S. 1972, {Gravitation and Cosmology: Principles and Applications of
  the General Theory of Relativity}

\bibitem[{{Yang} {et~al.}(2020){Yang}, {Banerjee}, \& {{\'O}
  Colg{\'a}in}}]{2020PhRvD.102l3532Y}
{Yang}, T., {Banerjee}, A., \& {{\'O} Colg{\'a}in}, E. 2020, \prd, 102, 123532

\bibitem[{{Yin} \& {Wei}(2019)}]{2019EPJC...79..698Y}
{Yin}, Z.-Y. \& {Wei}, H. 2019, European Physical Journal C, 79, 698

\bibitem[{{Young} {et~al.}(2009){Young}, {Elvis}, \&
  {Risaliti}}]{2009ApJS..183...17Y}
{Young}, M., {Elvis}, M., \& {Risaliti}, G. 2009, \apjs, 183, 17

\bibitem[{{Young} {et~al.}(2010){Young}, {Elvis}, \&
  {Risaliti}}]{2010ApJ...708.1388Y}
{Young}, M., {Elvis}, M., \& {Risaliti}, G. 2010, \apj, 708, 1388

\bibitem[{{Yu} {et~al.}(2018){Yu}, {Ratra}, \& {Wang}}]{2018ApJ...856....3Y}
{Yu}, H., {Ratra}, B., \& {Wang}, F.-Y. 2018, \apj, 856, 3

\bibitem[{{Zamora Mun{\~o}z} \& {Escamilla-Rivera}(2020)}]{2020JCAP...12..007Z}
{Zamora Mun{\~o}z}, C. \& {Escamilla-Rivera}, C. 2020, \jcap, 2020, 007

\bibitem[{{Zhang} \& {M{\'e}sz{\'a}ros}(2001)}]{2001ApJ...552L..35Z}
{Zhang}, B. \& {M{\'e}sz{\'a}ros}, P. 2001, \apjl, 552, L35

\bibitem[{{Zhang} {et~al.}(2017){Zhang}, {Li}, \& {Xia}}]{2017EPJC...77..434Z}
{Zhang}, M.-J., {Li}, H., \& {Xia}, J.-Q. 2017, European Physical Journal C,
  77, 434

\bibitem[{{Zhao} \& {Xia}(2021)}]{2021EPJC...81..948Z}
{Zhao}, D. \& {Xia}, J.-Q. 2021, European Physical Journal C, 81, 948

\bibitem[{{Zheng} {et~al.}(2021){Zheng}, {Cao}, {Biesiada}, {Li}, {Liu}, \&
  {Liu}}]{2021SCPMA..6459511Z}
{Zheng}, X., {Cao}, S., {Biesiada}, M., {et~al.} 2021, Science China Physics,
  Mechanics, and Astronomy, 64, 259511

\end{thebibliography}

\begin{appendix}
        \section{ 1$\sigma$ and 2$\sigma$ contours in the 2D parameter space utilizing different expansion orders and datasets }
        
        \begin{figure*}
                \centering
                \includegraphics[width=0.3\textwidth]{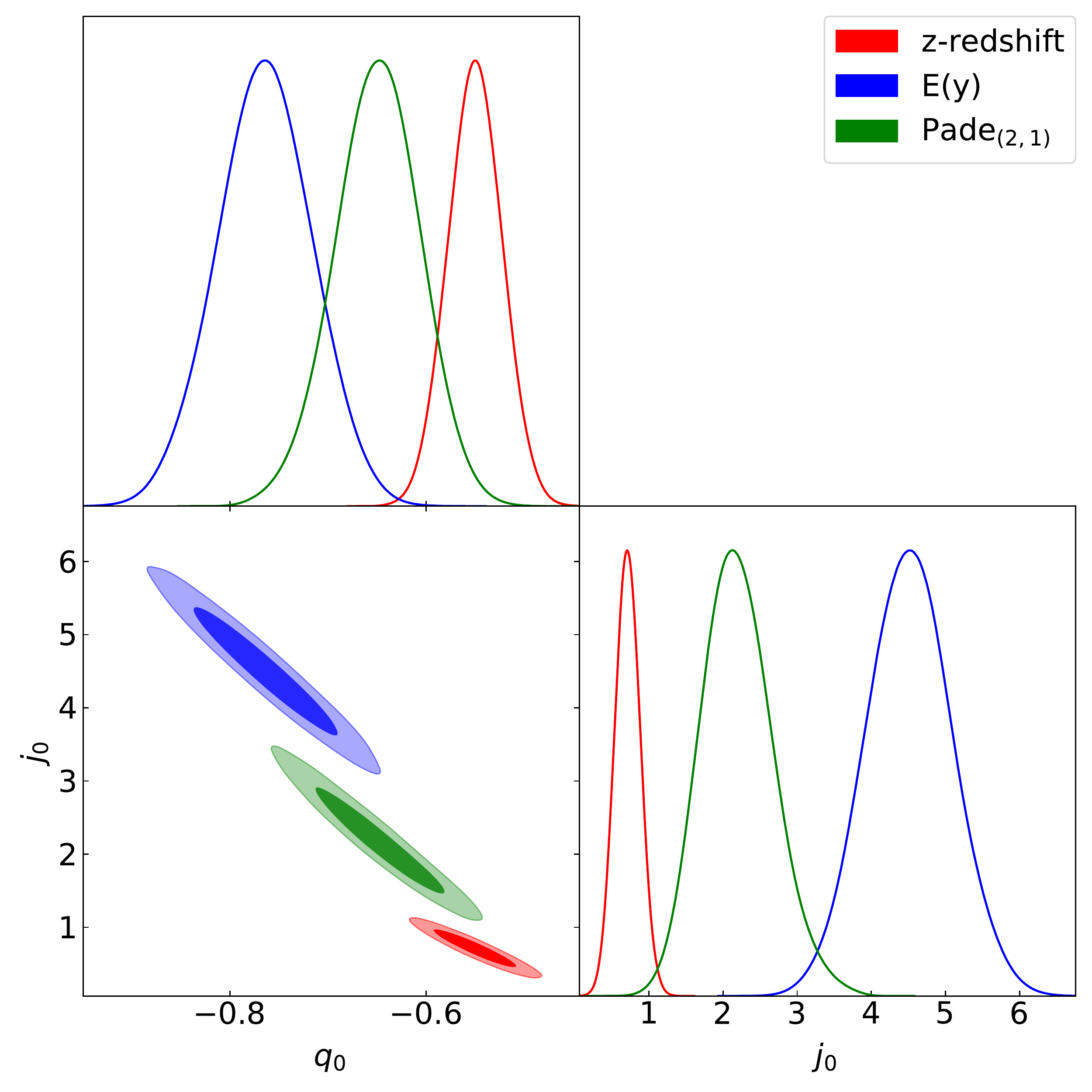}
                \includegraphics[width=0.3\textwidth]{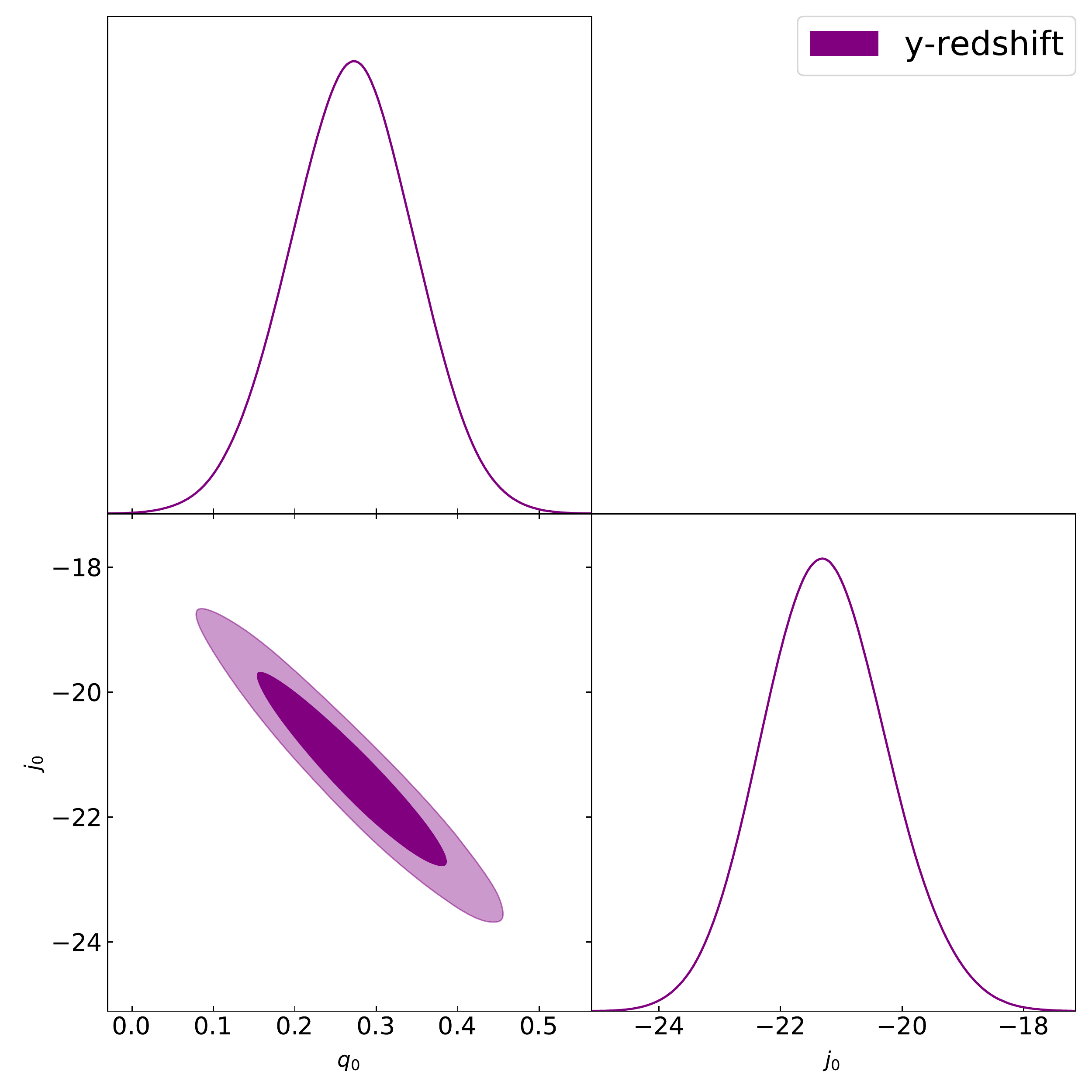} 

                \includegraphics[width=0.3\textwidth]{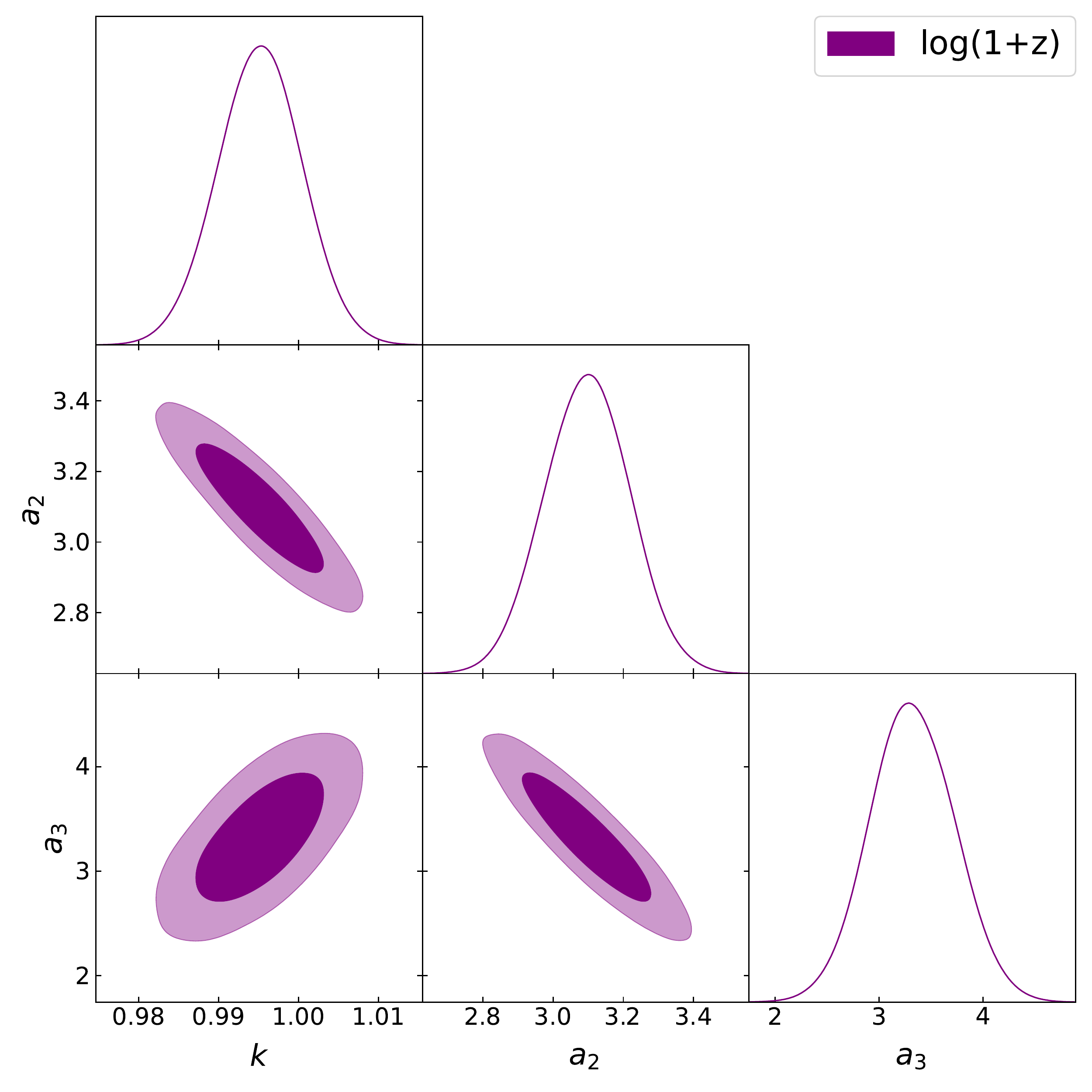} 
                \includegraphics[width=0.3\textwidth]{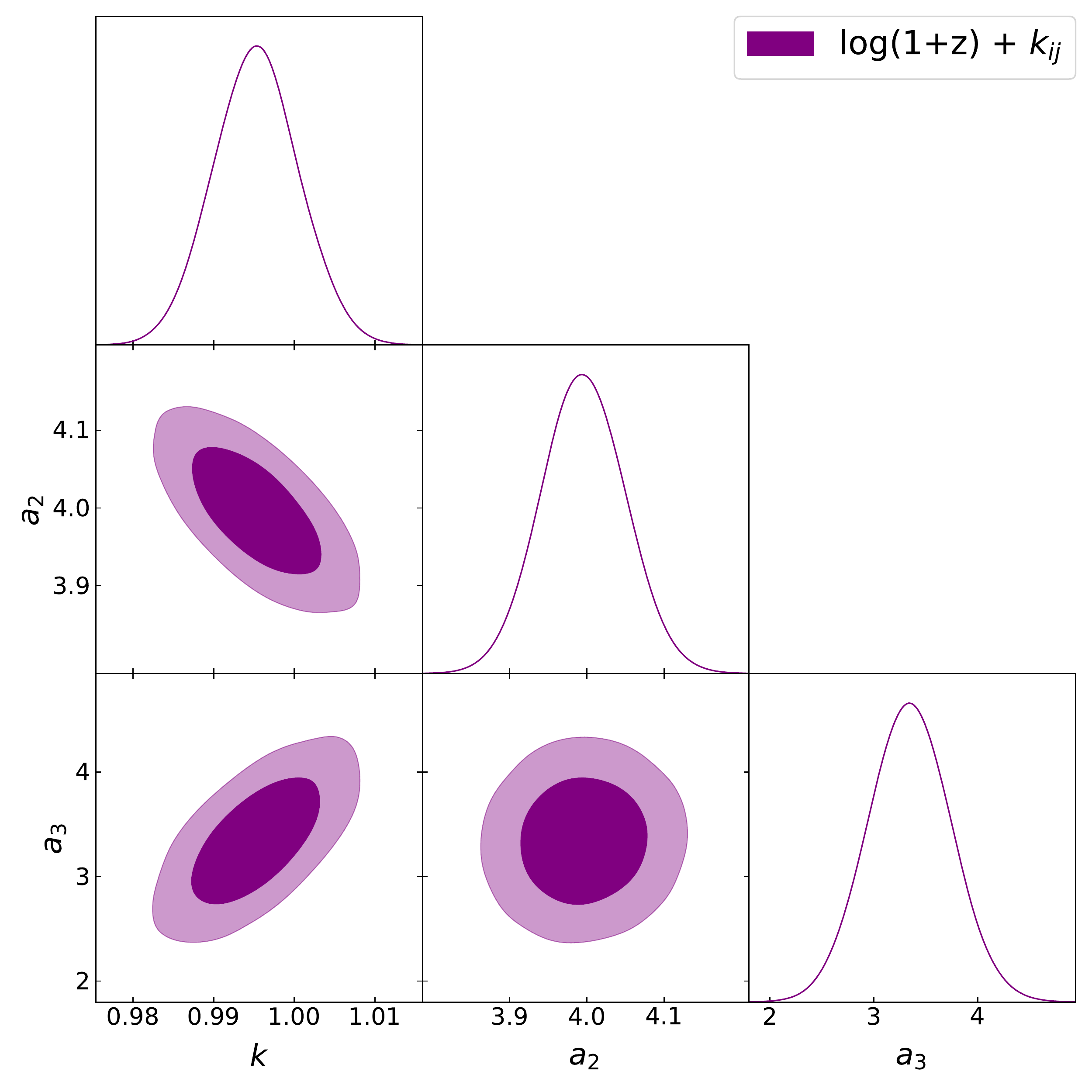}
                \caption{ Confidence contours ($1\sigma$ and $2\sigma$) for the cosmographic parameters space using the Pantheon sample and the third-order expansion. }
                \label{F8}       
        \end{figure*}

        \begin{figure*}
                \centering
                \includegraphics[width=0.3\textwidth]{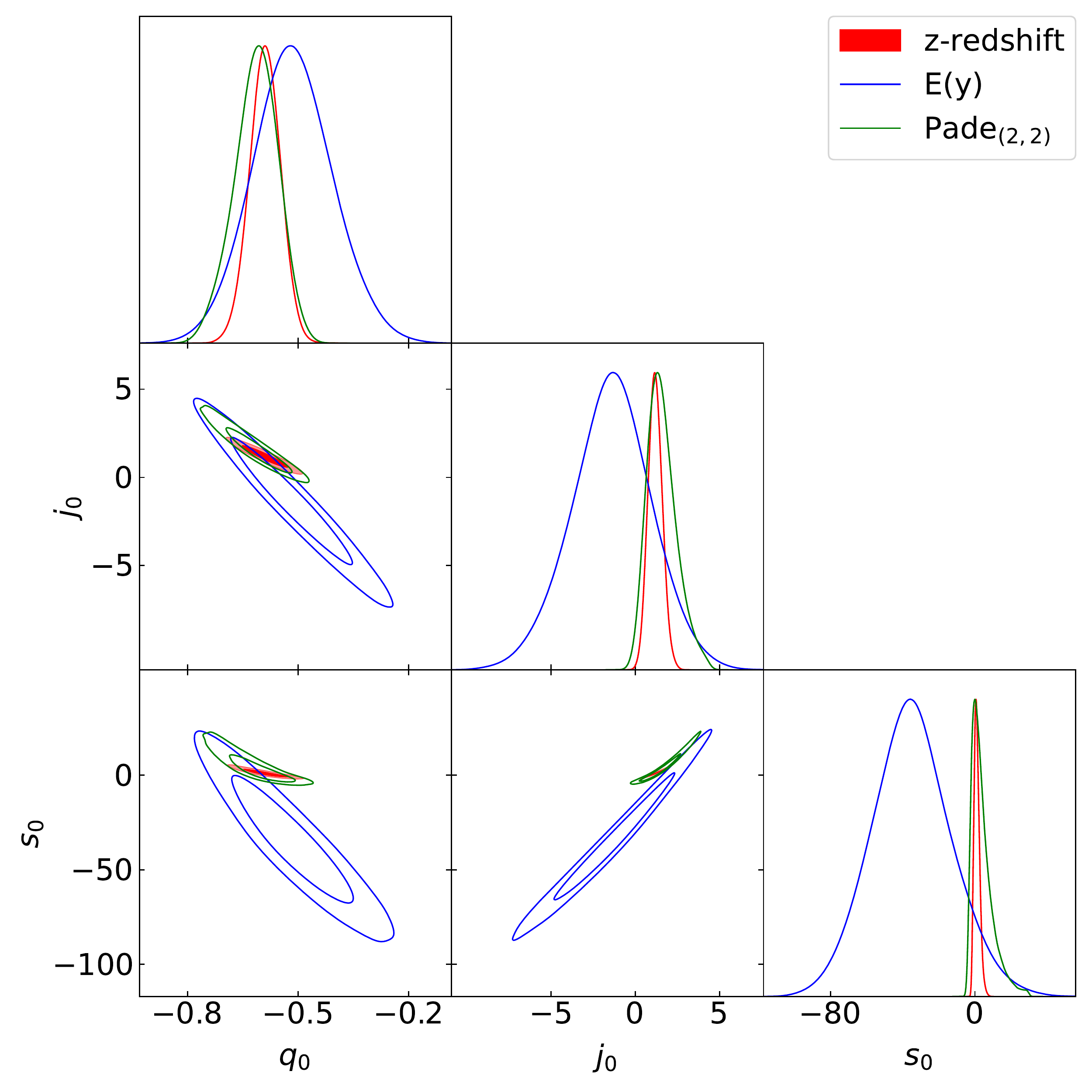}
                \includegraphics[width=0.3\textwidth]{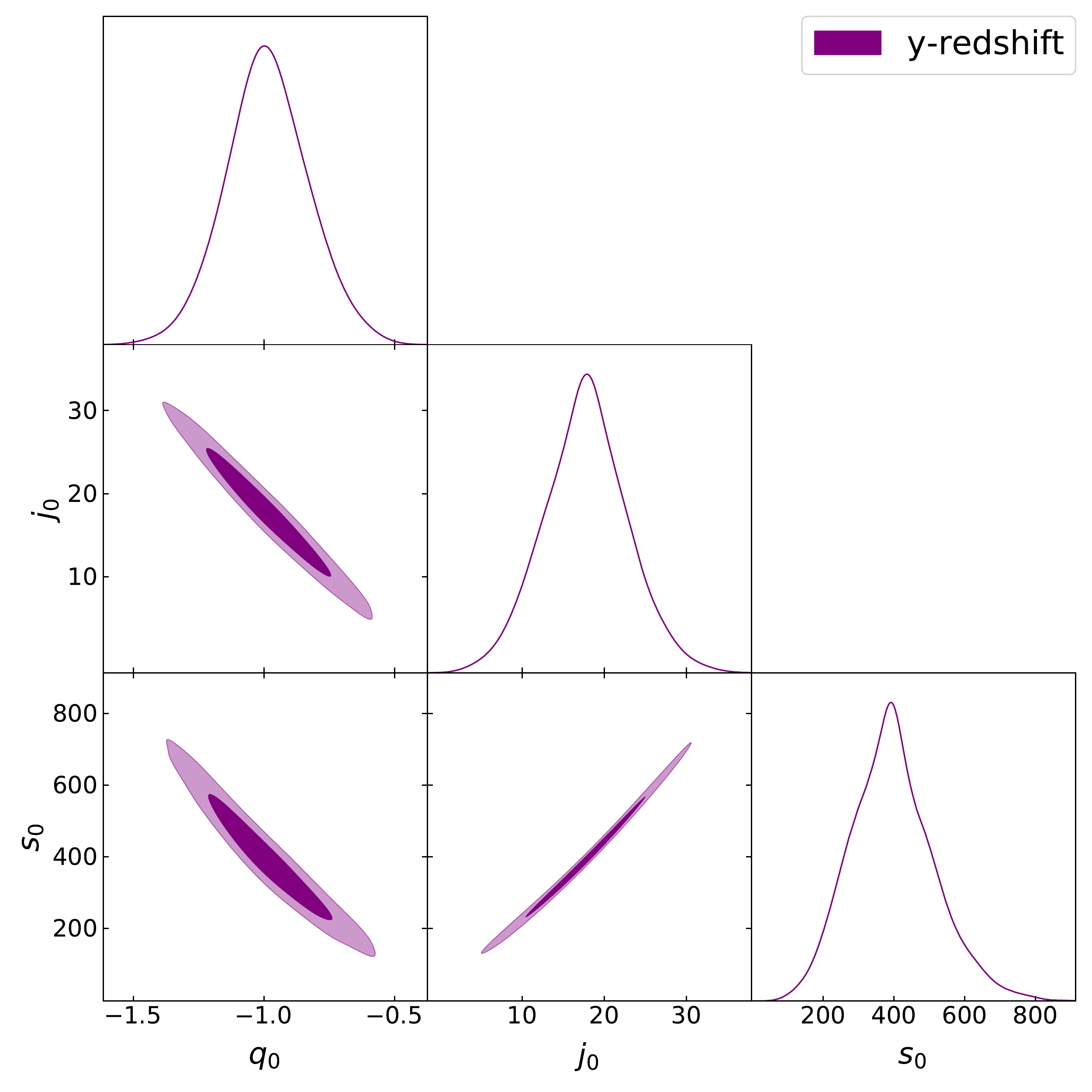} 

                \includegraphics[width=0.3\textwidth]{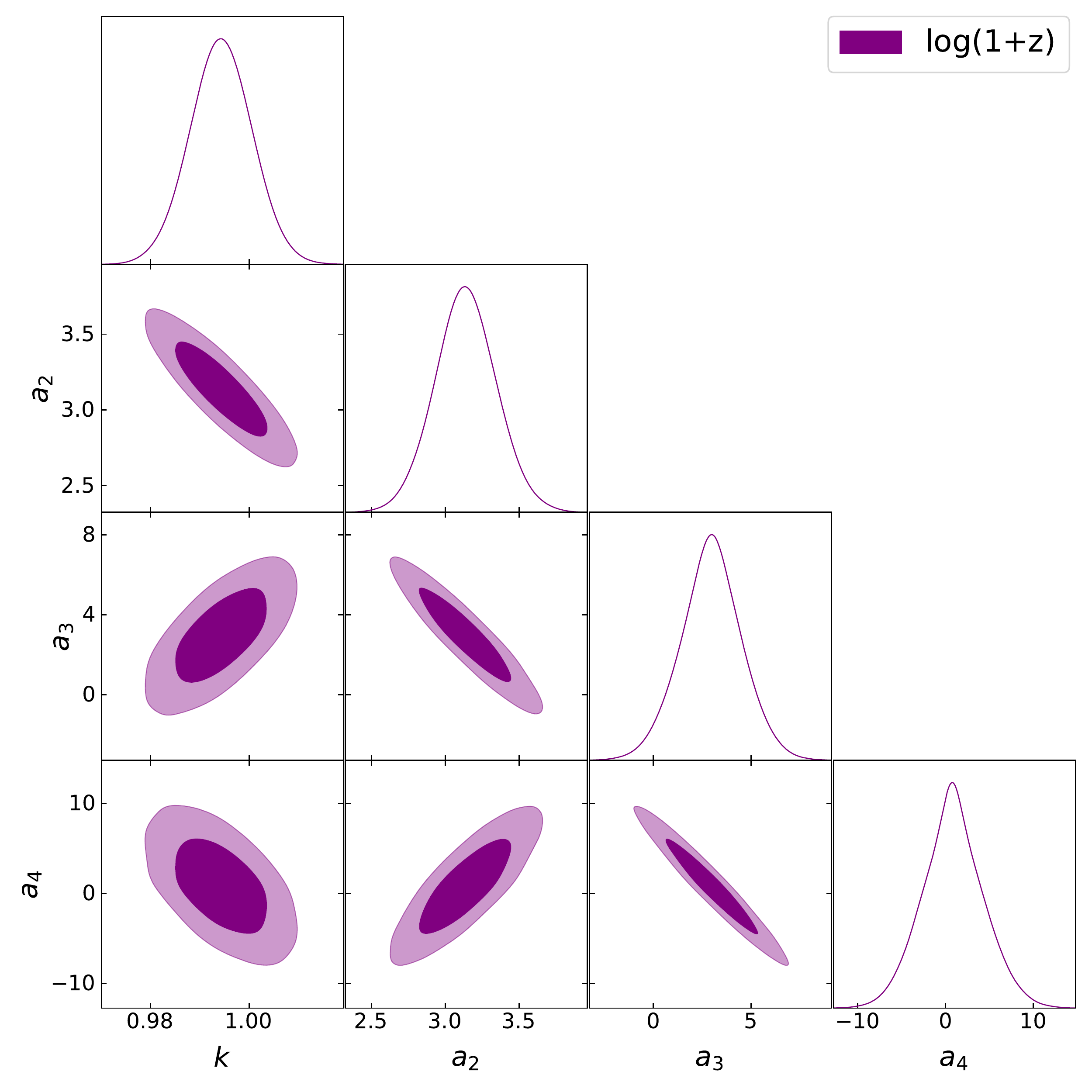} 
                \includegraphics[width=0.3\textwidth]{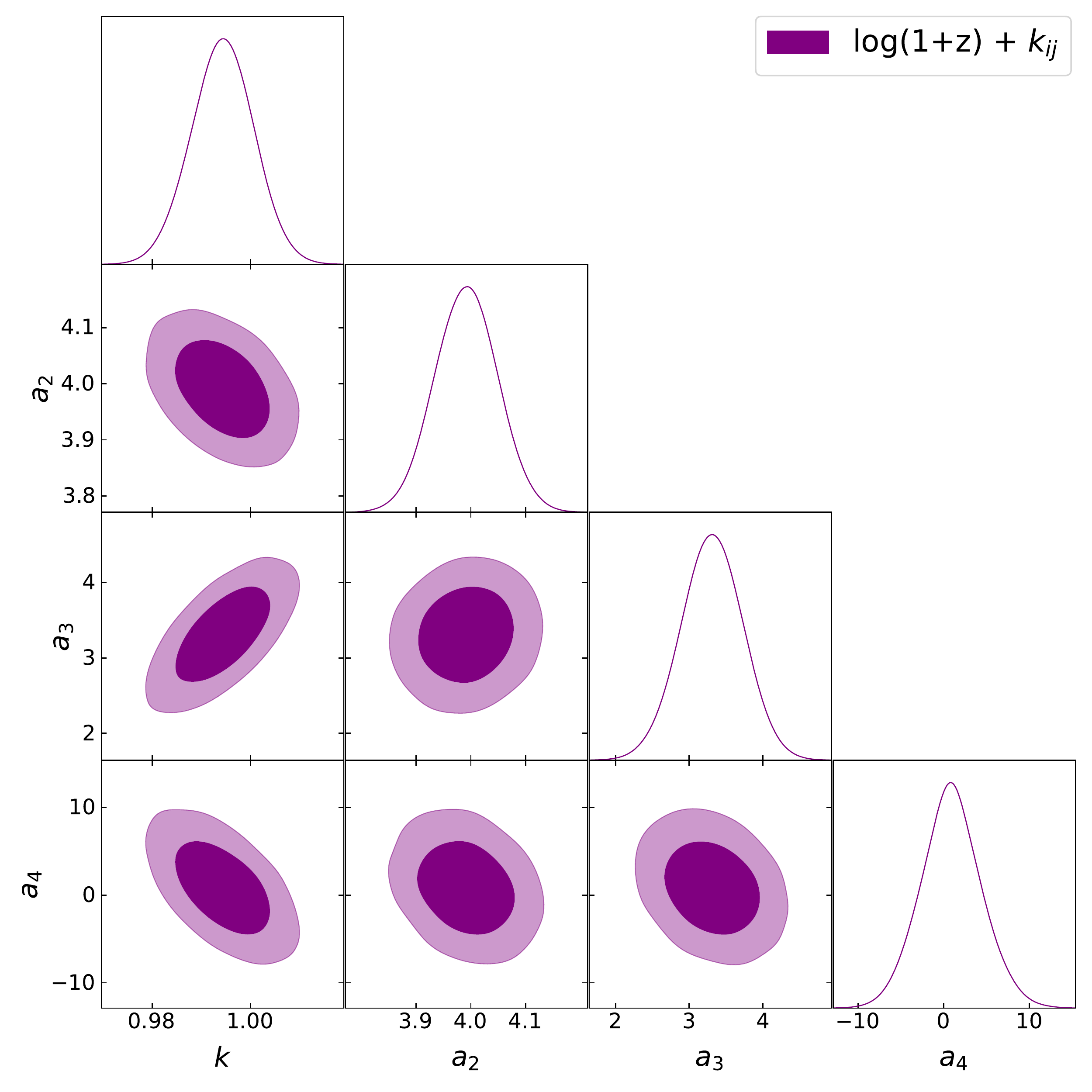}
                \caption{ Confidence contours ($1\sigma$ and $2\sigma$) for the cosmographic parameters space using the Pantheon sample and the fourth-order expansion. }
                \label{F9}       
        \end{figure*}

        \begin{figure*}
                \centering
                \includegraphics[width=0.3\textwidth]{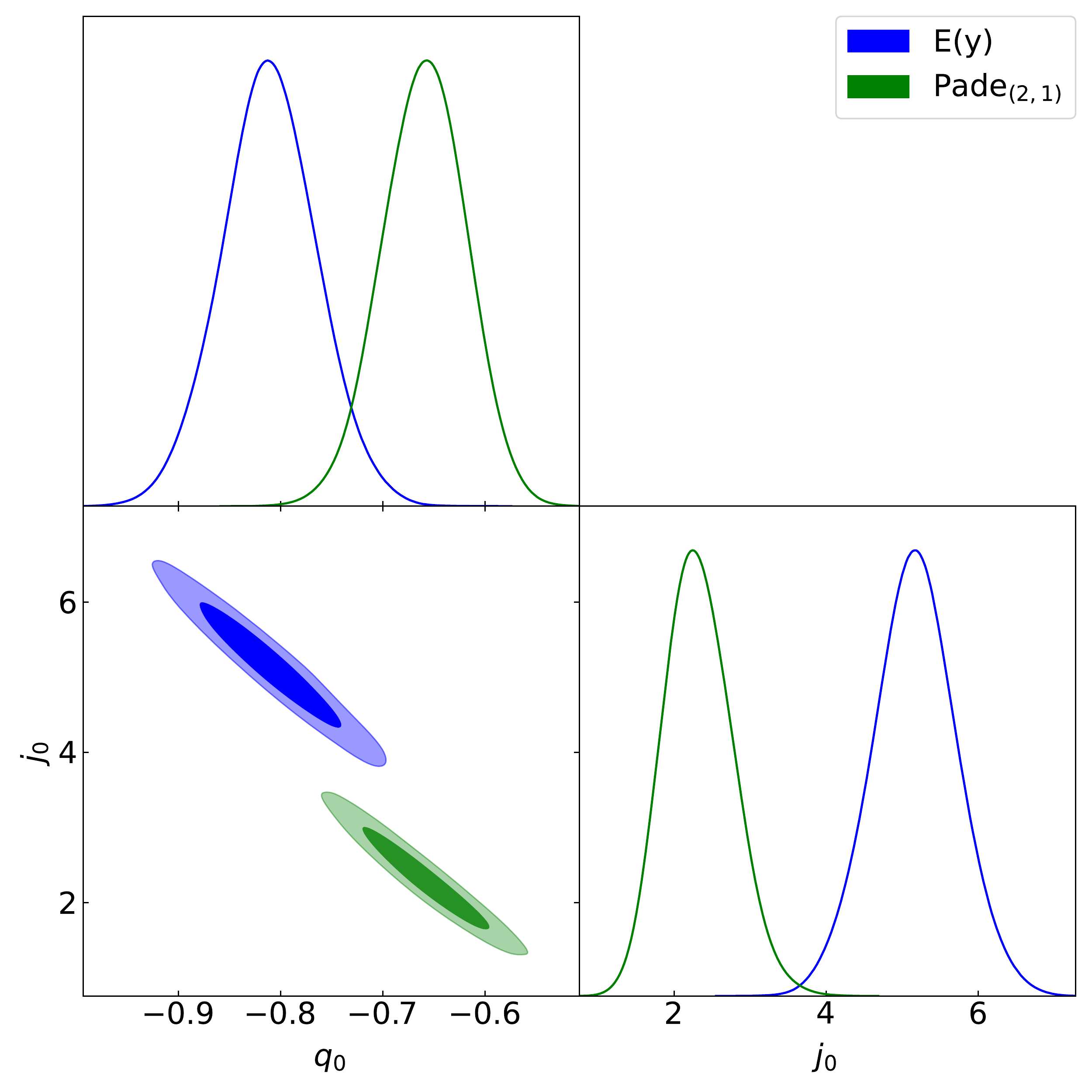} 
                \includegraphics[width=0.3\textwidth]{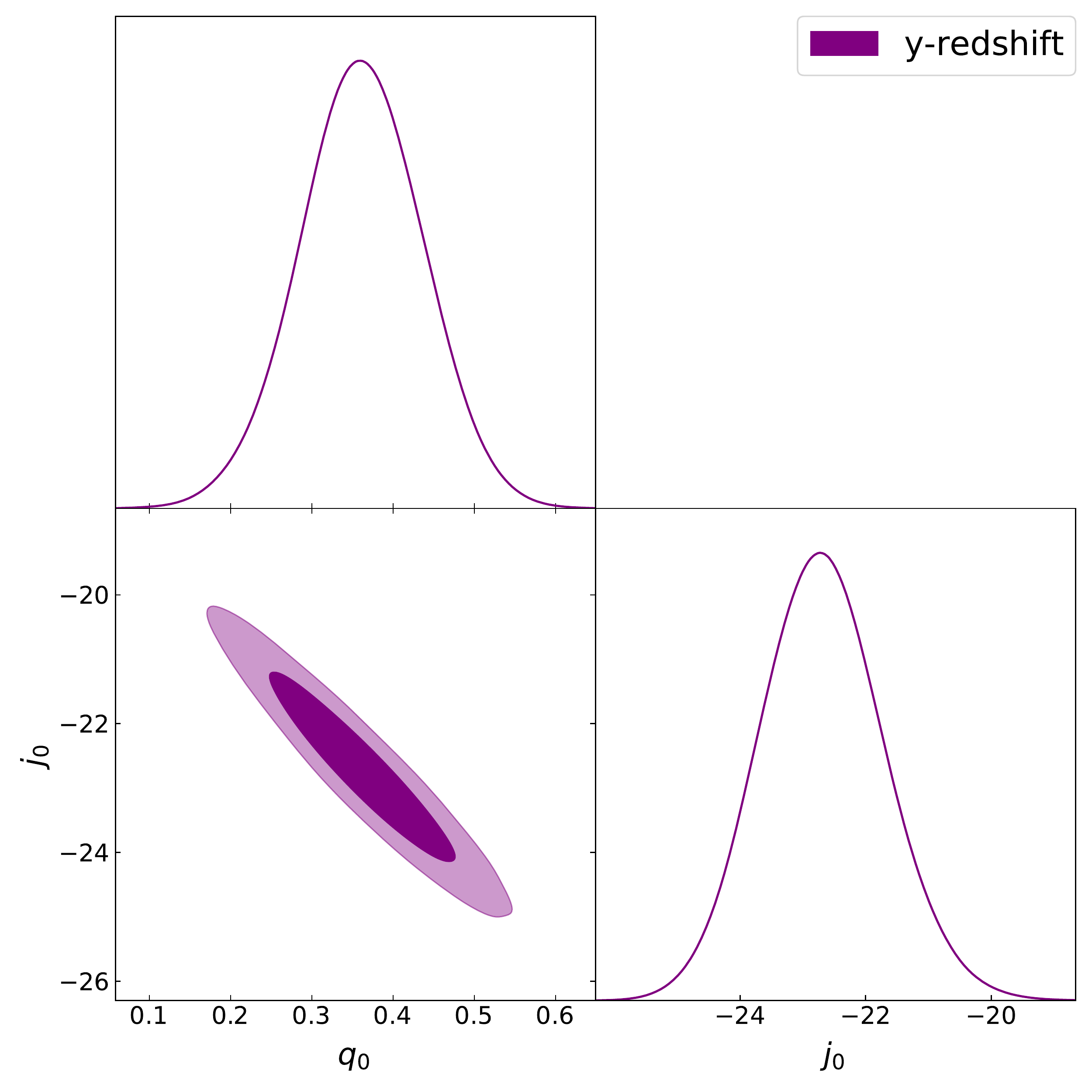}
 
                \includegraphics[width=0.3\textwidth]{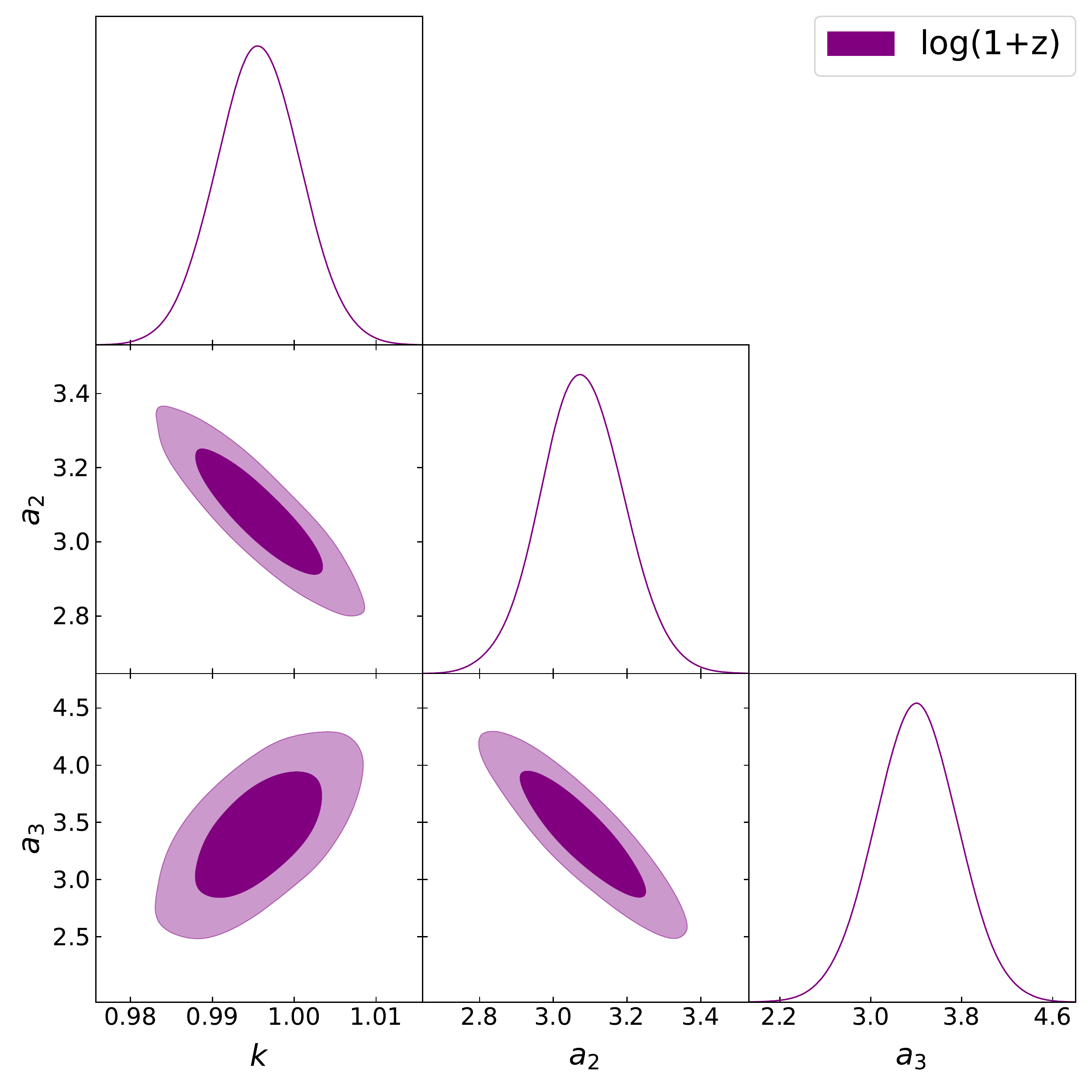} 
                \includegraphics[width=0.3\textwidth]{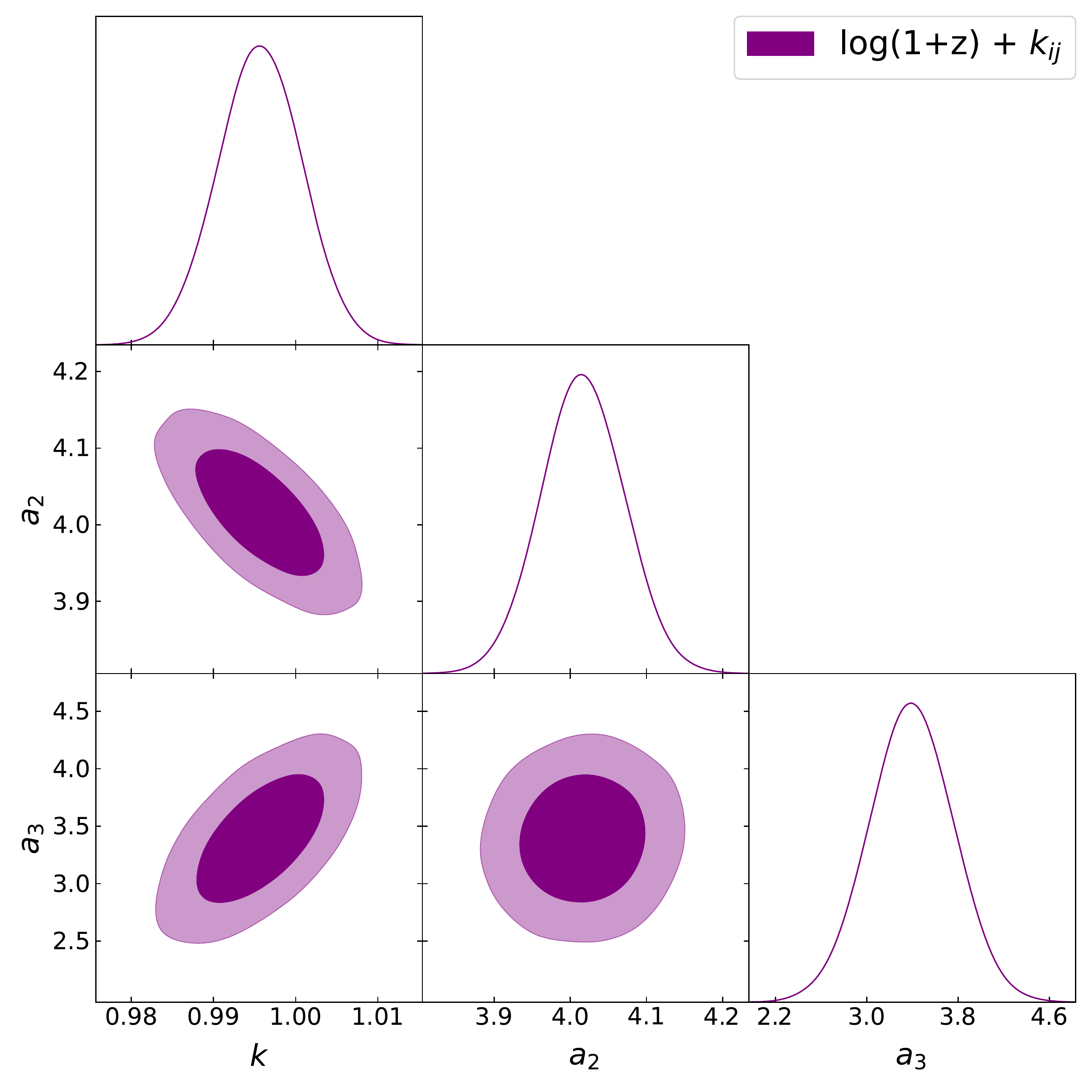}
                \caption{ Confidence contours ($1\sigma$ and $2\sigma$) for the cosmographic parameters space using the combined sample and the third-order expansion. }
                \label{F10}       
        \end{figure*}

        \begin{figure*}
                \centering
                \includegraphics[width=0.3\textwidth]{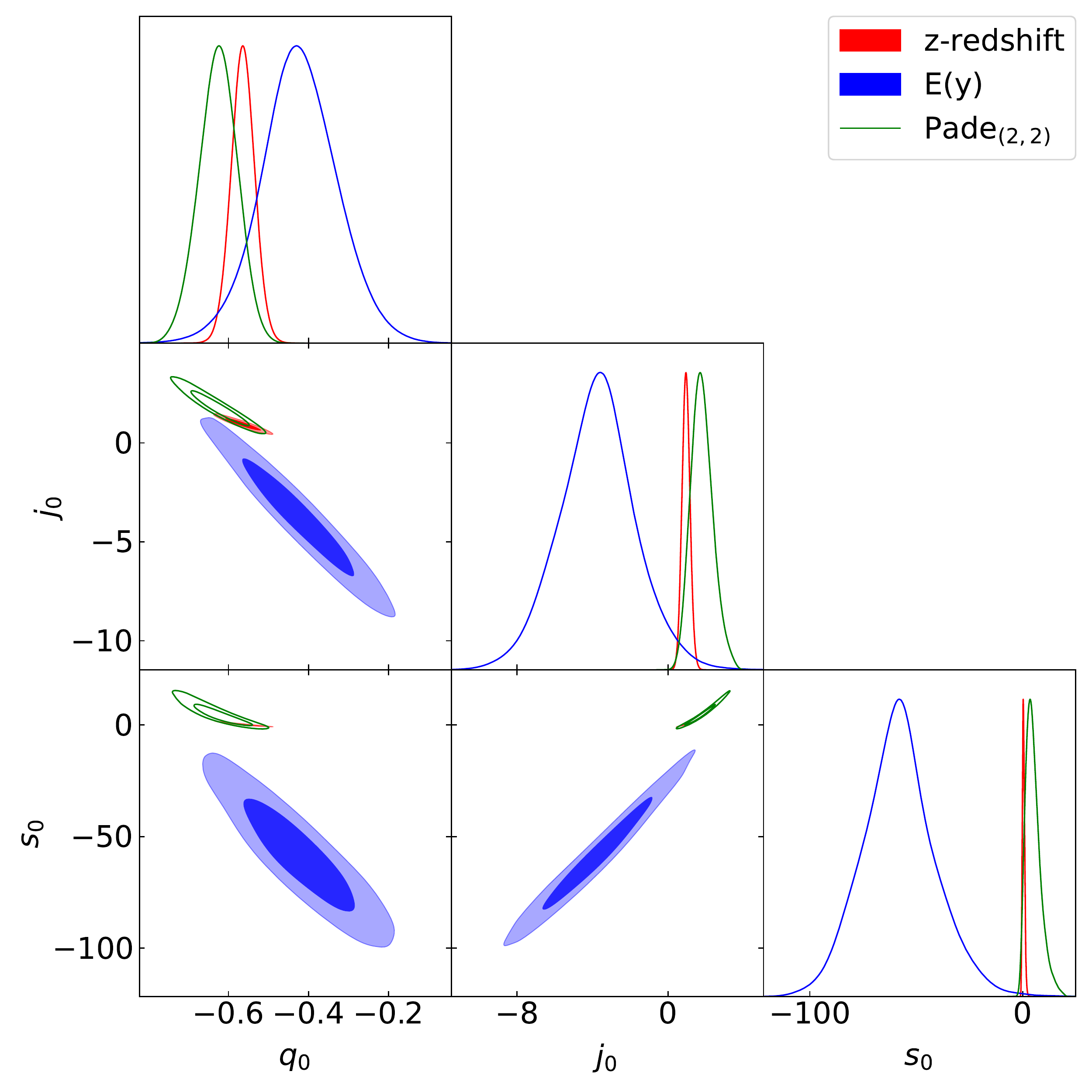}
                \includegraphics[width=0.3\textwidth]{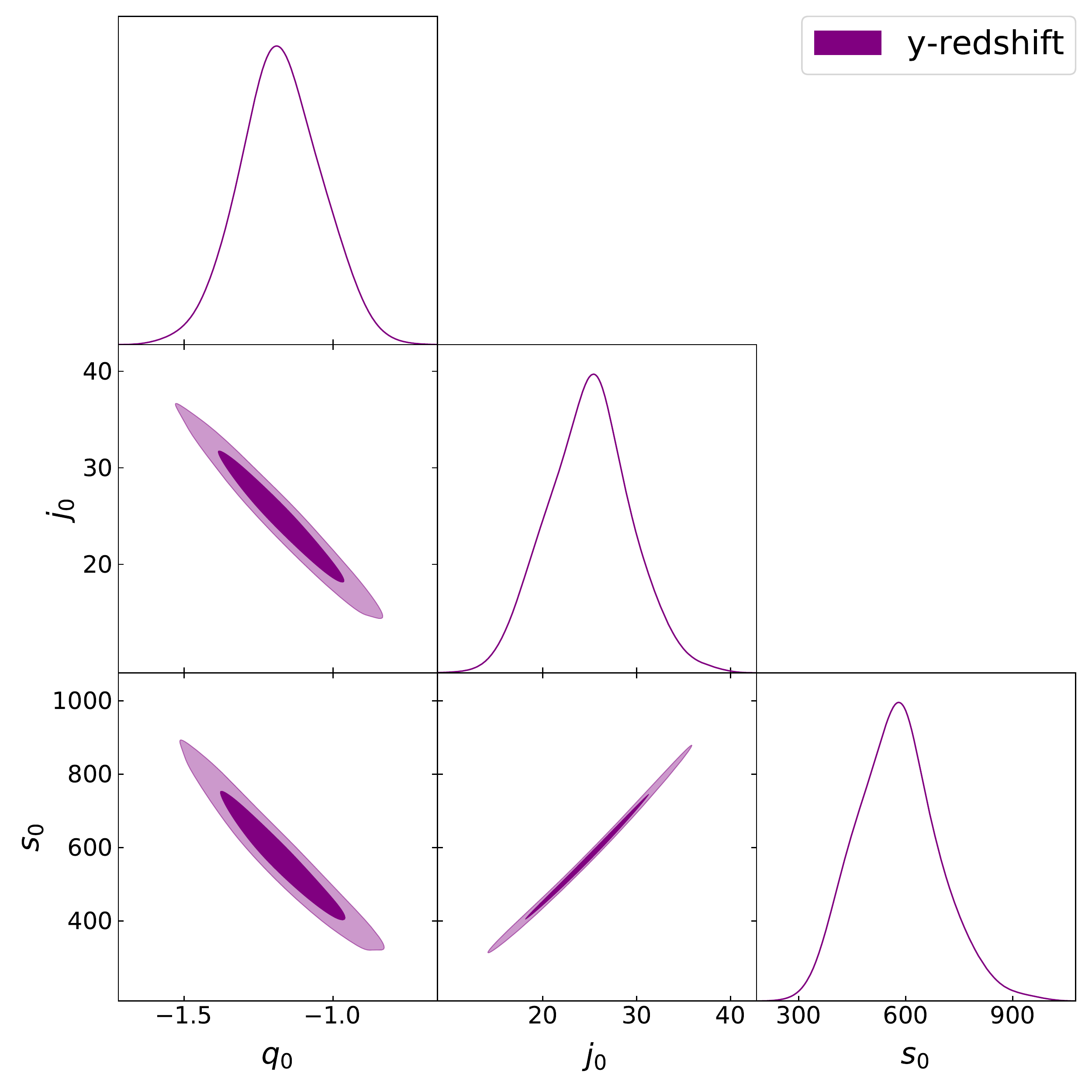} 

                \includegraphics[width=0.3\textwidth]{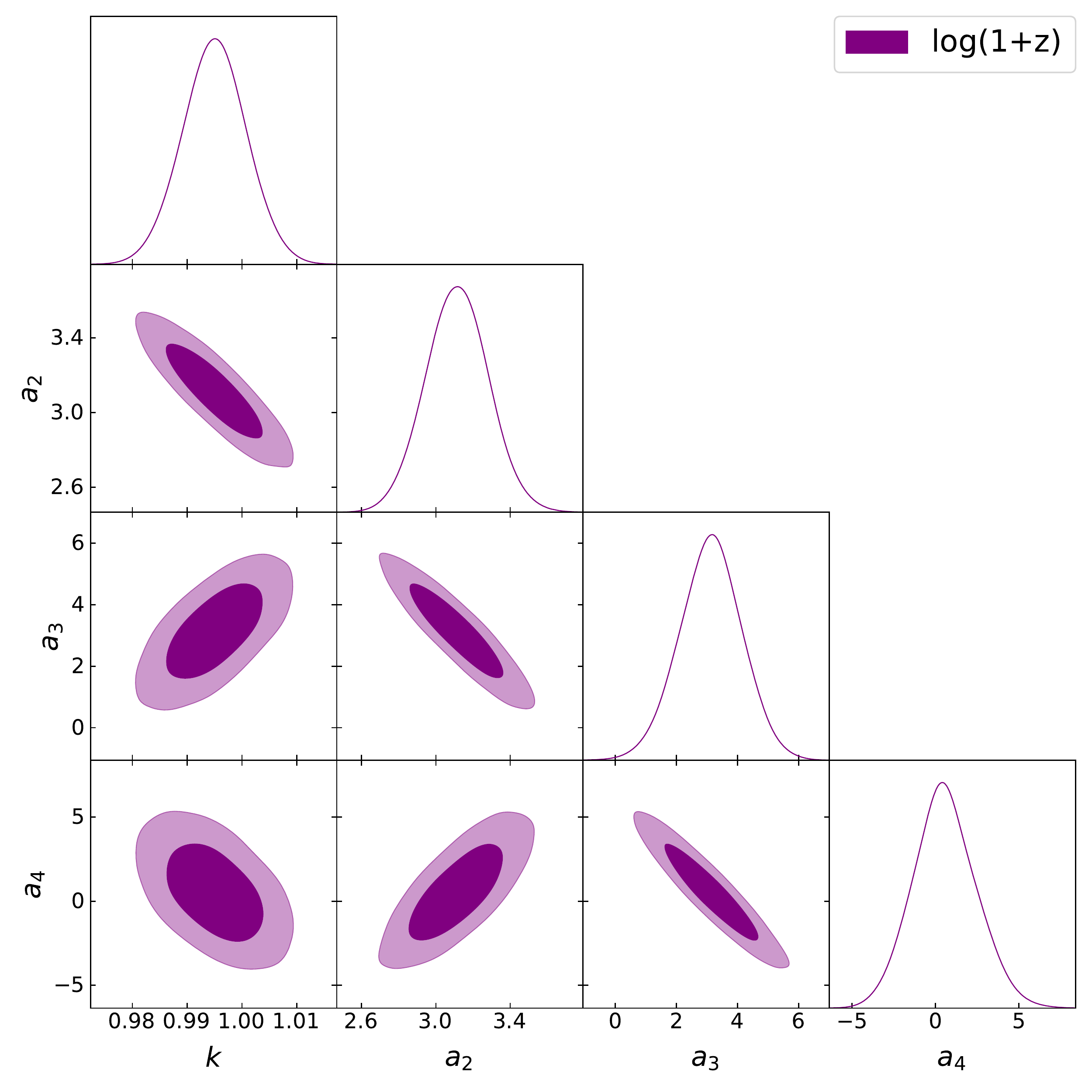}
                \includegraphics[width=0.3\textwidth]{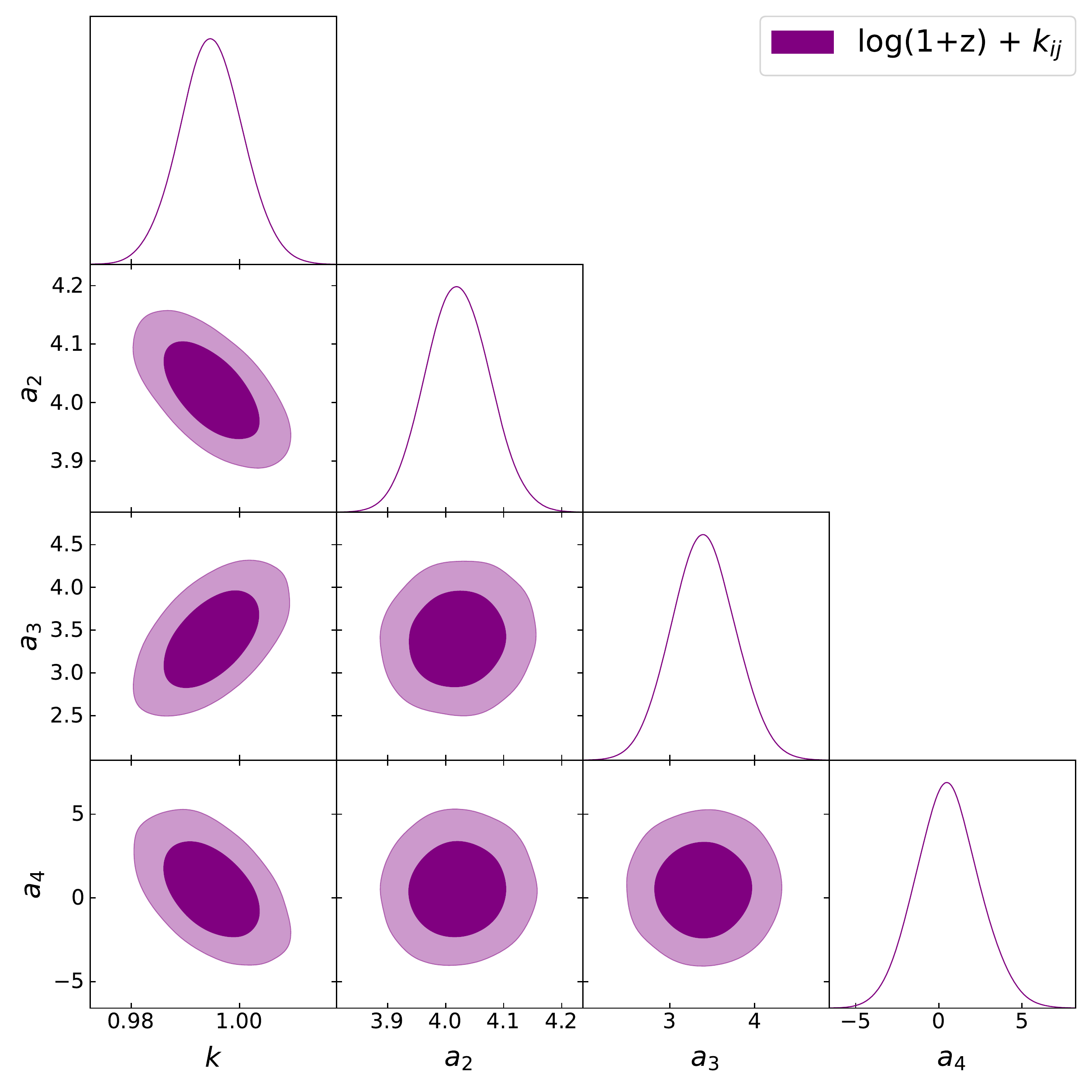}
                \caption{ Confidence contours ($1\sigma$ and $2\sigma$) for the cosmographic parameters space using the combined sample and the fourth-order expansion. }
                \label{F11}       
        \end{figure*}
        
\end{appendix}

\label{lastpage}
\end{document}